\documentclass[onecolumn]{emulateapj}
\usepackage{graphicx}
\usepackage{amsmath}
\newcommand{\bfig}{\begin{figure}}
\newcommand{\efig}{\end{figure}}

\date{December 11, 2006}
\shorttitle{3-Dimensional MHD Simulations of SNRs}
\shortauthors{Raley,  Shelton \& Plewa}
\begin{document}

\title{A Study of the Vertical Motion of Supernova Remnant Bubbles in the Interstellar Medium Drawn from the Results of Three-Dimensional MHD Simulations}

\author{E. A. Raley,\footnotemark[1]~  R. L. Shelton\footnotemark[1] ~\& T. Plewa\footnotemark[2]$^{~,}$\footnotemark[3]}
\affil{\footnotemark[1]~Department of Physics and Astronomy, 
       University of Georgia,
	Athens, GA 30602}
\affil{\footnotemark[2]~Center for Thermonuclear Flashes,  
       University of Chicago,5640 South Ellis Avenue,
	Chicago, IL 60637}
\affil{\footnotemark[3]~Department of Astronomy \& Astrophysics,  
       University of Chicago,5640 South Ellis Avenue,
	Chicago, IL 60637}
\email{raley@physast.uga.edu, rls@physast.uga.edu, tomek@uchicago.edu}

\begin{abstract}
In order to determine the circumstances under which isolated SNRs are capable of rising into and enriching the thick disk and galactic halo, simulations of supernova remnants are performed with the FLASH magnetohydrodynamic code.  
We performed simulations in which the interstellar magnetic field is parallel to or perpendicular to the galactic plane as well as a simulation without a magnetic field.  The ambient gas density distribution and gravitational potential are based on observations of our galaxy.  
We evolve the remnants to ages of roughly $10^7$~years.
For our simulation without a magnetic field, we compare the evolution of the hot bubble's velocity with the velocity evolution calculated from the buoyant and drag accelerations.  We found surprisingly small vertical velocities of the hot gas, from which we estimated the drag coefficient to be ten for the non-magnetic simulation.  Although we found little buoyant motion of the hot gas during the remnant's lifetime, we found rapid vertical motion of the associated cool dense gas near the end of the remnants life.  This motion deformed the remnant into a mushroom cloud structure similar to those found in previous simulations.  The simulation in which we have a 4 $\mu$G magnetic field parallel to the galactic mid-plane shows a dramatically elongated bubble parallel to the magnetic field.  The magnetic field pins the supernova remnant preventing it from rising.  
 In the simulation with the 4 $\mu$G magnetic field perpendicular to the midplane the hot bubble rises more, indicating that having the magnetic field in the same direction as the gravitational force enhances the rise of the bubble.

\end{abstract}

\newpage
\newpage
\newpage
\eject\eject
\section{Introduction}
High temperature gas exists far from the galactic mid-plane in the thick disk and halo.   \citet {s&f} first suggested that gas heated in the plane would rise to heat these upper regions of the galaxy.
One way that this is thought to happen is that stellar winds and supernovae from OB clusters form superbubbles (SBs) of hot gas.  Perhaps these superbubbles (SBs) could burst open spewing gas high into the galaxy (examined by \citet{tomisaka_ikeuchi}, \citet{MacLow_McCray}, and \citet{tomisaka98}) or create worm-like structures funneling gas into the thick disk and halo (simulated by \citet{avillez_berry}).  
Preferential expansion could also push hot gas upward since the galaxy's ambient density decreases with height above the midplane making it is easier for hot gas to move away from the midplane.   \cite{ferriere95,ferriere98_2} has analyzed the effects of preferential expansion on SBs and also on single supernova remnants (SNRs). 

Isolated SNRs differ from SBs in that they possess much less energy since their only source is a single supernova explosion.  Because their energy input is a one time occurrence, they are not pinned to the source of their energy in the way that a superbubble is pinned to the OB cluster that continuously feeds it energy.  Thus, it may be possible for SNRs to move vertically.  \cite{jones} looked at buoyancy of individual SNRs, as a mechanism for explaining horseshoe-like structures seen in observations.  
Motivated by observations of a mushroom-like structure \cite{English} simulated azimuthally symmetric single supernova remnants in a non-magnetic ambient medium, finding a mushroom-shaped structure at 8 Myrs in the cool dense region outside of the bubble.  \cite{avillez_maclow} confirmed their results by finding mushroom-shaped clouds in their late time non-magnetized simulations of a galactic disk incurring many SN explosion.  It was thought that the mushroom shaped clouds were shaped by buoyant forces.

The purpose of this paper is to explore whether isolated SNRs rise out of the thick disk to enrich the halo.  This phenomenon has been analyzed by \citet{shelton2006} who combined results from a one dimensional hydrodynamic SNR simulation with analytical calculations to predict the buoyancy of these remnants.  However, the frictional drag constant was not known a priori, preventing her from unequivocally predicting the SNRs' rise due to buoyancy. \cite{jones,ferriere98,avillez_maclow} simulated multiple-dimensional SNRs without a magnetic field. Here we add to their studies by adding a magnetic field and tracking the SNRs rise with respect to time. The galactic field may line up along the spiral arms of a galaxy \citep{Zweibel and Heiles}, or be distorted by SN explosions and stellar winds such that the field lines bend around the bubbles or superbubbles \citet{heiles}.  
In order to accommodate the variation in magnetic field geometries and in order to compare with previous work, we analyze three simulations: one without a magnetic field, one with a magnetic field parallel to the galactic midplane, and one with a magnetic field perpendicular to the midplane.

  We use the FLASH magnetohydrodynamics code to simulate SNRs in an ambient background representative of the ISM in the vicinity of the solar circle, i.e. with a stratified density and gravitational potential with respect to height above the galactic midplane.  In each case we set off a SN explosion at 400~pc above the plane and follow the evolution of the bubble for 8 to 12~Myrs.  For the simulation without a magnetic field, we evolve the SNR until it dissipates.

Section 2 discusses the computational method and physical parameters.  
Section 3 describes the results of the simulations for our three cases, no magnetic field (subsection 3.1), a 4 $\mu$G parallel to the galactic mid-plane (subsection 3.2.1), and a 4 $\mu$G magnetic field perpendicular to the galactic mid-plane (subsection 3.2.2).  Section 4 discusses our conclusions about buoyancy of the hot gas, the vertical motion of cool gas, and SNR morphology. In section 4 we also discuss why galactic shear would not have much of an effect on such bubbles.
\section{Numerical Methods}
Version 2.5 of
the FLASH magneto-hydrodynamics code was used for this project.  
FLASH simulates magneto-hydrodynamic flows in multiple dimensions on an 
Eulerian grid.  For non-magnetic flows it uses the Piecewise Parabolic Method 
(PPM, \citet{c&w}), and for magnetic flows it uses (by default) a Roe-type solver \citep{powell}.   The code uses dynamic
Adaptive Mesh Refinement (AMR) techniques and is
parallelized.
The FLASH code is described further in \citet{FLASH}.

FLASH version 2.5 does not include full radiative transfer that is compatible with the MHD modules,
but does include a radiative cooling algorithm.
We replaced the default radiative cooling algorithm  
with a more temperature-sensitive
algorithm derived from the \citet{Gaetz} tables. 
In order to hold the undisturbed medium within the simulation grid at a
constant temperature, we universally prevented radiative 
cooling in cells whose 
temperatures were less than $150\%$ of the cell's initial temperature.
This constraint also prevented disturbed gas from 
radiatively cooling to a temperature far below its initial temperature.
Our constraint did not prevent adiabatic cooling.    

The simulation grid was oriented with the $\hat{z}$ direction 
perpendicular to the galactic midplane and
the $\hat{x}$ and $\hat{y}$ directions parallel to the midplane.  In our simulations with a magnetic field, Model B has the 
magnetic field along the $\hat{y}$-direction such that the three directions are unique; the gravitational potential is in the $\hat{z}$-direction, the magnetic field is in the $\hat{y}$-direction, and the $\hat{x}$-direction has neither a gravitational acceleration, nor a magnetic field parallel to it.  Model C has the magnetic field along the $\hat{z}$-direction such that the forces in the $\hat{x}$ and $\hat{y}$ direction are not qualitatively different; the gravitational acceleration and magnetic field are both along the $\hat{z}$-direction.
For the simulations presented in this paper, we placed the supernova
explosion at $(x,y,z) = (0,0,400$~pc), thus at the intersection of x=0 (y-z plane) and y=0 (x-z plane) and midway up the $\hat{z}$ axis within the simulation grid.  The SNRs are symmetric across x=0 (y-z plane) and y=0 (x-z plane).  Therefore, it is possible to determine the properties of an entire remnant from a simulation of only one fourth of the remnant.  (See Section 2.1.)   
For the purpose of describing the boundary conditions required for a partial remnant simulation, we define the inner $x$ boundary as the boundary with the smallest constant $x$ value, and outer $x$ as the boundary with the largest constant $x$ value.  The inner $y$, outer $y$, inner $z$, and outer $z$ boundaries are defined analogously.
By setting the explosion in this location and setting the boundary
conditions along the inner $x$ and $y$ planes to simulate reflection, we were able to simulate quarter
remnants and hence reduce the needed CPU time and memory by 75\%.
As a check on the accuracy of this technique, we also simulated 
full remnants to several hundred thousand years, which we found to have indistinguishable temperature, density,
and pressure structures
as those of the quarter remnants.

As for the other boundary conditions, we used
fixed boundary conditions in the $\hat{z}$ direction and
outflow boundary conditions in the $\hat{x}$ and $\hat{y}$ directions
at the far $x$ and $y$ sides.  By fixed boundary conditions we mean that we fix the values at the boundary such that they are constant in time and sufficient for the ambient medium to be in hydrostatic equilibrium.  
Outflow boundary conditions in the $\hat{z}$ directions would allow the remnant of the shock to leave the grid.  However, it is difficult to maintain HSE at the boundaries of the simulation grid under normal circumstances (i.e. without a shock) when the boundary conditions are set to ``outflow''.  In that case, ambient material leaks off the grid, and this disturbs HSE.  For this reason we have used fixed boundary conditions at the upper and lower $\hat{z}$ boundaries.

If the whole grids were at the finest level of refinement, the size of the AMR grids would be 256$\times$256$\times$512 cells for the simulation with no magnetic field and 256$\times$384$\times$384 cells for the simulation with the magnetic field parallel to the midplane and 160$\times$160$\times$800 cells for the simulation with the magnetic field perpendicular to the midplane.
Where the density or pressure gradients were large (greater than 10$\%$ difference in density, or 33$\%$ in pressure between adjacent cells),the AMR subdivided cells down.  It will subdivide cells to volumes of $\stackrel{<}{\sim} 1.25^3$~pc$^{3}$. For the simulation with no magnetic field, we divide the $\hat{y}$ and $\hat{z}$ directions into 3 blocks and divide the $\hat{x}$ direction into 2 blocks (a total of 18 blocks); we then allow 5 levels of refinement for each block, and 8 cells per sub-block.  For the simulation with a magnetic field parallel to the midplane, we divide the $\hat{z}$ direction into 2 blocks; within each of these blocks we allow 6 levels of refinement, and 8 cells per sub-block.   For the simulation with a magnetic field perpendicular to the midplane, we divide the $\hat{z}$ direction into 5 blocks, and allow 5 levels of refinement in each block; within each refined sub-block we have 10 cells.  This yields a resolution (at the highest level of refinement) of approximately 1.25 $pc$ in each direction.
\cite{avillez_breit} showed that such resolution is necessary and sufficient for models of supernova heated gas in the interstellar medium.
We compare our FLASH2.5 simulation with no magnetic field with an earlier simulation we did with ZEUS-MP \citep{zeusmp,zeus2d_1,zeus2d_2,zeus2d_3} 
with no magnetic field and an explosion height of 400 pc.  Our ZEUS-MP simulation had a density gradient that was a function of our gravitational potential and an isothermal background at 2.3$\times 10^4$ K.  Despite these differences, we still see a mushroom shaped cloud at 8 Myrs, and we see a buoyant rise of the SNR of less than 60 pc.

In each of the three magnetic field cases, the remnant is symmetric about two planes, the x-z plane bisecting the bubble and the y-z plane bisecting the bubble.  The remnant is not symmetric about the x-y plane bisecting the bubble due to the gravitational force and density gradients in the $\hat{z}$-direction.  Because of the existing symmetry, we need only simulate a quarter of the remnant. 

Our quarter remnant simulations required setting up the code so that it would simulate reflection across both the x-z plane (also called the y=0 plane) and the y-z plane (also called the x=0 plane).  Both ZEUS-MP (Version 1) and FLASH (Version 2.5) include a reflective boundary algorithm that works 'out of the box' for simulations in which $B=0$.  However if $B\neq 0$, since the boundary conditions for $\bf{B}$ are a function of the direction of the initial field, the user must choose how to set up the boundary conditions.  In ZEUS-MP, boundary conditions for $\bf{B}$ are set indirectly by setting the boundary conditions for electromotive force.  There are two options for the boundary conditions, one corresponding to a magnetic field that is perpendicular to the boundary and one corresponding to a magnetic field that is parallel to the boundary \citep{zeus2d_2}.  For FLASH2.5, the user must define the boundary conditions for the magnetic field.
These magnetic boundary conditions are non-intuitive so we describe them in the appendix for the most general case, that of having symmetry across the x-z, y-z, and x-y planes.

\subsection{Initial Conditions} 

\begin{figure}
\centering
\includegraphics[height=0.3\textwidth, angle=0]{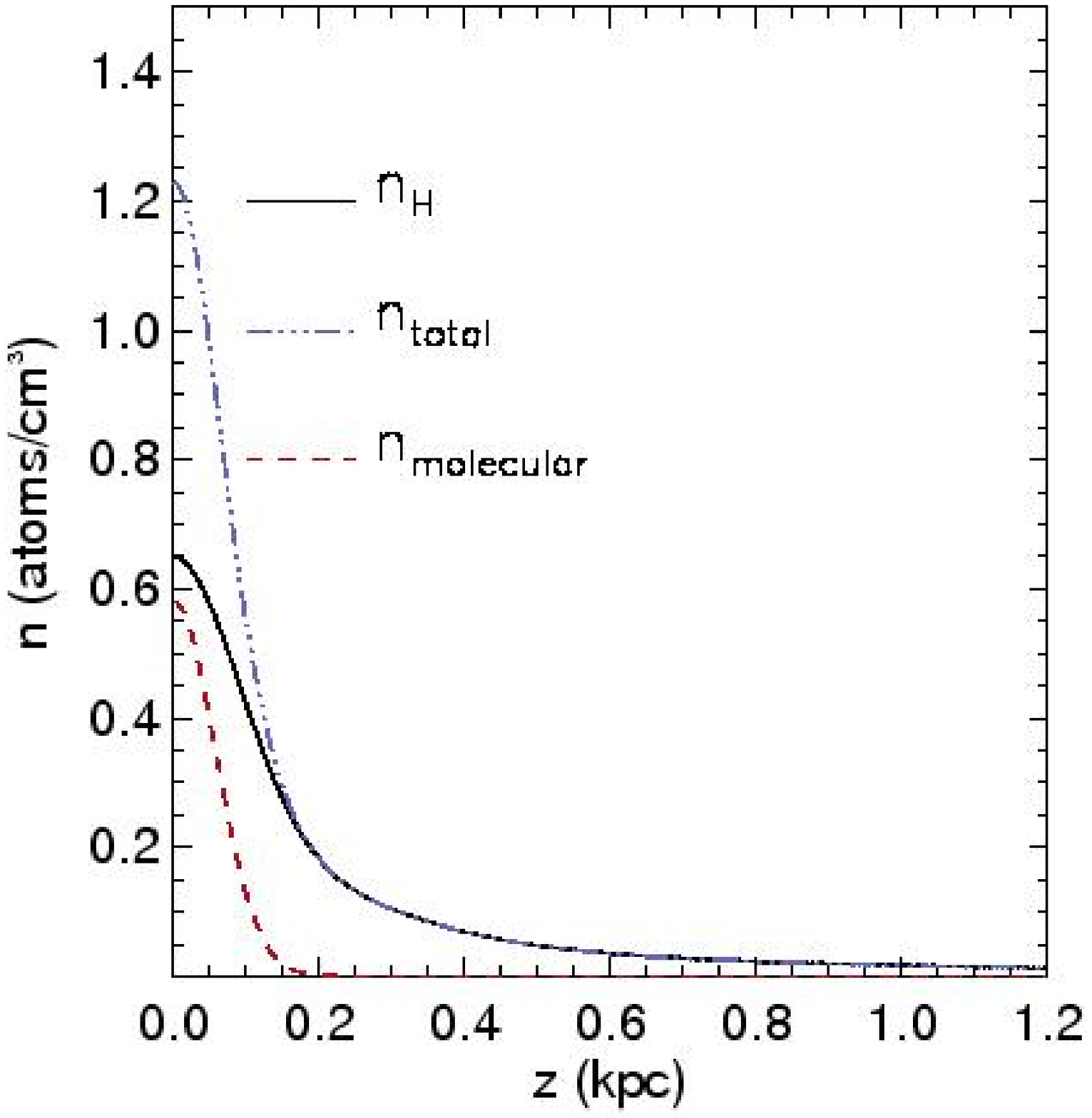}
\includegraphics[height=0.3\textwidth, angle=0]{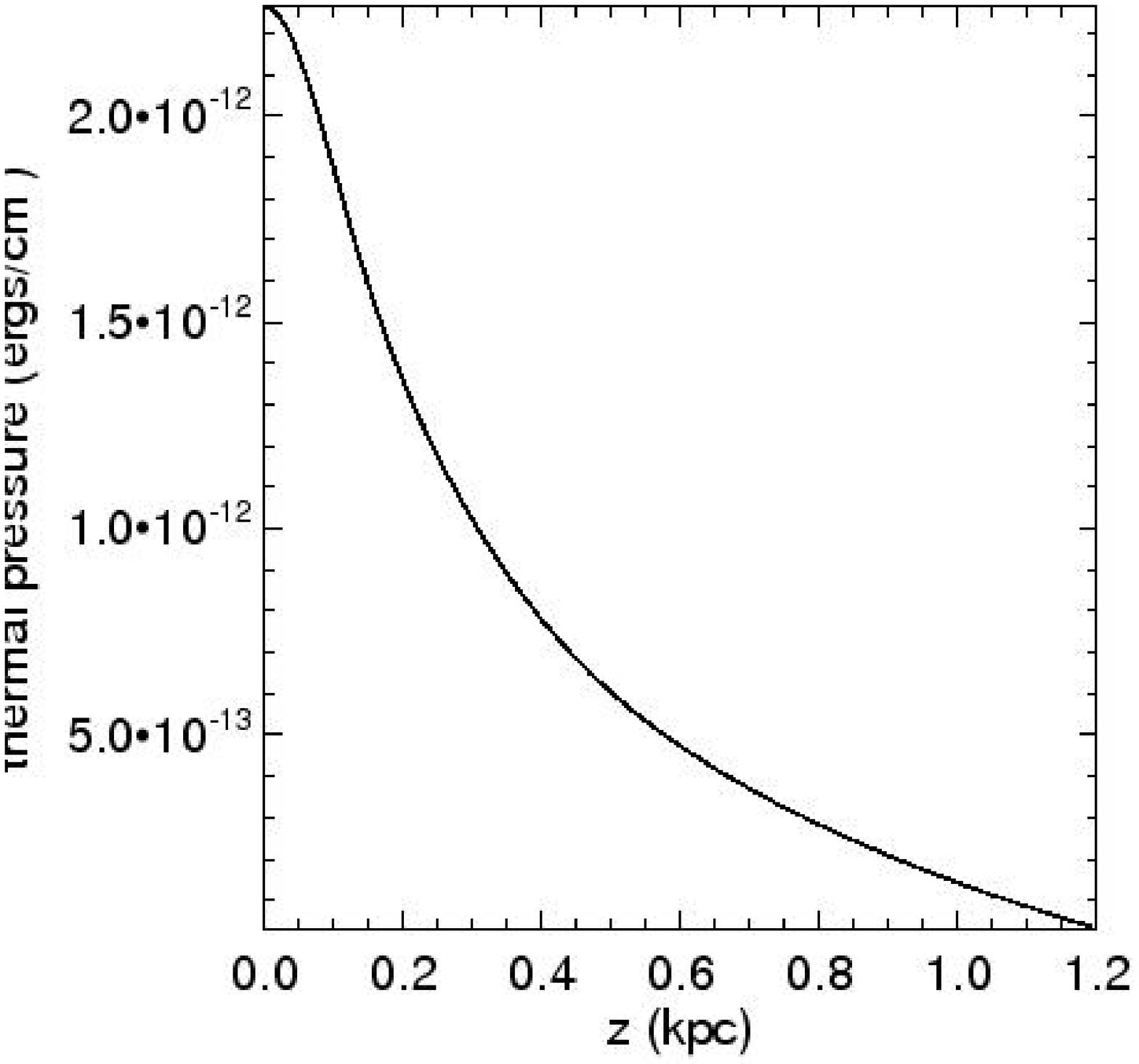}
\includegraphics[height=0.3\textwidth, angle=0]{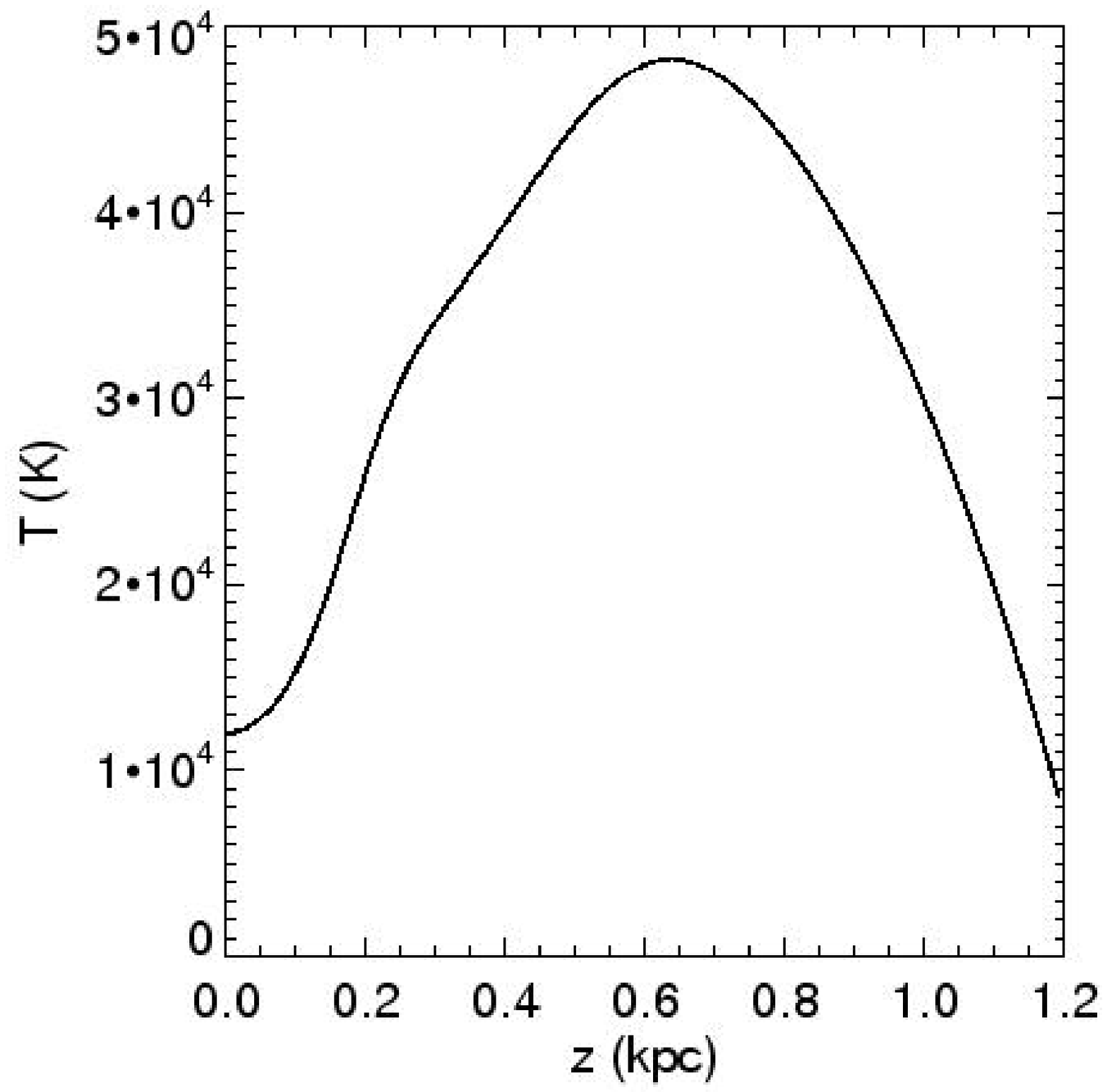}
\caption{(left) Density of non-molecular hydrogen, $n_H$, as a function of height above the midplane \citep{ferriere98}.  Also shown are the molecular and total (i.e. molecular plus non-molecular hydrogen) densities.  (center) Thermal pressure as a function of height above the midplane before the SN explosion. (right) Temperature versus distance above the galactic midplane before the SN explosion. }
\label{ambient}
\end{figure}

The modeled region of the galaxy is located at the Sun's galactocentric radius, $R_{\odot}$, and within 1~kpc of the galactic midplane.  In this region the gravitational acceleration in the vertical direction is:
\begin{equation}
-g(R_z)=(4.4\times 10^{-9} {\rm cm~s^{-2}})\frac{z}{\sqrt{z^2+(0.2~{\rm kpc})^2}} +(1.7\times 10^{-9}{\rm~cm~s^{-2}})\left(\frac{z}{1~{\rm kpc}} \right)
\label{gravity}
\end{equation}
\citep{ferriere98}.

The density of non-molecular hydrogen in the interstellar medium, $n_H(z)$, includes contributions from the cold neutral medium ($n_c$), the warm neutral medium ($n_w$), the warm ionized medium ($n_i$), and the hot ionized medium ($n_h$).  We neglect molecular gas, because it resides near the midplane, well below our region of interest.  We adopt Ferriere's formulas for the hydrogen gas density distributions, resulting in
\begin{eqnarray}
n_H(z)  =  n_c(z)+n_w(z)+n_i(z)+n_h(z)
 =  0.395~e^\frac{-z^2}{\left(127{\rm pc}\right)^2} + 0.107~e^\frac{-z^2}{{\left(318{\rm pc}\right)}^2} 
\\
+ 0.064~e^\frac{-z}{403{\rm pc}} + 0.0237~e^\frac{-z}{1000{\rm pc}} 
+ 0.0013~e^\frac{-z}{150{\rm pc}} + 0.00048~e^\frac{-z}{1500{\rm pc}}.
\end{eqnarray}
Figure \ref{ambient} (left) shows the density of non-molecular hydrogen for our ambient background as a function of height above the galactic midplane.
Accounting for the 1 to 10 abundances of helium relative to hydrogen in the interstellar medium, the density of atoms is $n_a(z)={\frac{11}{10}}~n_H(z)$.
The large flux of ultraviolet photons emitted by SNRs is thought to pre-ionize the surrounding gas.  Thus we take the ambient medium to be fully ionized and the density of particles to be ${\frac{23}{10}}~n_H(z)$.

We assume that the simulated interstellar gas is at rest and in Hydrostatic Equilibrium (HSE) before the SN explosion.  In other words, the upward forces due to thermal and magnetic pressure gradients exactly balance the downward force due to gravity.  Since we have set the initial magnetic field to be constant with respect to $\hat{z}$, the only pressure gradient is that due to thermal pressure.  By solving the hydrostatic equilibrium equation with $n_H(z)$ and $g(z)$ as constraints, we are able to determine the thermal pressure gradient and therefore the temperature gradient.

   It must be noted however that the real ISM is not in HSE, so the dependent variable in these calculations has constraints not found in nature.  Because the interstellar temperature distribution is not well known, isothermal simulations are often done.  
If we choose isothermal temperature and gravity as constraints the density distribution needed for HSE has a much different profile than Ferriere's equations yield.  Because the density distribution is better known and we believe more important to the evolution of the bubble, we choose density and gravity as our constraints.  We set the temperature distribution such that the thermal pressure gradient balances the gravitational force.  The thermal pressure and temperature profiles are shown in figures \ref{ambient} (center and right).  Note that including non-thermal pressure gradients would give a flatter and probably more physical temperature gradient.  This would involve including a stratified magnetic field and including cosmic rays in the calculations, which we have not included here.

This procedure establishes the functional form of the pressure distribution but not the value of the pressure at any given height.  However, some choices of midplane pressure result in unphysical temperatures within the simulation region.  To avoid unphysical temperatures within the simulation region, we set the galactic mid-plane pressure such that the temperature at the mid-plane is 12,000 K.  The thermal pressure and temperature as a function of height above the midplane are shown in Figures \ref {ambient} (center and right).

According to \citet{ferriere98}, the magnetic field strength decreases mildly with height at the height of our explosions.  It drops from 4.25 $\mu$G to 4.05 $\mu$G between heights of 300 pc and 500 pc. (This is the region in which our bubble resides throughout its lifetime.)  If we were to model the magnetic field gradient, in the two simulations having a magnetic field (Models B and C), then our equation for hydrostatic equilibrium would have to be adjusted, hence our thermal pressure gradient would have to be adjusted for those two cases.  It is preferable to use a single thermal pressure distribution for all of our simulations (so that they may be compared with each other) and to sacrifice the magnetic field gradient.  Thus we use a constant magnetic field of 4$\mu$G, parallel to the midplane for Model B, and perpendicular to the plane for Model C.

For our SN explosion we create an over-pressurized region with radius of 10~pc at 400~pc above the galactic midplane.  The thermal energy in the overpressurized region is $1.0\times10^{51}$~ergs more than in an equivalent volume in the ambient gas.  This is how we simulate an explosion with an energy of $1.0\times 10^{51}$~ergs.

\section{Results}

\subsection{Model A: No Magnetic Field}
We tracked Model A, the remnant evolving in a non-magnetized ambient medium, until all that remained was a faint toroidal structure having a temperature T of less than $3\times 10^6$~K at 12 million years.  The remnant's temperature and density structure are displayed in Figures \ref{mushroom2} and \ref{mushroom1}. The entire SNR is hot and rarefied during its early evolution, as can be seen in  the top left panels of Figures \ref{mushroom2} and \ref{mushroom1}.  But, by 300,000 years, the gas behind the shock has cooled to $4.7\times10^4$~K.  The shock front expands faster than the hot interior causing the gap filled by cool material to expand and the gas to decompress.  By five million years, the previously dense and narrow ``cool shell'' has been extended from the hot bubble (r=~95~pc) to the shock front (r=~260~pc) and its density has dropped to about 1.1 times that of the ambient medium.  

Also by 5 Myr, the shock front has become so weak  that when it collides with the ambient medium, it can only heat the gas to $5.2\times10^4$~K, which is not much higher than the temperature of the background.  At such a low temperature, the recently shocked gas quickly cools by radiation down to the ambient temperature.  Because the shock front no longer contributes heat to the hot bubble and the recently shocked material is no longer sufficiently dense or unified as to be easily observed, the shock front is now irrelevant.  In the future, it will slow to the sound speed and so evolve into a sound wave.  In the Milky Way, such shocks and sound waves contribute to the turbulent motion of the interstellar medium.  In our simulation the shock wave degrades into a sound wave before it 
reaches the grid's outer boundaries.  The sound wave bounces off of the fixed z-boundary at the top of the grid at around six million years and reflects off the fixed z-boundary at the bottom of the grid at about eight million years. It is allowed to propagate through the outer x and y boundaries at x= 315~pc and y= 315~pc, respectively.  Neither of the reflected sound waves come back to hit the bubble within the simulation time.  If they had reached the bubble within the simulation time, they would be too weak to significantly disturb the bubble.

The simulated remnant begins life with a nearly spherically symmetric shape.  The shock front remains so throughout the remnant's life, but the hot interior develops structure before it is 300,000 years old (see second panels of Figures \ref{mushroom2}, and \ref{mushroom1}).  Eddies start to form in the hot bubble by 300,000 years and by 600,000 years they have started to form a cauliflower-like structure within the bubble. The velocity of the low density material in these eddies is mostly circular and often goes against the radial direction of mean fluid flow.  The eddies strengthen throughout the first three million years, until which point the lowest density material has been mixed into the rest of the bubble by these eddies, so that the eddies die down.  The exterior of the bubble (and hence the interior of the cool shell) stays spherically symmetric until about 5 million years, when the bubble cools and shrinks to the vicinity of the waning eddies.  (There are two reasons the bubble shrinks: (a) the edges cool and are then no longer part of the bubble, and (2)the interior cools and thus has less thermal pressure; when it has less thermal pressure than the surrounding material, it will shrink by compression.) At this time the outer edge of the bubble takes on the cauliflower-like structure of the eddies.  The inner edge of the cool shell coincides with the outer edge of the bubble, giving it the same structure.  

The dense region surrounding the bubble develops into a mushroom shaped structure by the age of nine million years (Figure \ref{mushroom1}).  The cap of the mushroom is outlined by a cool dense shell and is filled with the hot rarefied gas that makes up the bubble.  The faint stem is made up of dense, coolish material below the bubble.  At this time, the bubble is vertically rising, as can be seen in the positive velocity in the z-velocity plot (Figure \ref{mushroom_vel}).  Earlier events preconditioned the bubble to rise. 
Through radiative cooling, the bubble and the cool region around it had become depressurized by 2 million years.  Over the next several million years they progressively become more depressurized.  Starting by 2 Myrs, the material below the bubble and above the shock (the lower part of the cool shell) has developed a pressure gradient such that the pressure decreased with distance from the midplane much more steeply than the initial pressure gradient for the ambient background in HSE.  In other words $|\frac{dp_{th}}{dz}| > |mg|$.   Hence, the cool material below the bubble feels an upward force.  Having initially been pushed downward by the shock, this material slows, stops and then rebounds.  By 2 Myrs a significant portion of it is moving upwards.  This material accelerates over the next several million years and by about 7 million years, relatively high pressure has built up in the material immediately below the bubble.  Between 7 and 9 Myrs this region has grown significantly into a high pressure, high density 'stem' below the bubble.  By 9 million years the 'stem' has begun to rise into the bubble carving the bubble into a 'mushroom cap'.  By 10 million years the mushroom shaped structure is more developed.  By 11 million years the stem has risen even further into the bubble reforming it into a torus of hot ($\sim$ 3 million Kelvin) gas by 12 million years.   

Surprisingly the bubble itself rises very little during the course of its life.    There are three reasons that we would expect to see the bubble rise vertically: (1) buoyancy; the hot bubble is much less dense than the surrounding gas and should therefore rise, (2) the ambient medium has a pressure gradient and it is easier for the shock front to expand in the direction of decreasing pressure, which is the direction away from the midplane, (3) the ambient medium has a density gradient, so the shock has the least material to push through when in travels away from the midplane.  The density gradient can also cause the bubble to be denser closer to the midplane which will make it cool much faster in that region.
The pressure and density gradients would cause the bubble to become larger above the explosion height than below it.  The vertical rise due to buoyancy can be counteracted early on by the weight of the cool shell, and at all times by frictional drag.  Although material as high as 130~pc above the initial explosion has been heated, the bubble itself, rose very little in our simulation.  In the course of 12 Myrs, the center of our simulated bubble moved, at most, 59 pc from the explosion height of 400~pc, as can be seen in the bottom right panels of Figures \ref{mushroom2} and Figures \ref{mushroom1}.  In section 4, we calculate the expected velocities due to buoyancy and drag and compare them to the average velocities of the simulated bubble.

\bfig
\centering

\includegraphics[height=0.35\textwidth, angle=0]{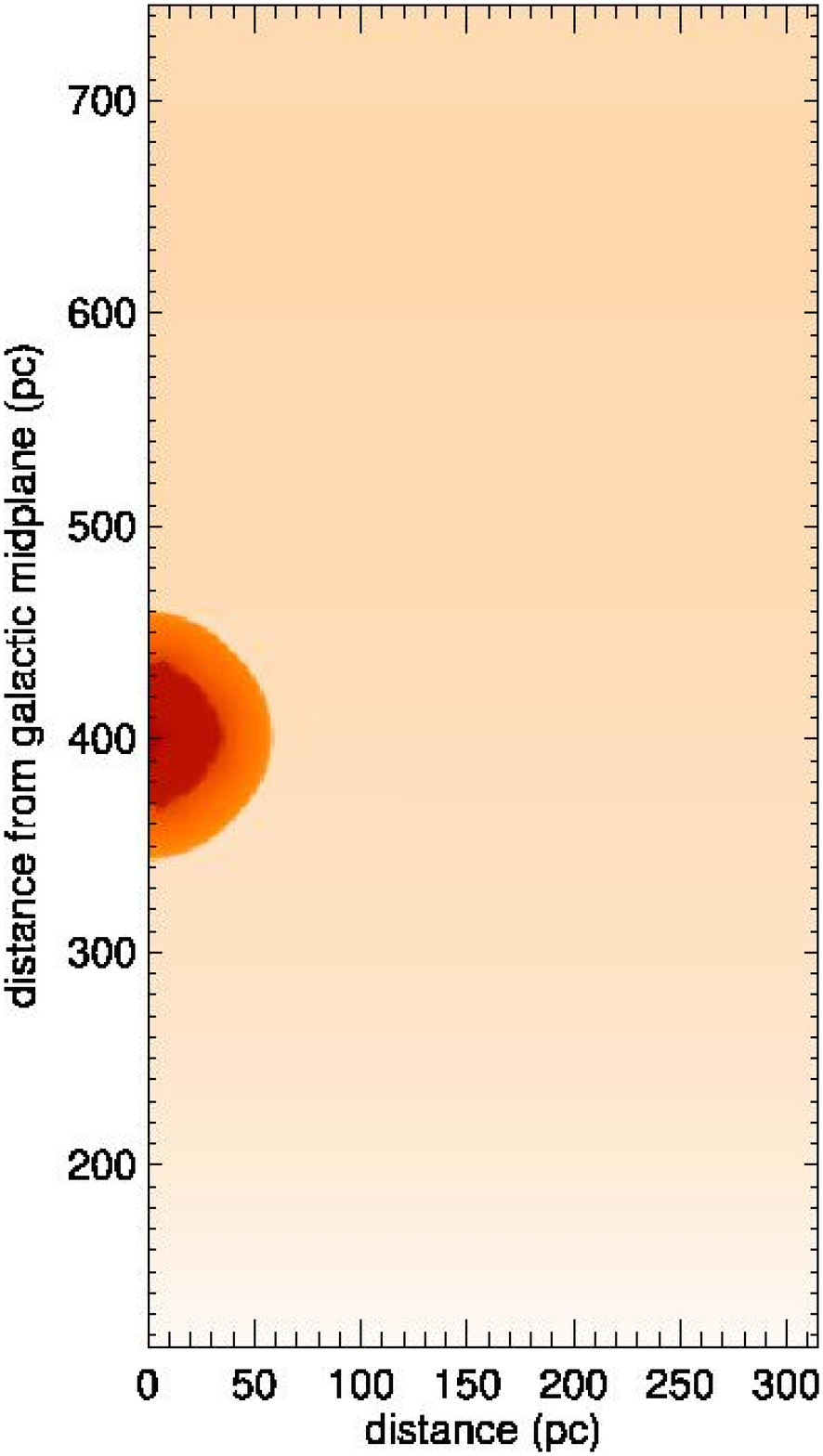}\hspace{0.5mm}\includegraphics[height=0.35\textwidth, angle=0]{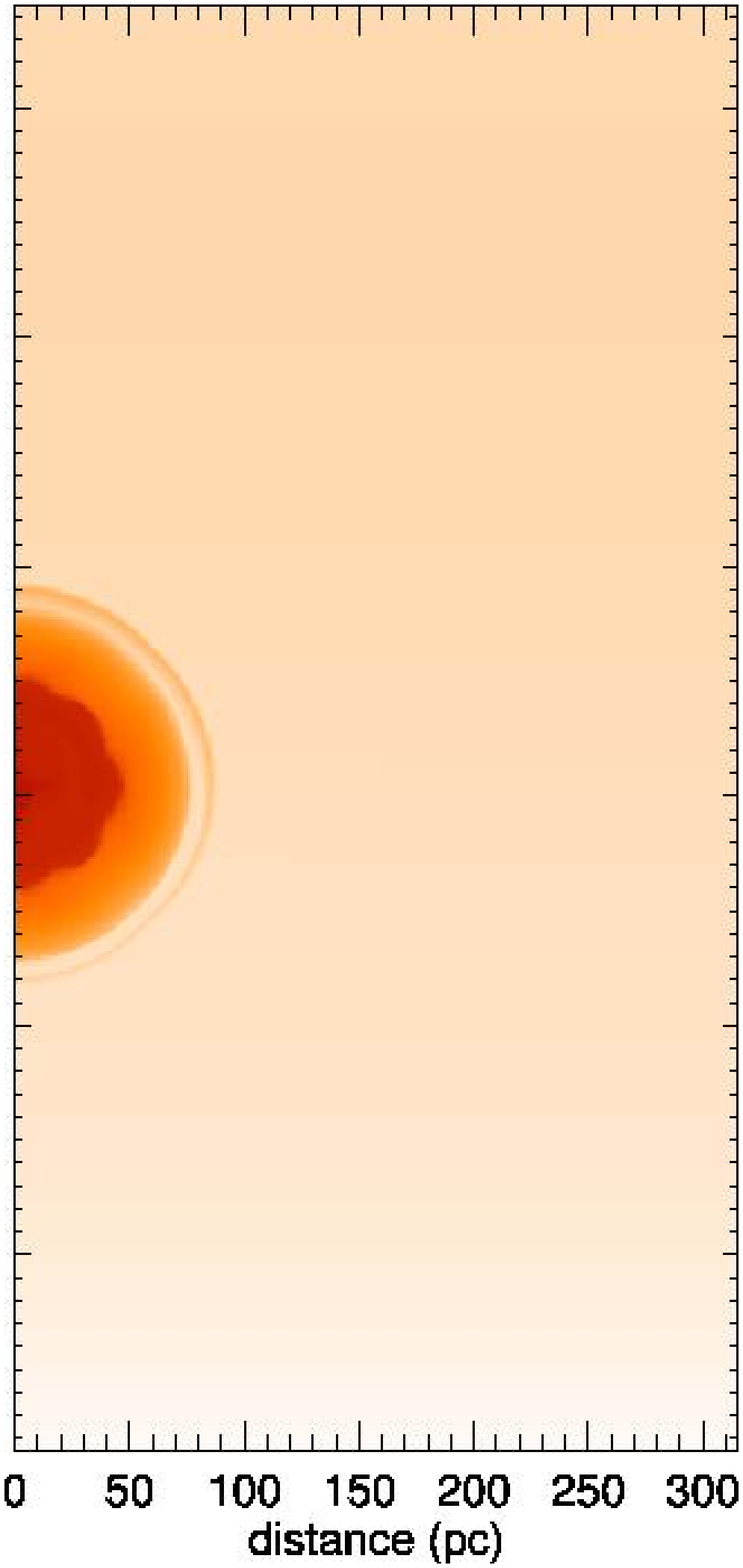}\hspace{0.5mm}\includegraphics[height=0.35\textwidth, angle=0]{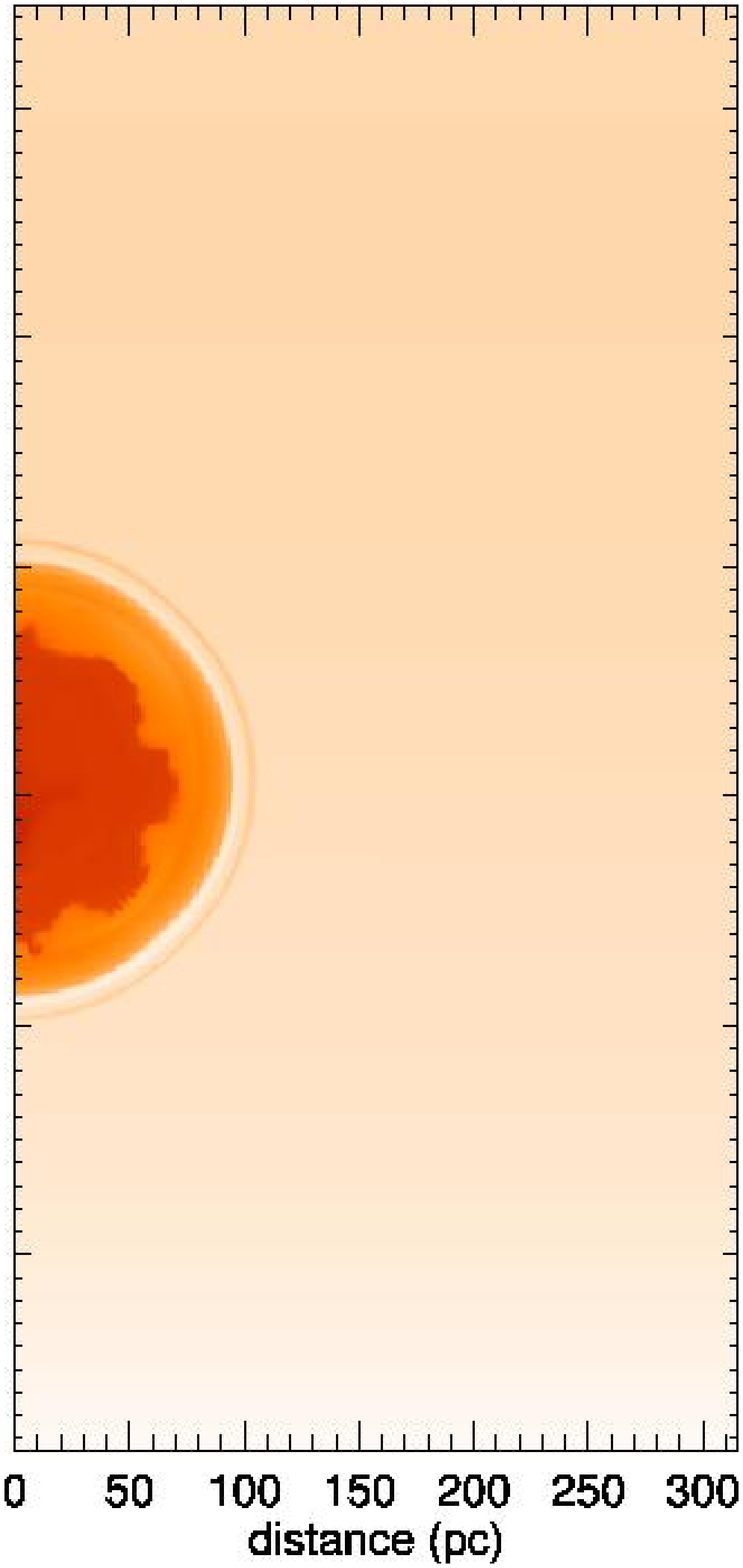}\hspace{0.5mm}\includegraphics[height=0.35\textwidth, angle=0]{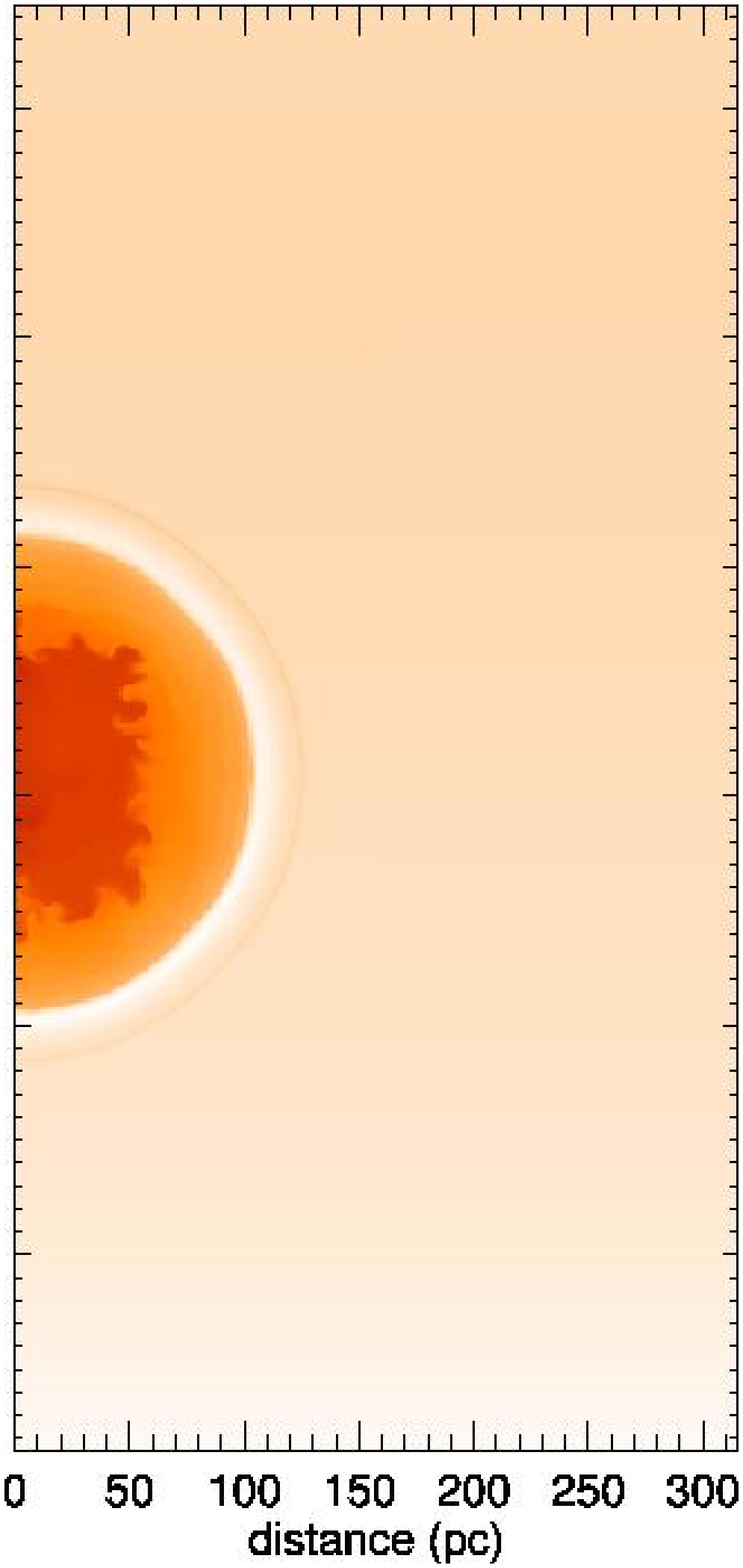}\hspace{0.5mm}\includegraphics[height=0.35\textwidth, angle=0]{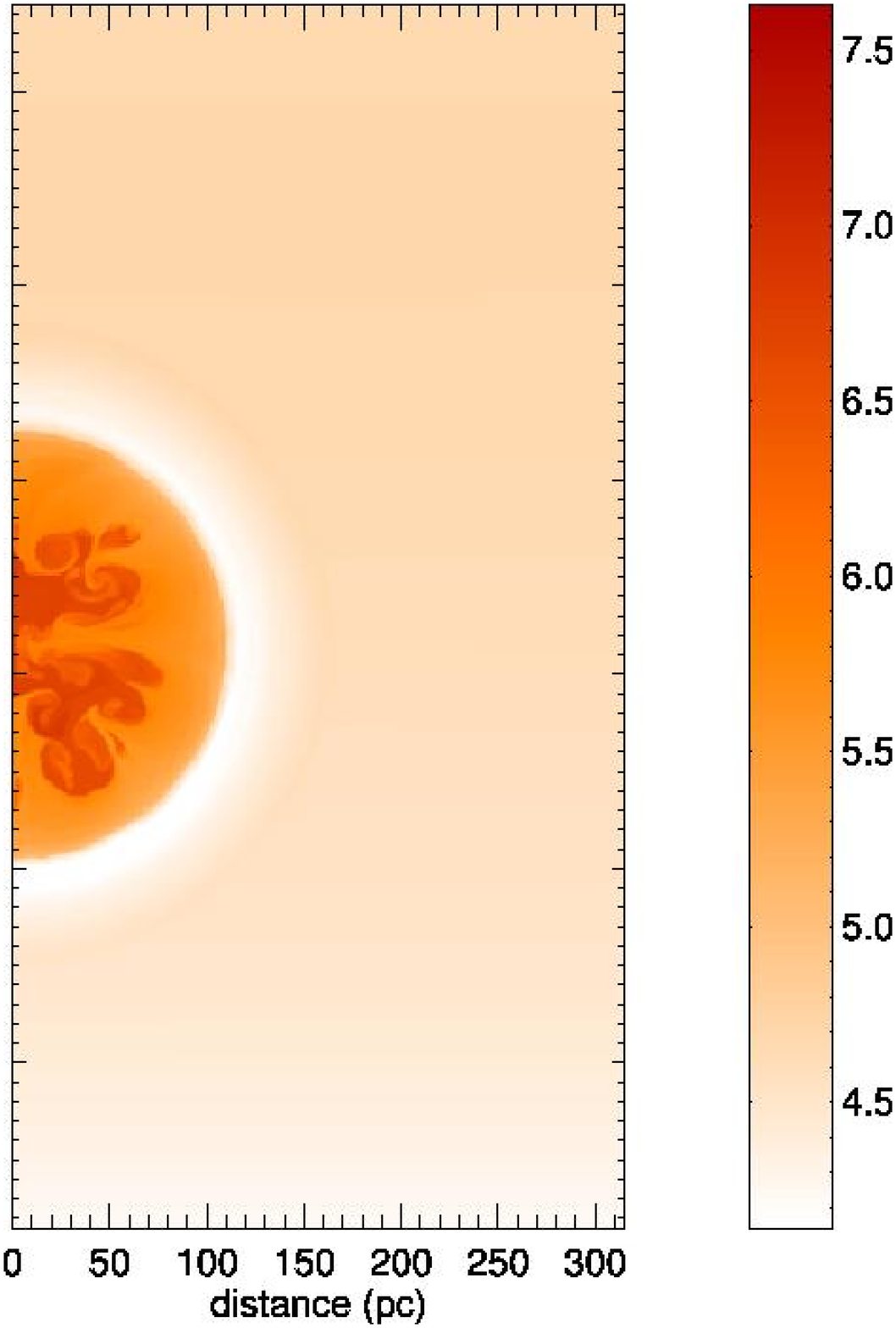}
\vspace{1.5mm}

\includegraphics[height=0.35\textwidth, angle=0]{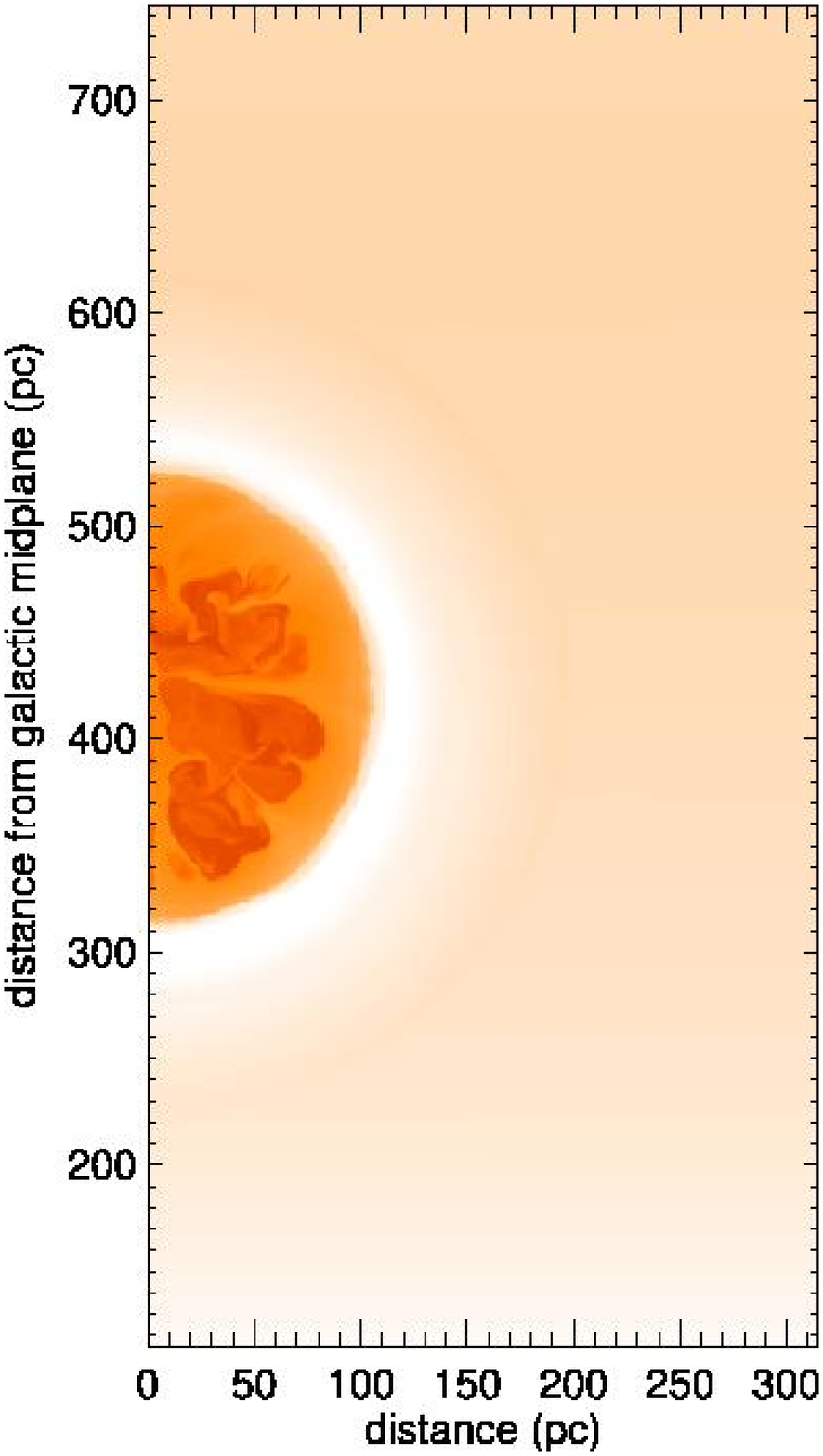}\hspace{0.5mm}\includegraphics[height=0.35\textwidth, angle=0]{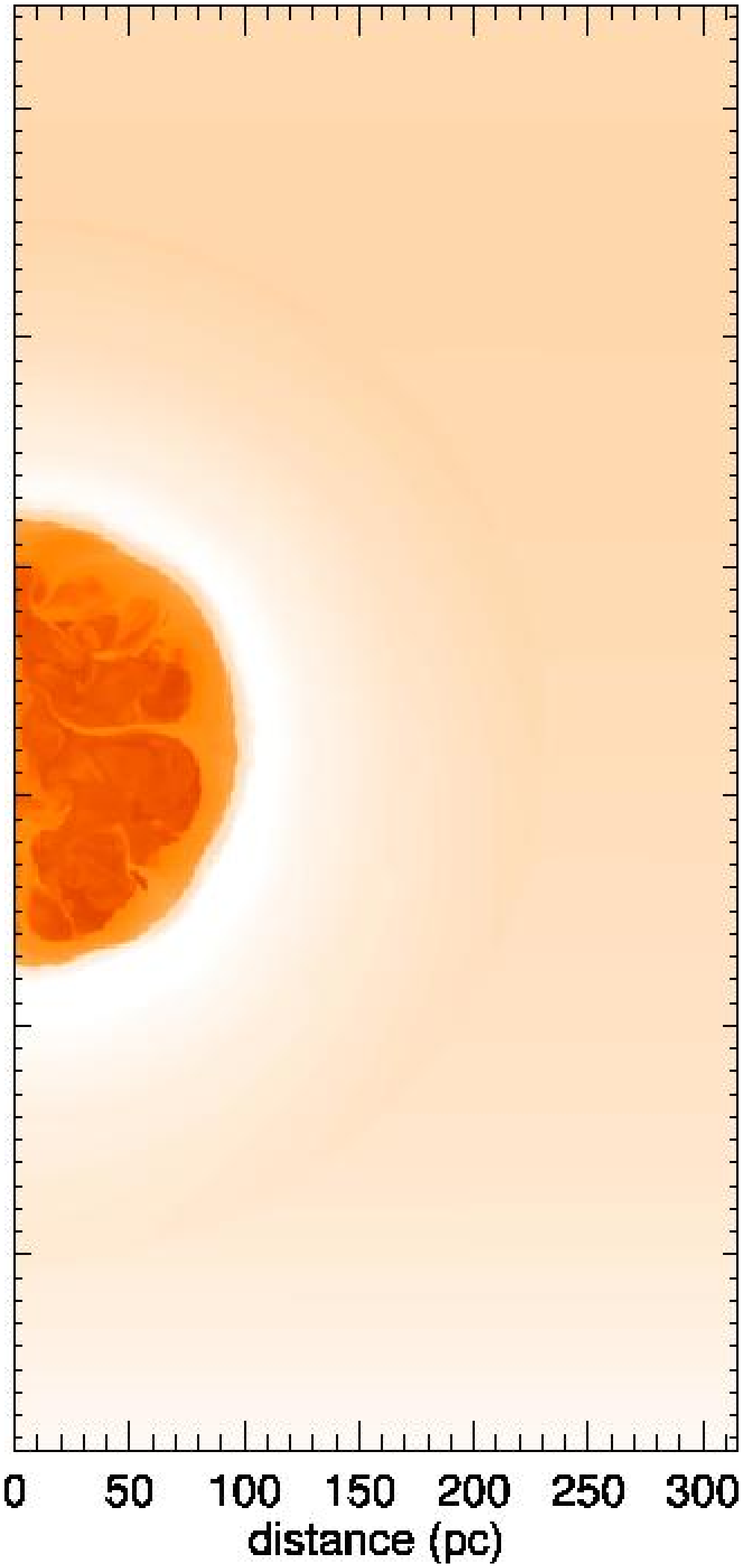}\hspace{0.5mm}\includegraphics[height=0.35\textwidth, angle=0]{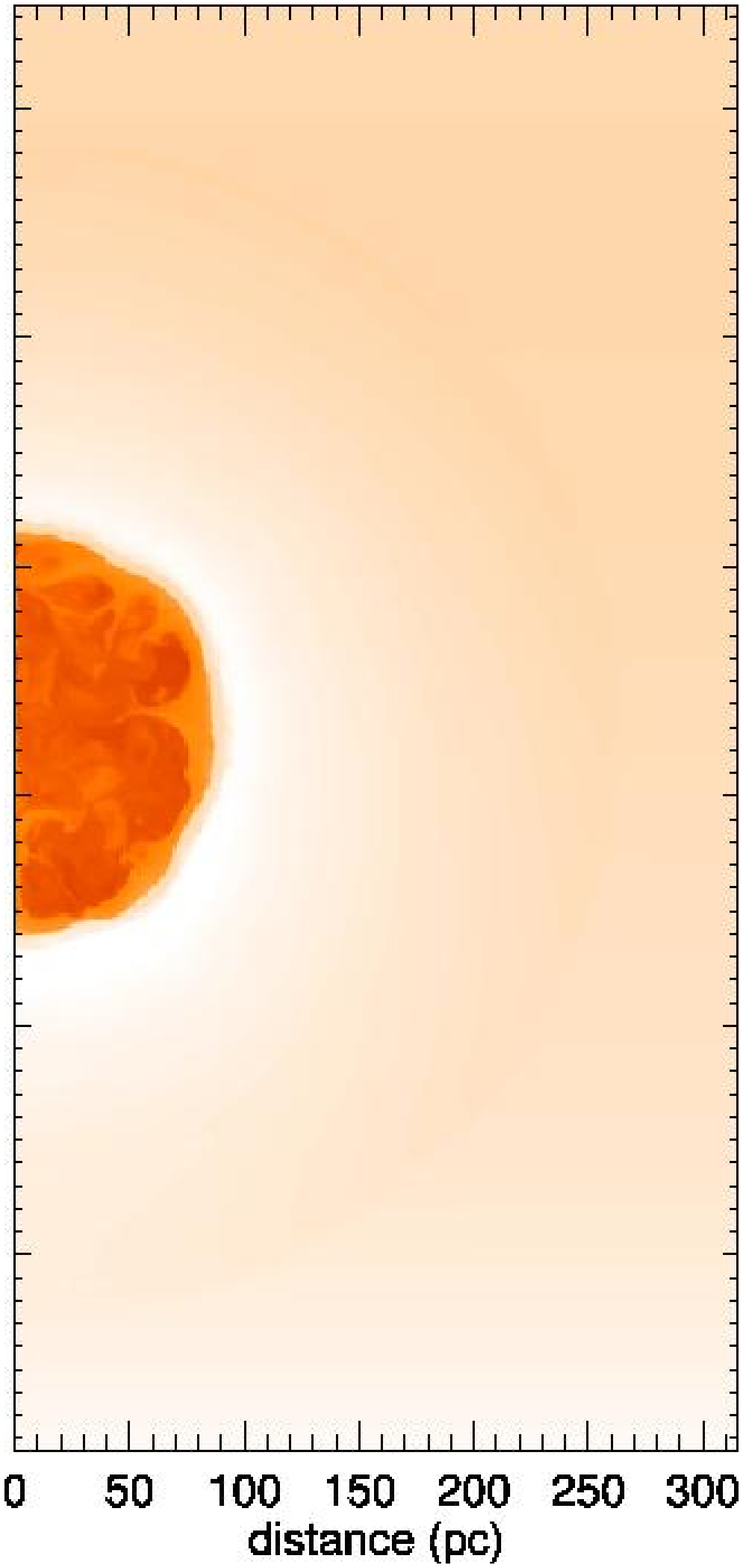}\hspace{0.5mm}\includegraphics[height=0.35\textwidth, angle=0]{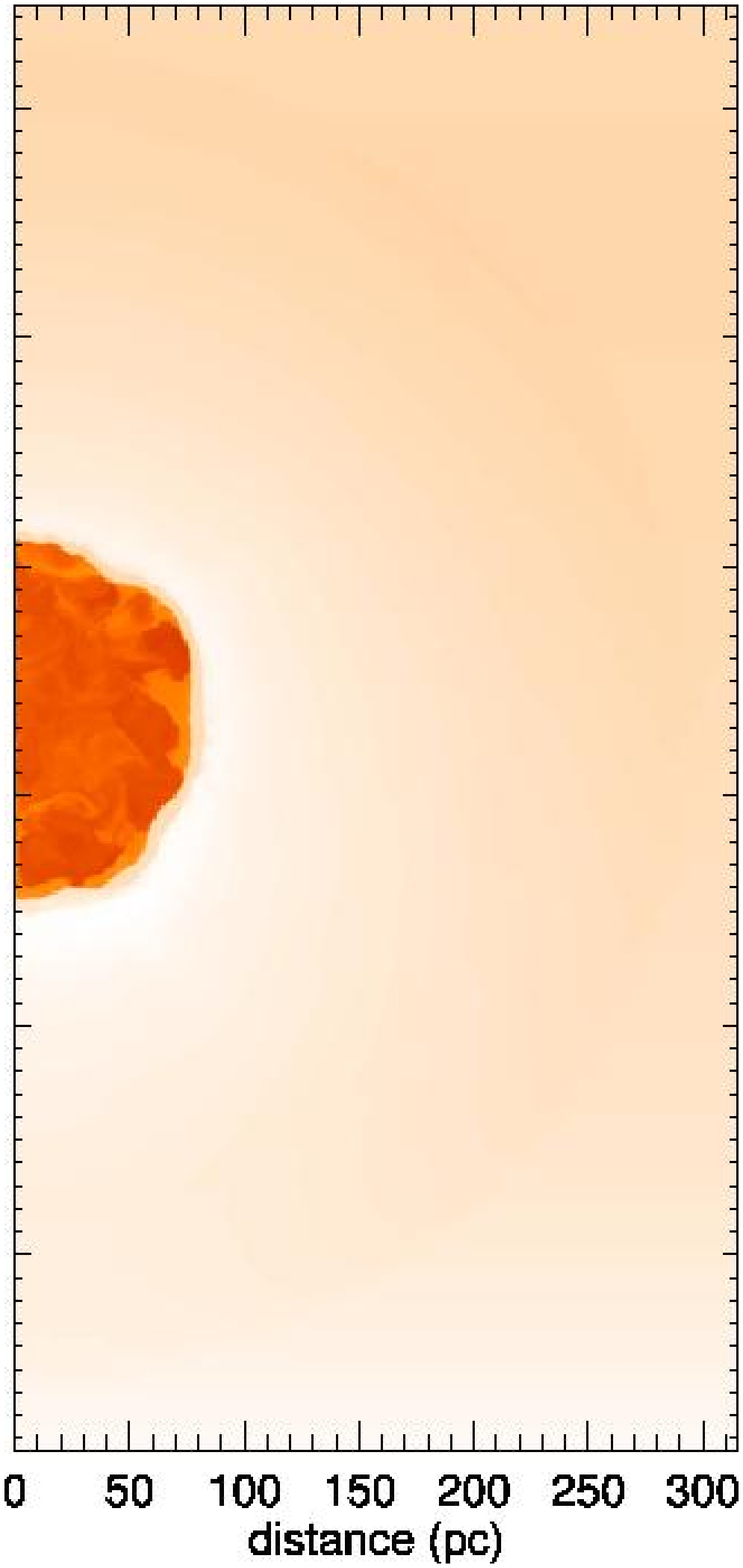}\hspace{0.5mm}\includegraphics[height=0.35\textwidth, angle=0]{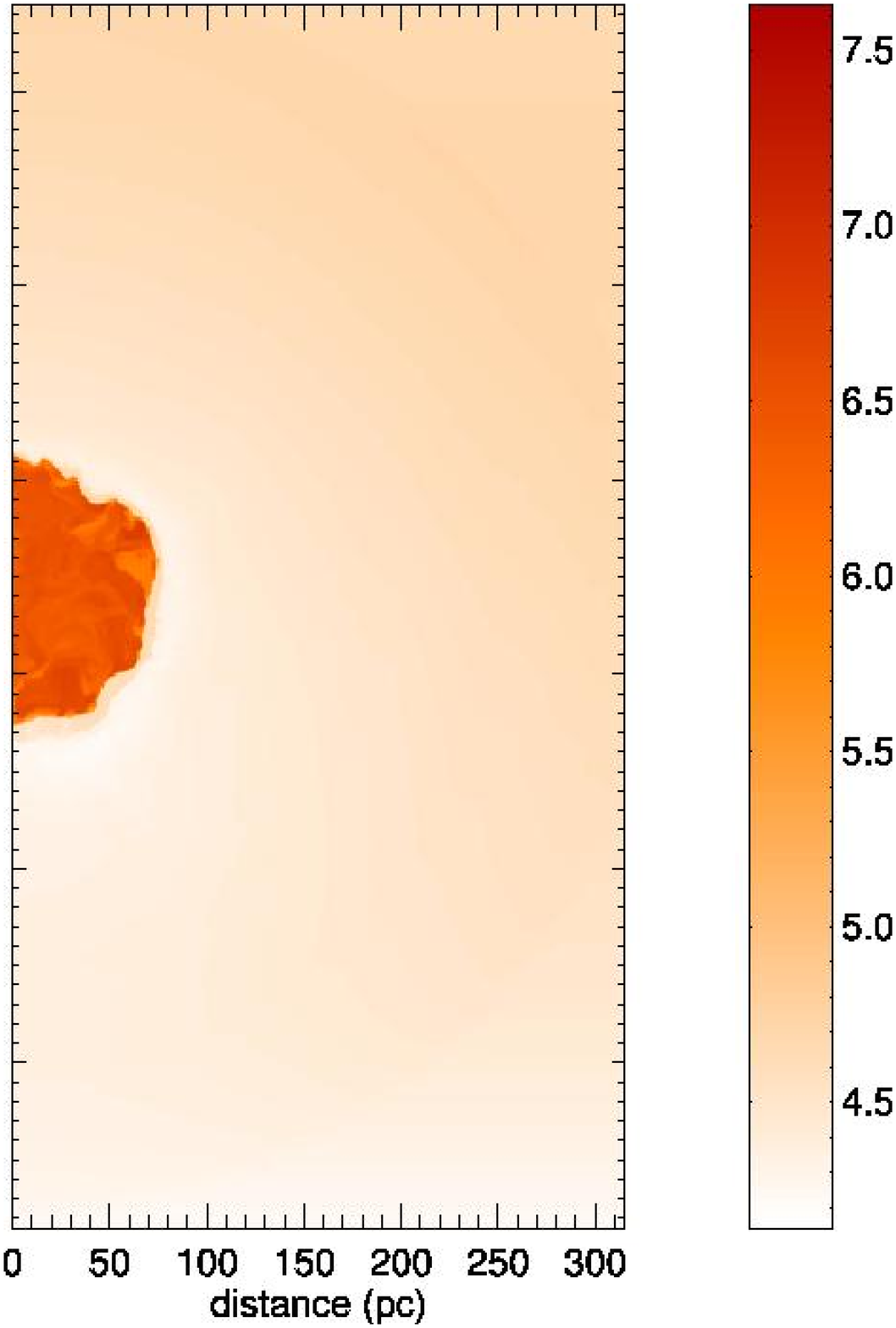}
\vspace{1.5mm}

\hspace{-3.mm}
\includegraphics[height=0.35\textwidth, angle=0]{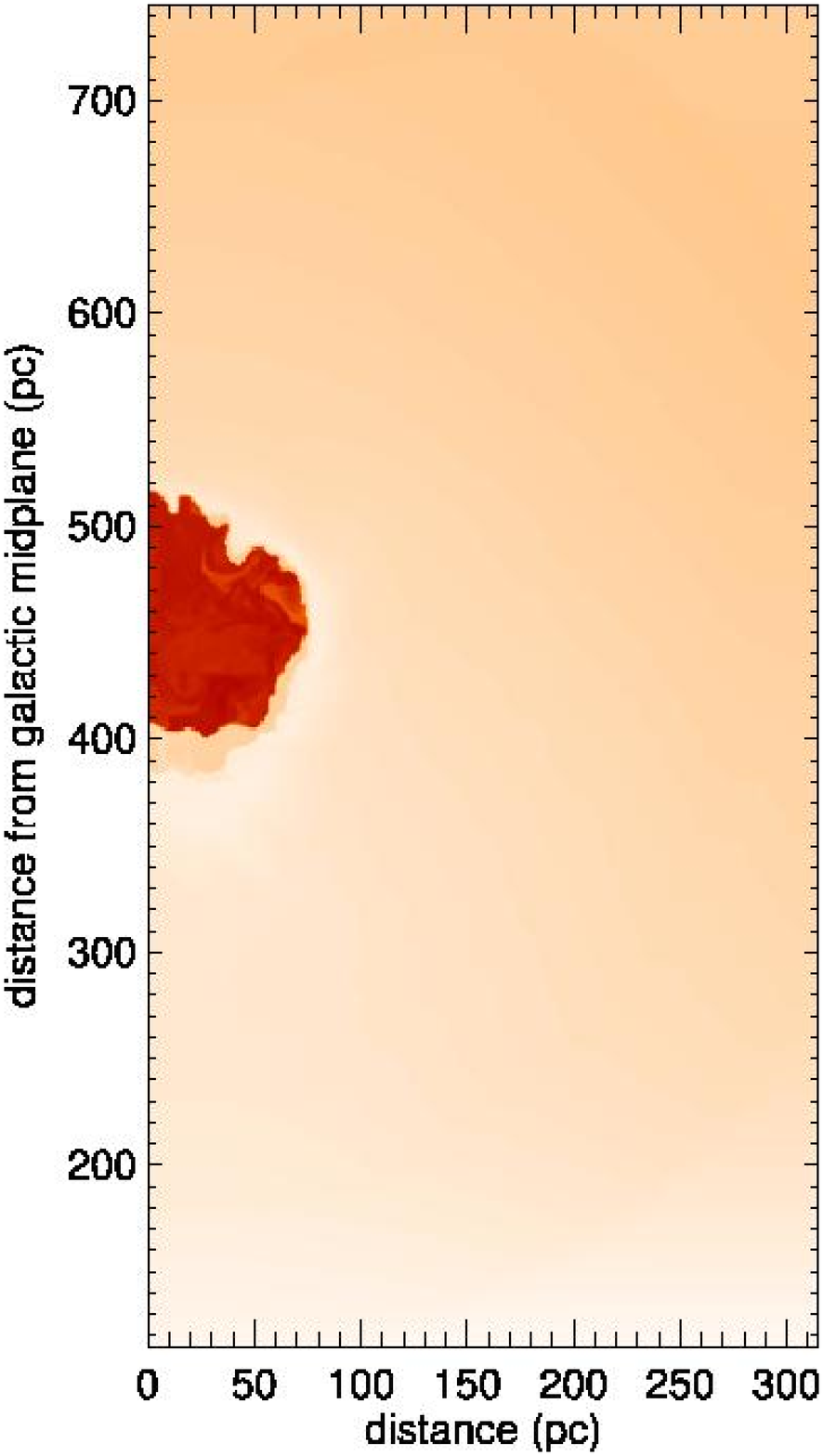}\hspace{0.5mm}\includegraphics[height=0.35\textwidth, angle=0]{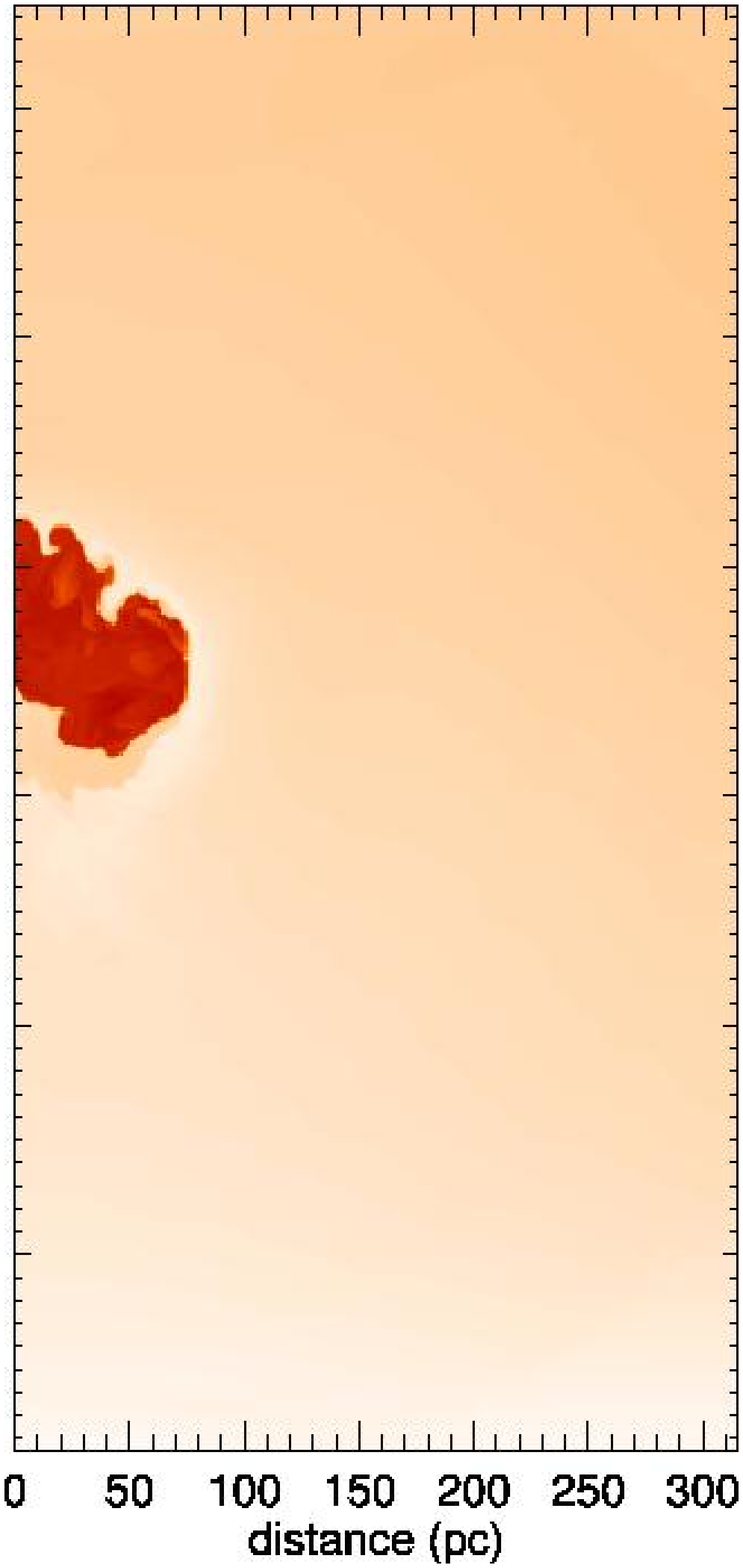}\hspace{0.5mm}\includegraphics[height=0.35\textwidth, angle=0]{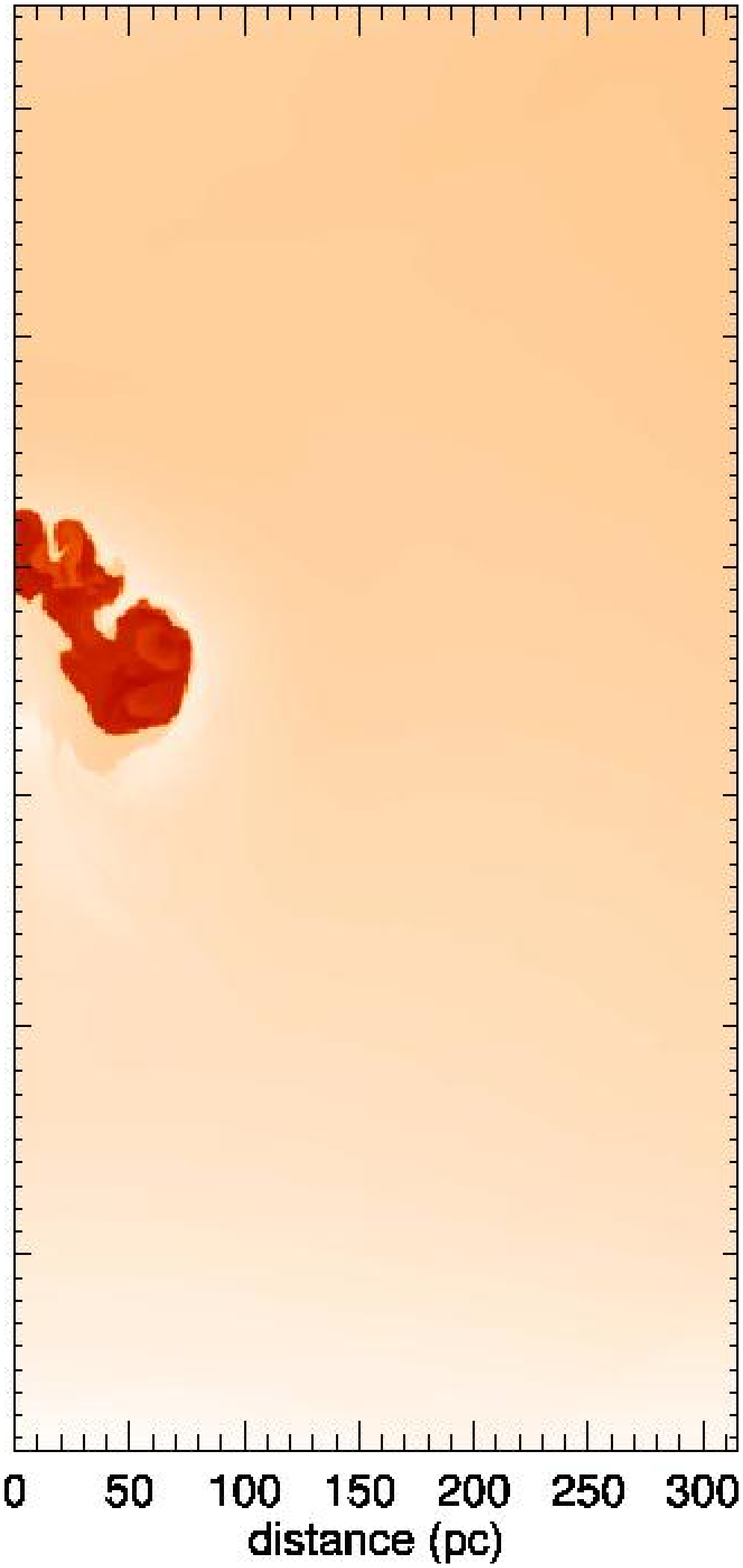}\hspace{0.5mm}\includegraphics[height=0.35\textwidth, angle=0]{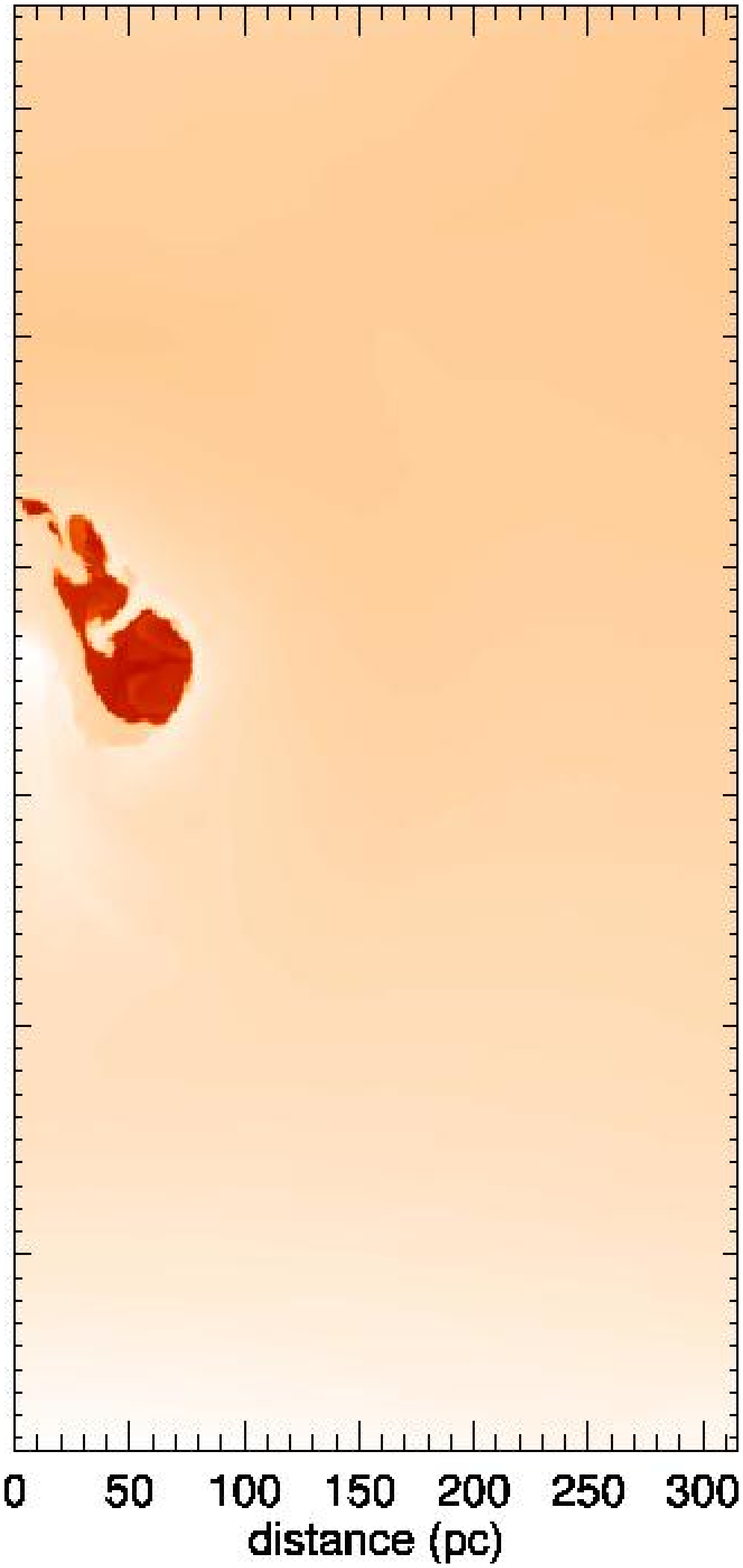}\hspace{0.5mm}\includegraphics[height=0.35\textwidth, angle=0]{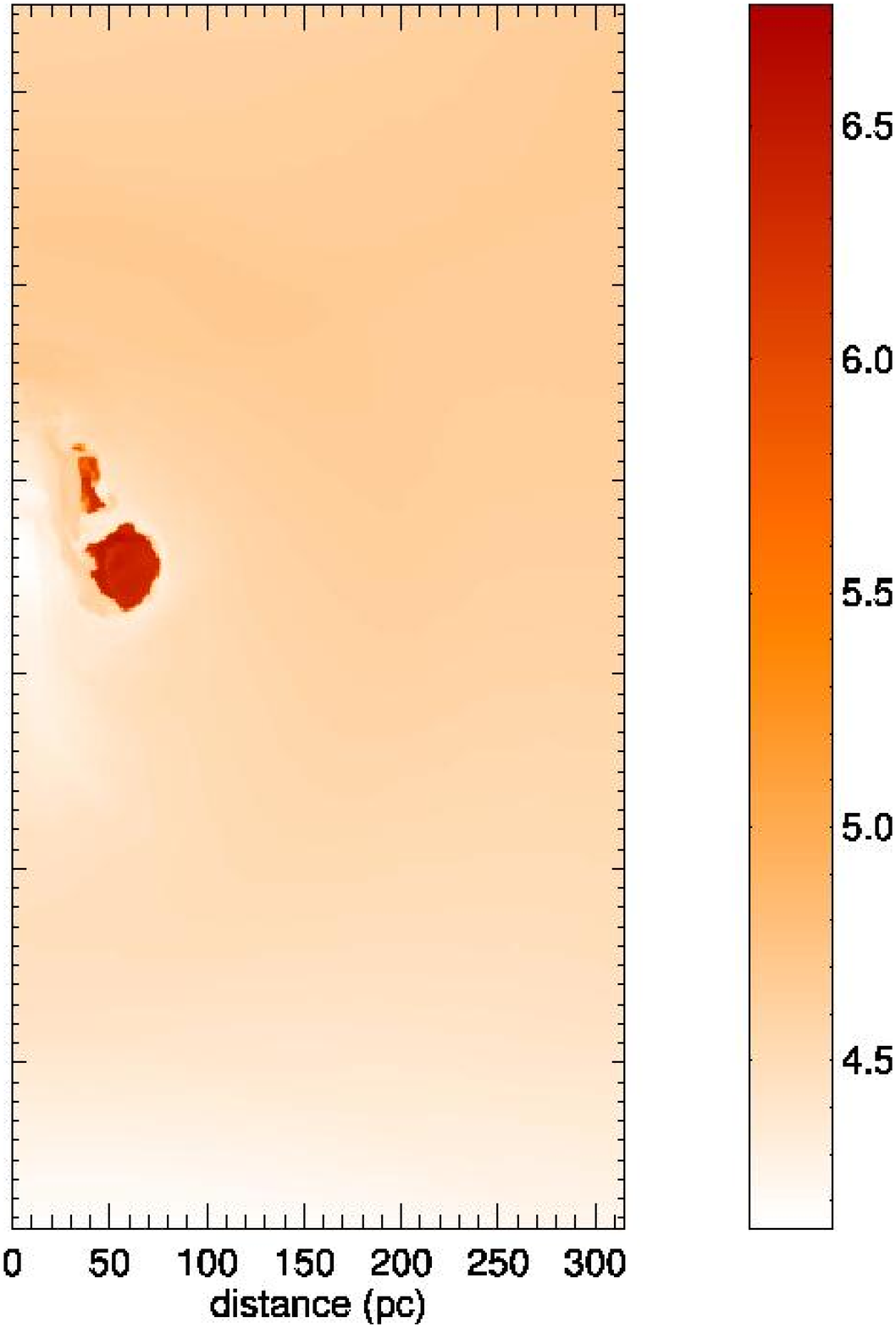}

  \caption{$Log_{10}$ of temperature (Kelvin) (left to right, top to bottom) for: 100 kyrs, 300 kyrs, 600kyrs, 1 Myr, 2 Myrs, 3 Myrs, 4 Myrs, 5 Myrs, 6 Myrs, 7 Myrs, 8 Myrs, 9 Myrs, 10 Myrs, 11 Myrs, and 12 Myrs, for Model A (no magnetic field). By 1 Myrs the cool shell has formed between the bubble and the shock.  By 8 Myrs a faint cool stem has appeared below the cap shaped bubble.  Between 9 and 12 Myrs this cool, low density stem is pushed into the high temperature, low density bubble. Because the two dimensions parallel to the plane are identical in this simulation, plots through the x-z or y-z (where $\hat{z}$ is the direction perpendicular to the midplane) are virtually indistinguishable.  These figures show the x=0 (y-z plane).}
\label{mushroom2}
\efig

\bfig
\centering
\includegraphics[height=0.35\textwidth, angle=0]{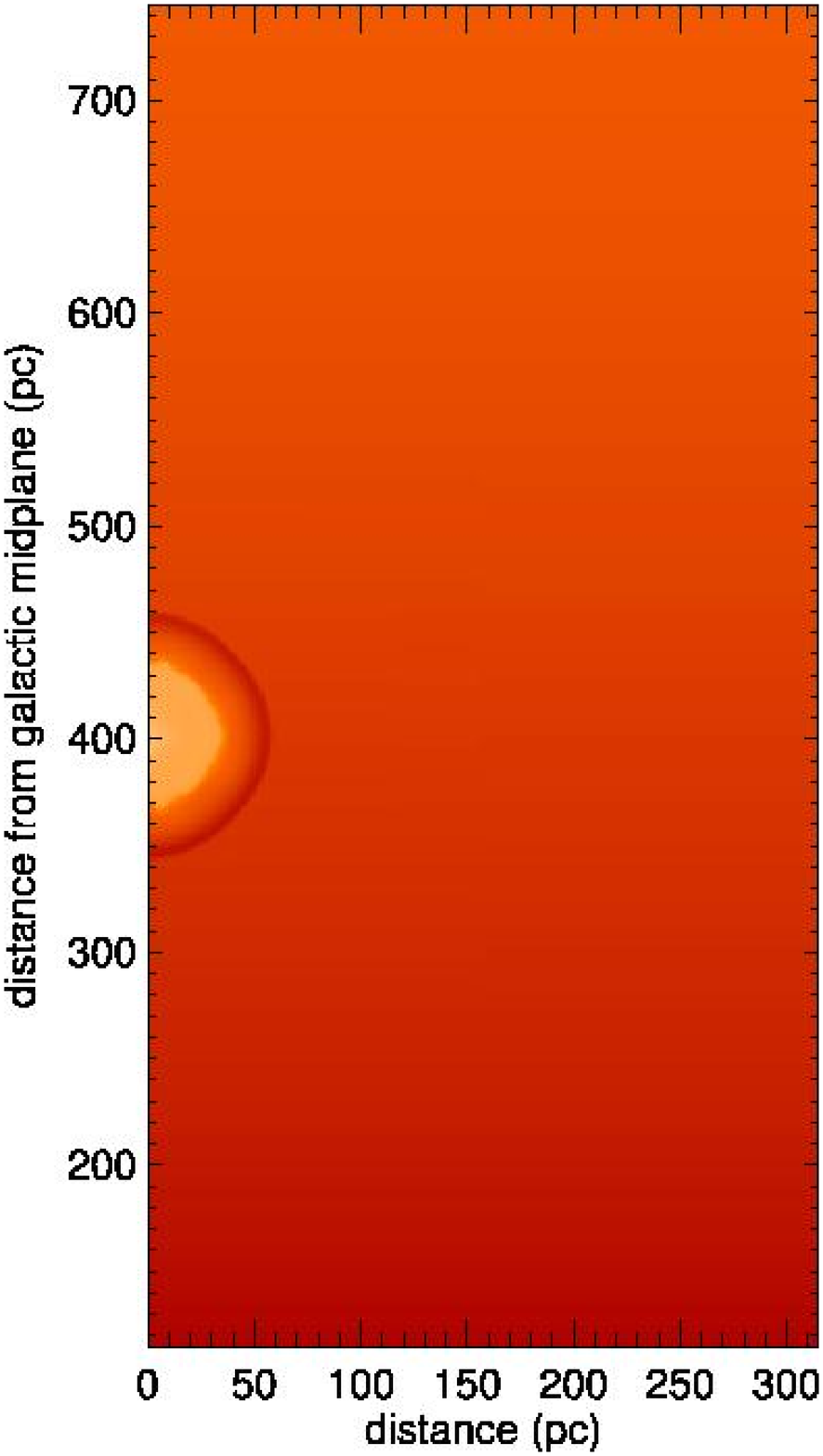}\hspace{0.5mm}\includegraphics[height=0.35\textwidth, angle=0]{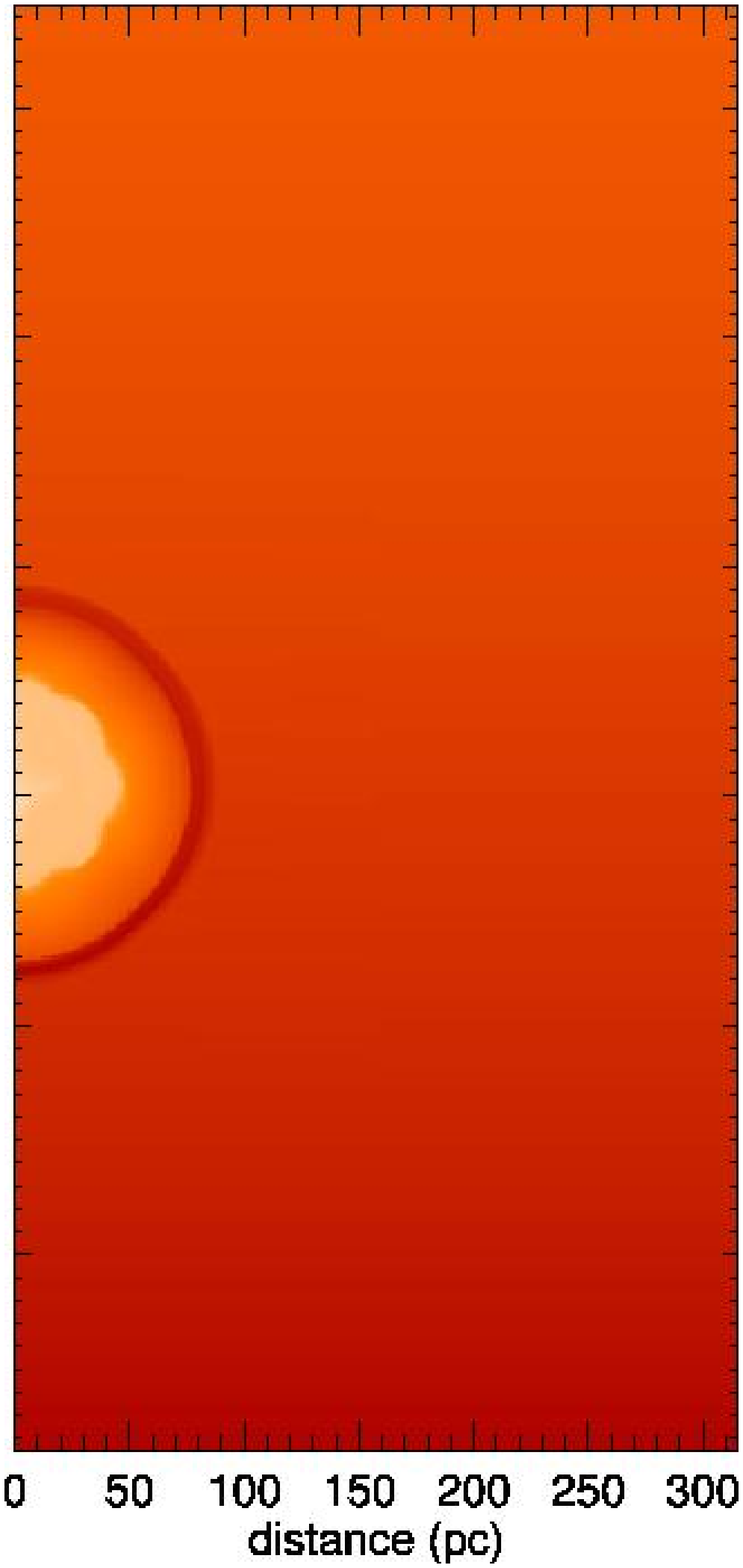}\hspace{0.5mm}\includegraphics[height=0.35\textwidth, angle=0]{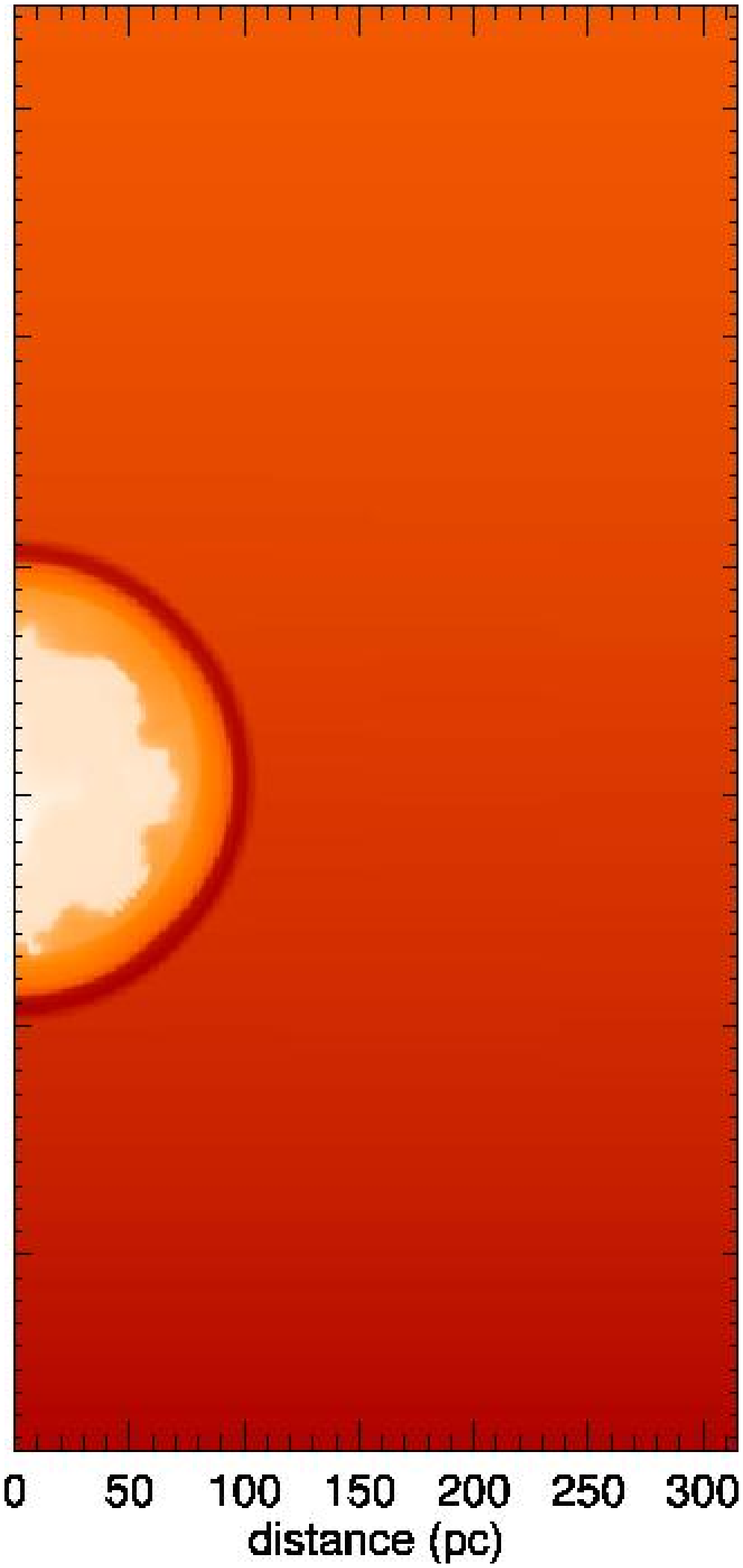}\hspace{0.5mm}\includegraphics[height=0.35\textwidth, angle=0]{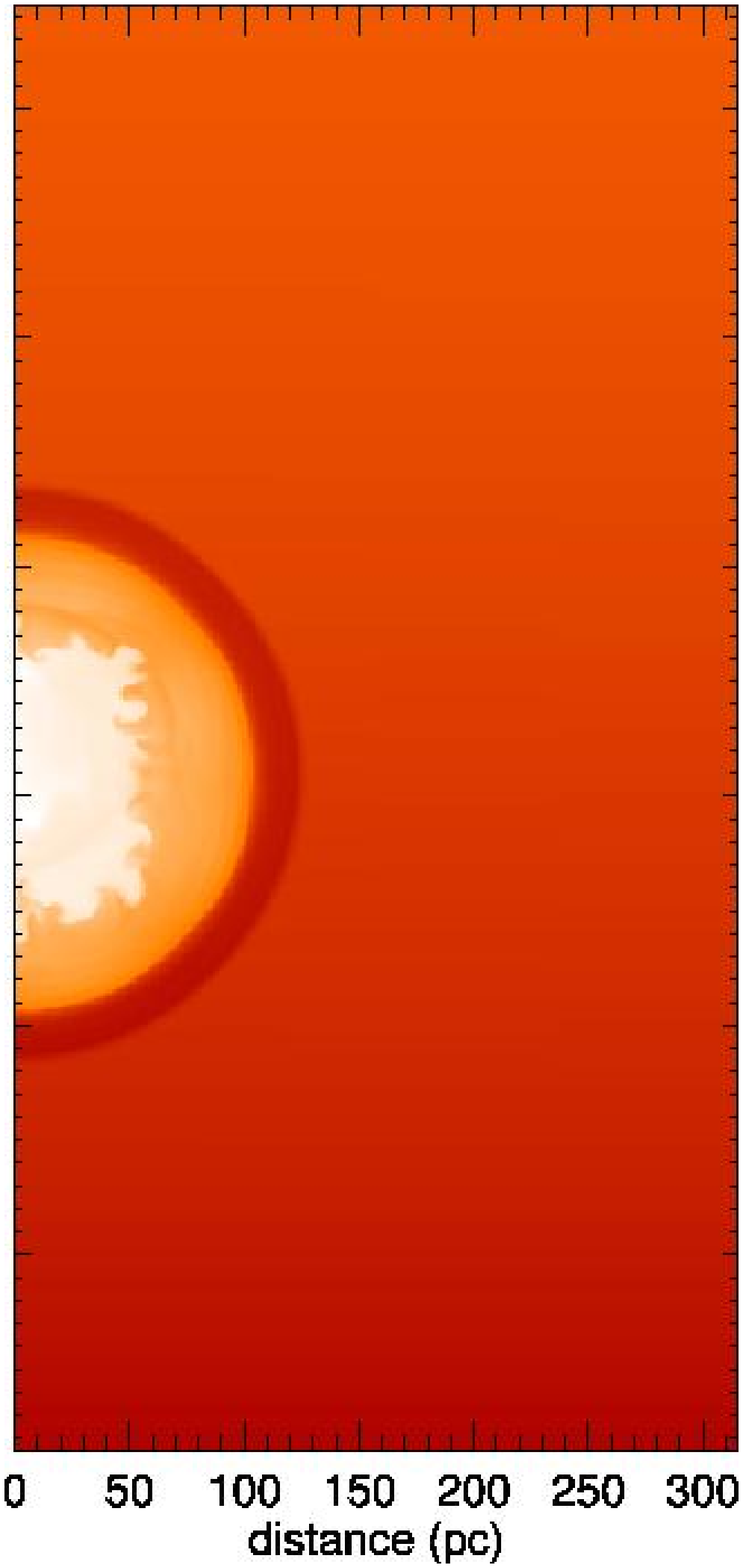}\hspace{0.5mm}\includegraphics[height=0.35\textwidth, angle=0]{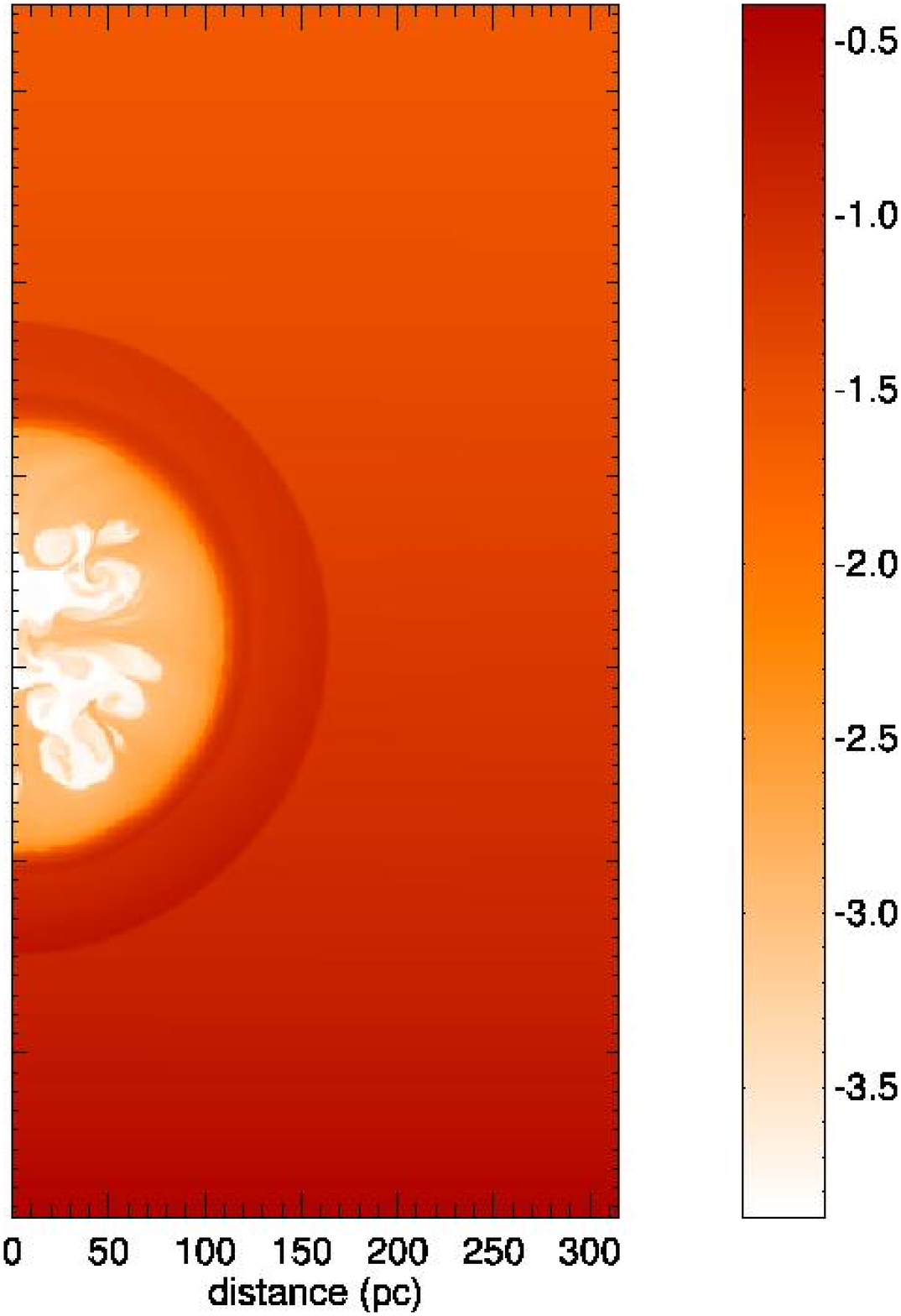}
\vspace{1.5mm}

\includegraphics[height=0.35\textwidth, angle=0]{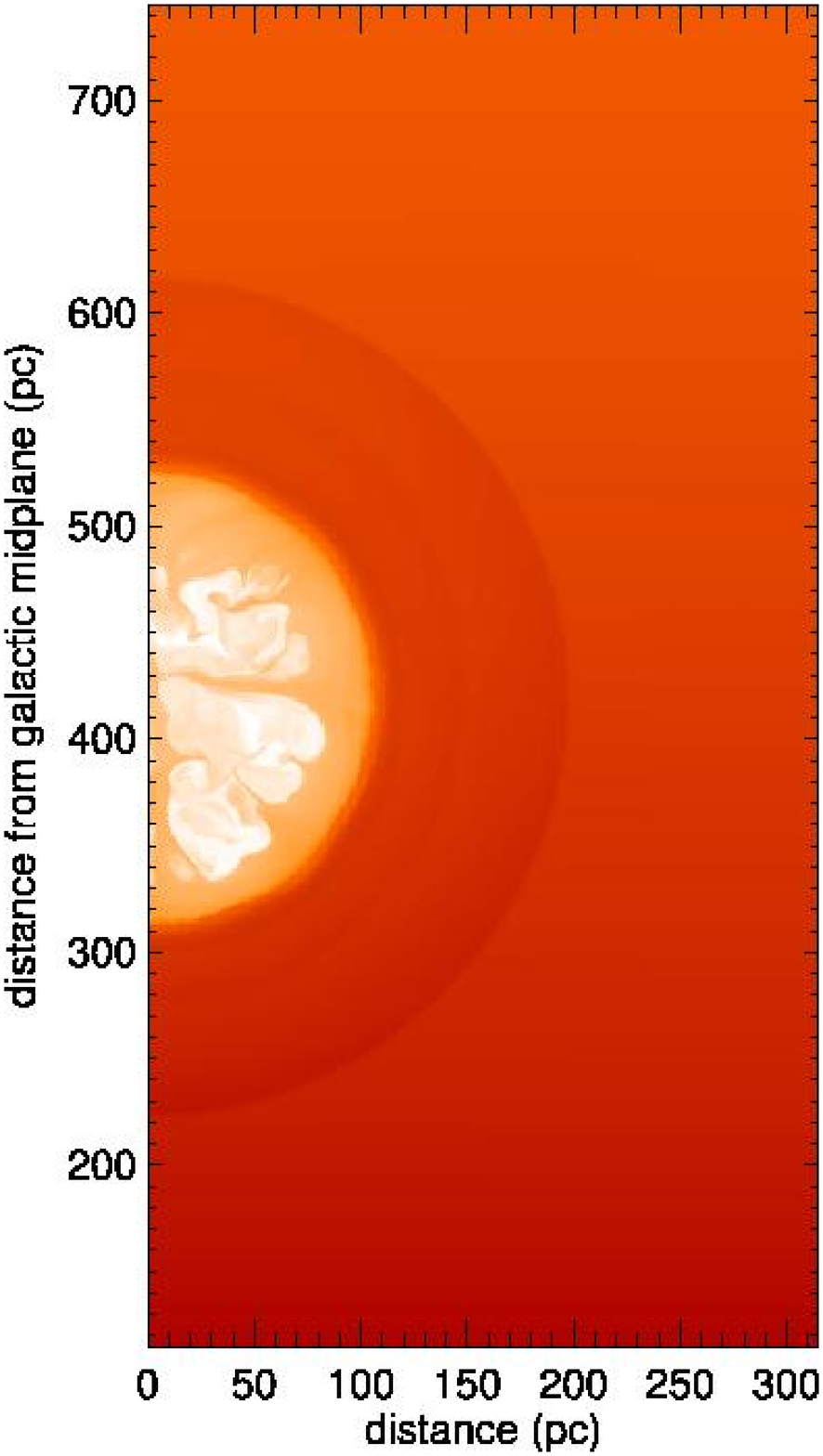}\hspace{0.5mm}\includegraphics[height=0.35\textwidth, angle=0]{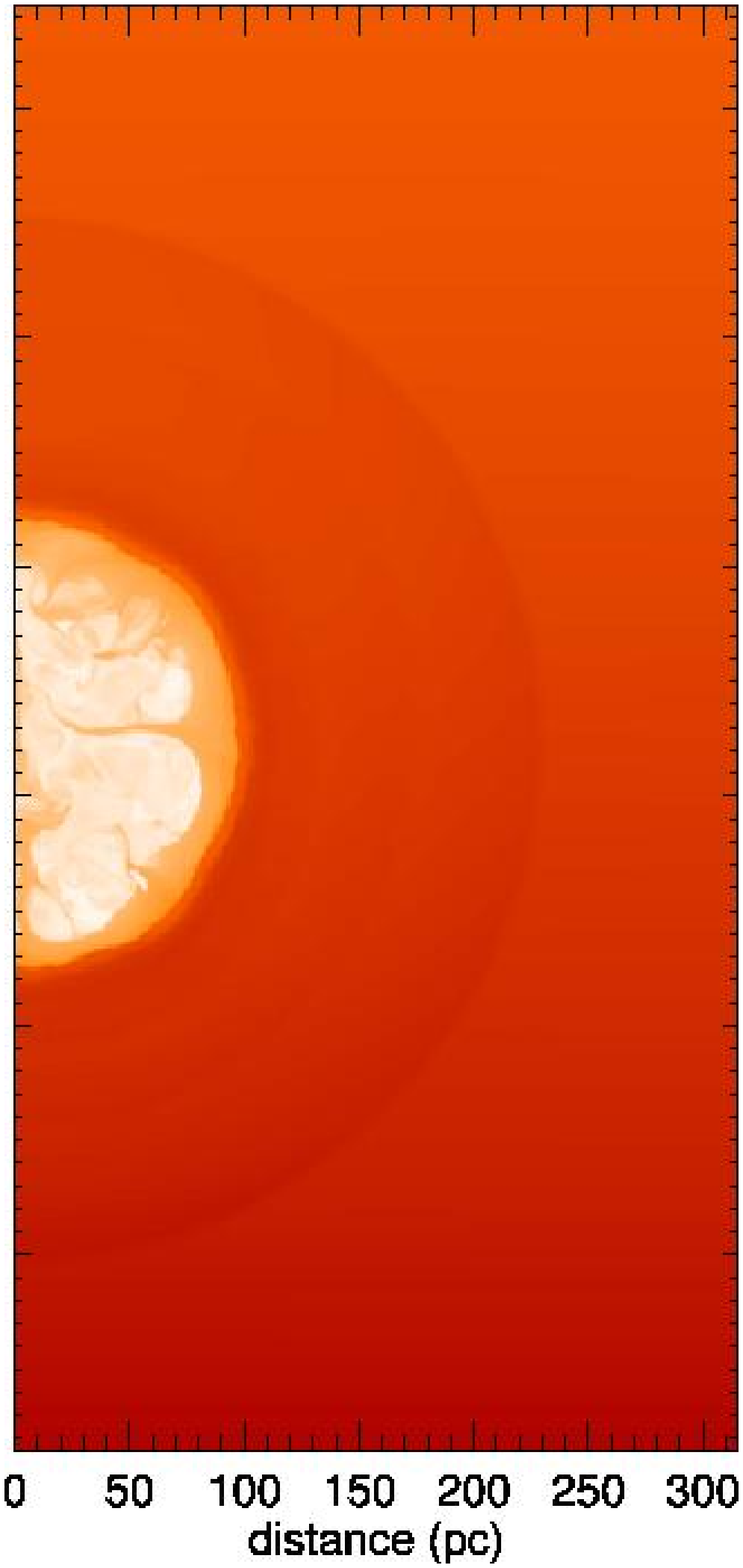}\hspace{0.5mm}\includegraphics[height=0.35\textwidth, angle=0]{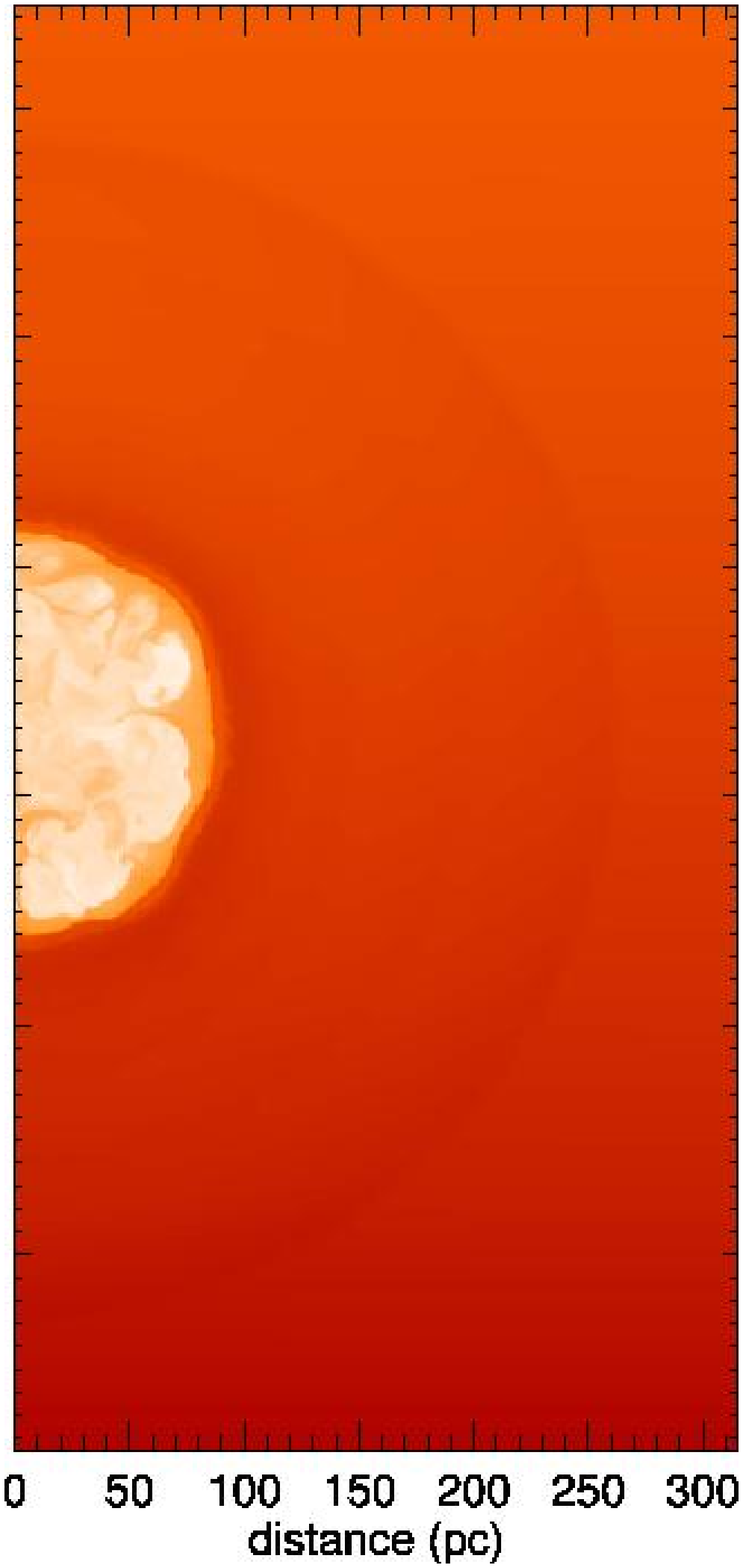}\hspace{0.5mm}\includegraphics[height=0.35\textwidth, angle=0]{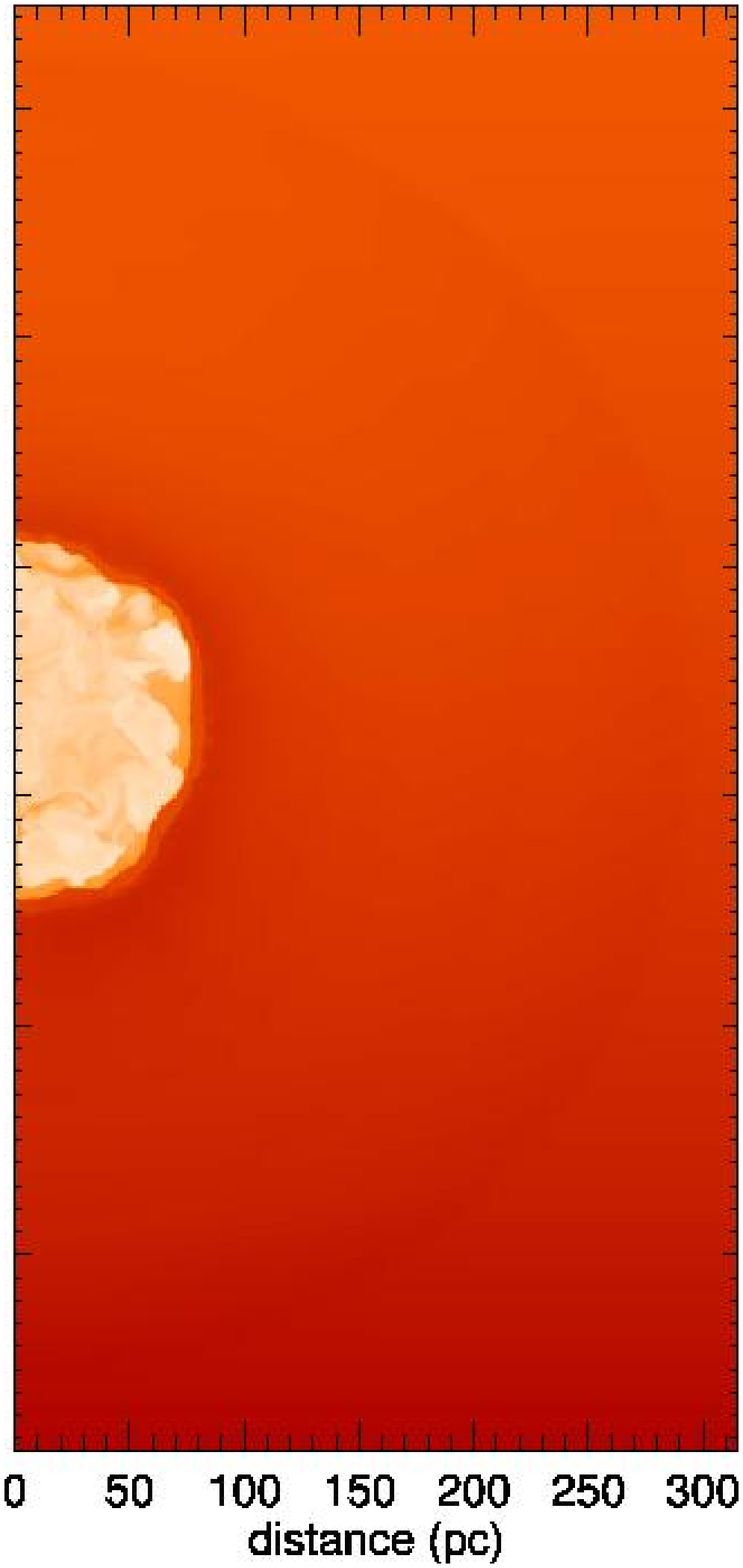}\hspace{0.5mm}\includegraphics[height=0.35\textwidth, angle=0]{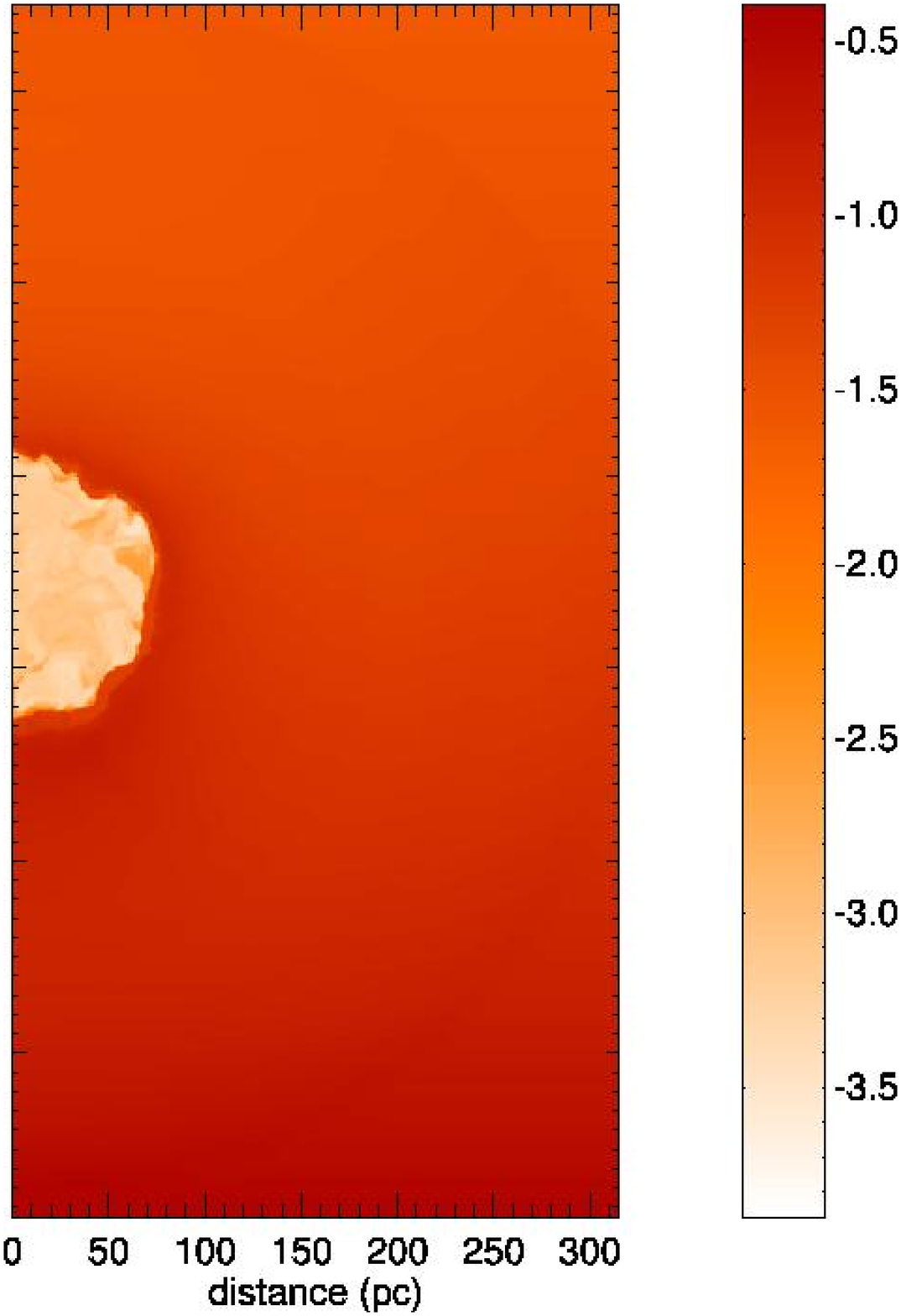}
\vspace{1.5mm}

\hspace{-3.5mm}
\includegraphics[height=0.35\textwidth, angle=0]{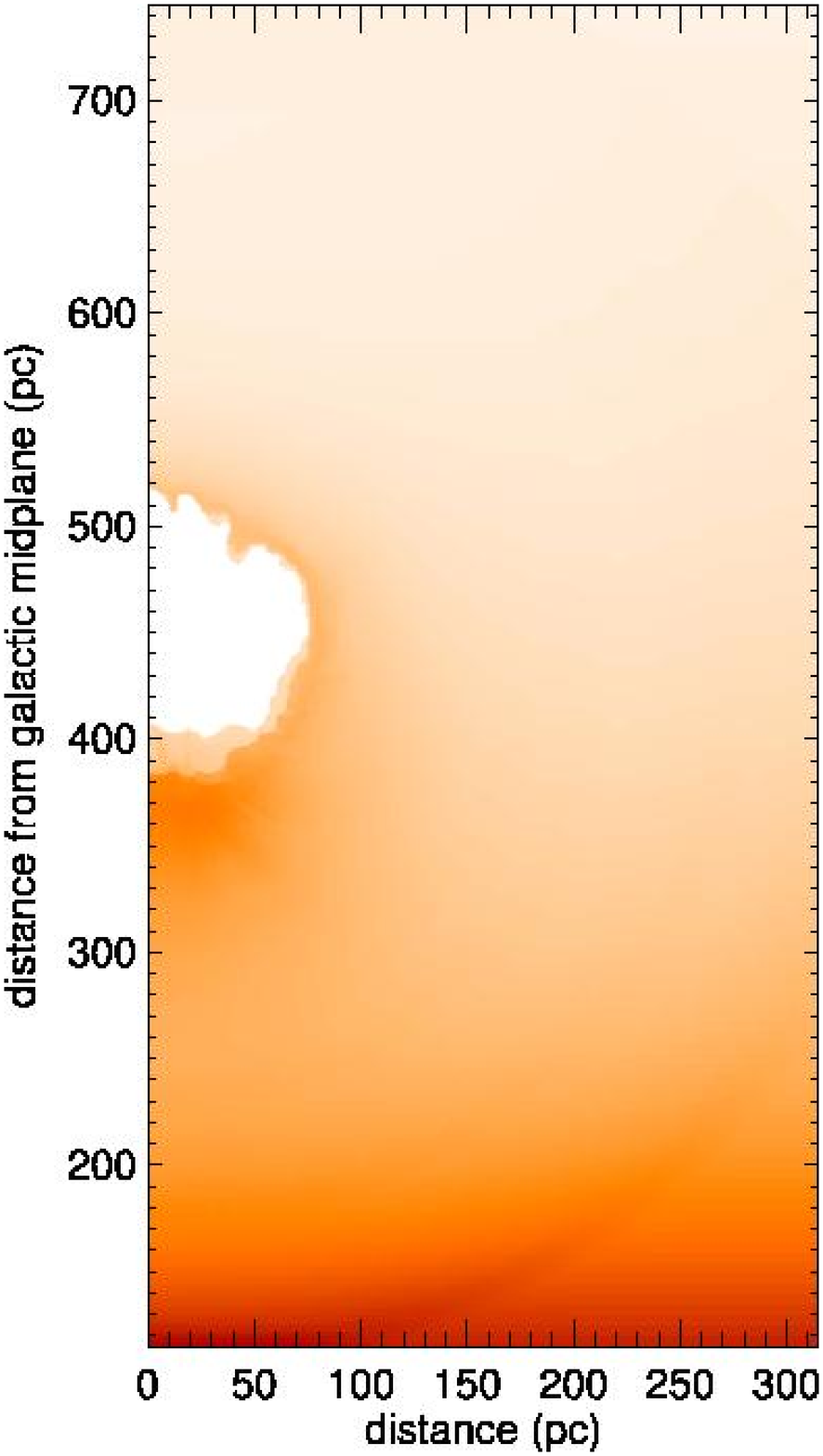}\hspace{0.5mm}\includegraphics[height=0.35\textwidth, angle=0]{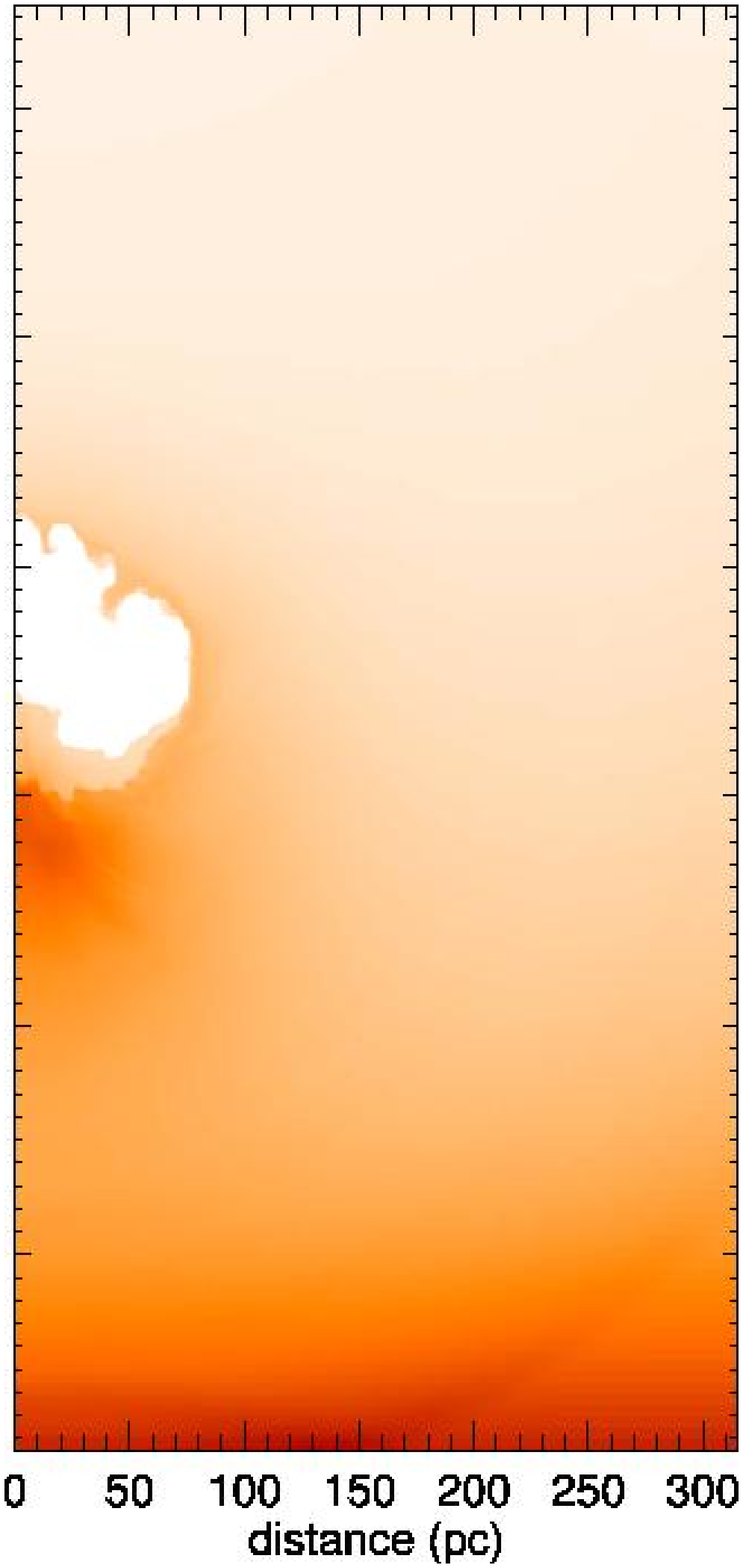}\hspace{0.5mm}\includegraphics[height=0.35\textwidth, angle=0]{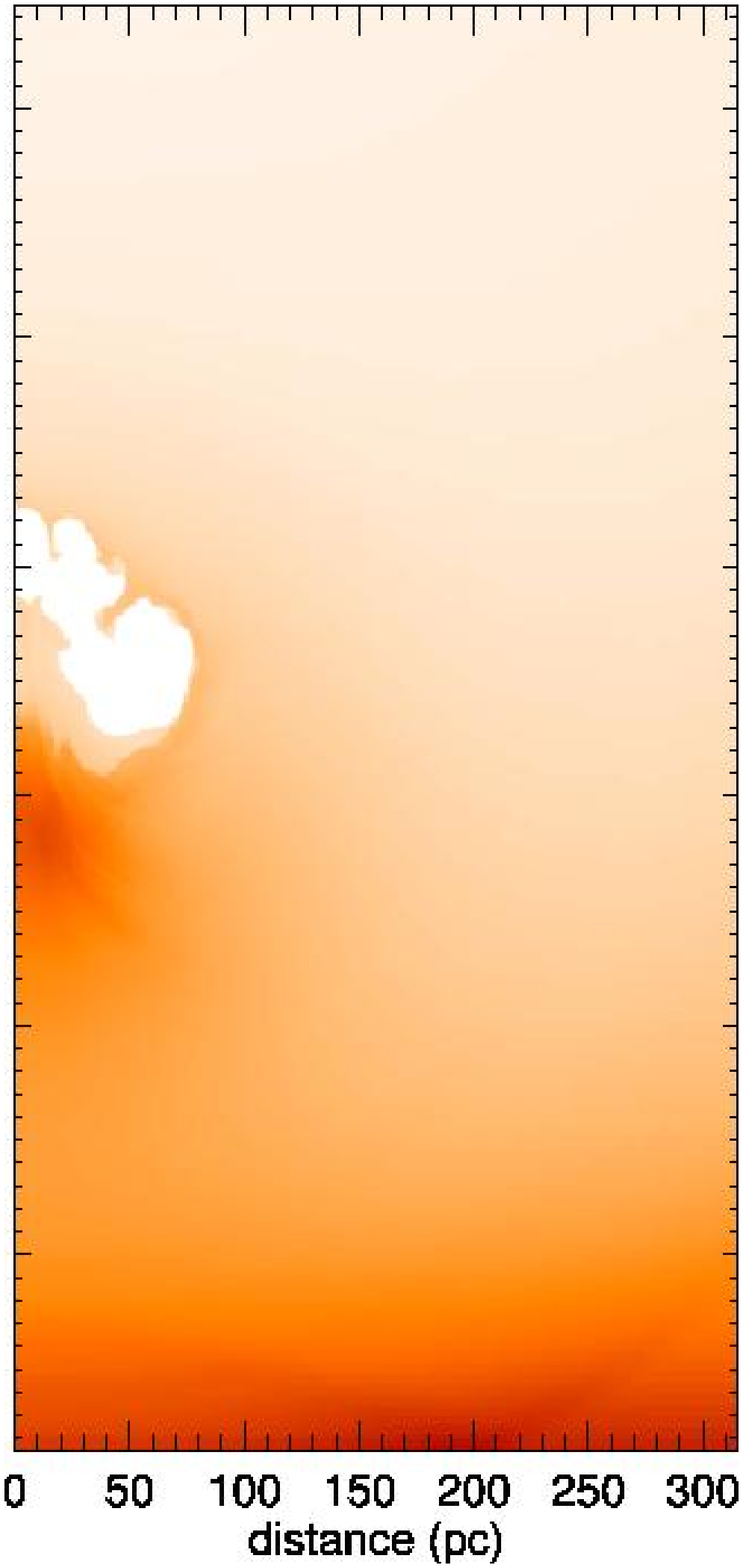}\hspace{0.5mm}\includegraphics[height=0.35\textwidth, angle=0]{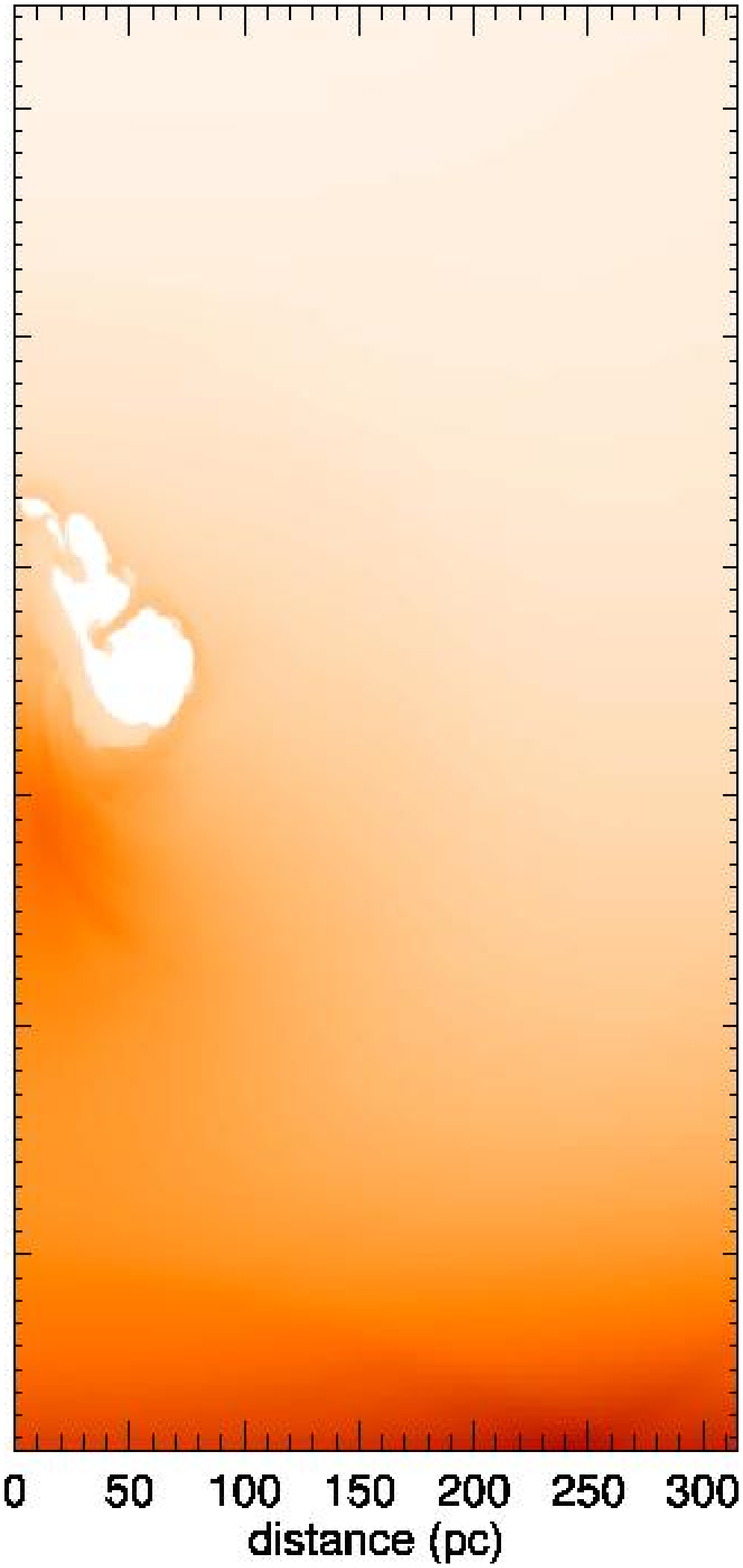}\hspace{0.5mm}\includegraphics[height=0.35\textwidth, angle=0]{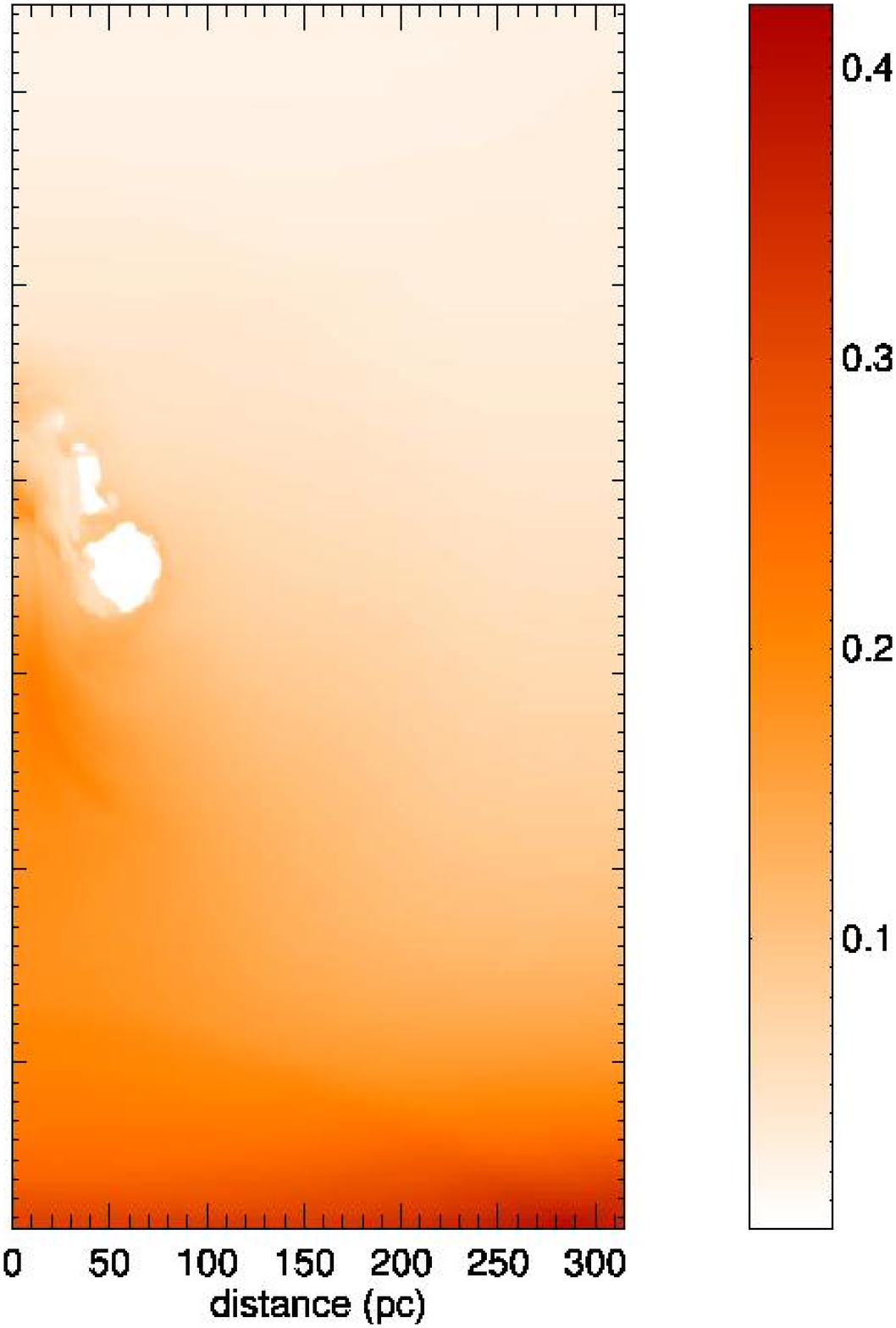}

  \caption{First two rows: $Log_{10}$ of density (atoms/cm$^3$) (left to right, top to bottom) for: 100 kyrs, 300 kyrs, 600 kyrs and every million years to 7 Myrs, for Model A (no magnetic field).  Last row: Density (atoms/cm$^3$) in linear scale (left to right, top to bottom) every million years from 8 to 12 Myrs. By 1 Myr we see that a dense shell has formed between the bubble and the shock.  By 6 Myrs the shock has weakened such the material just inside of the shock is no longer noticeably denser than the ambient background.  By 9 Myrs a mushroom shaped structure has developed in the high density region surrounding the bubble.  At 10 Myrs the mushroom structure has become more defined and after that it begins to dissipate as the dense stem is pushed up into the rarefied bubble. These figures show the x=0 (y-z plane).}
\label{mushroom1}
\efig

\bfig
\centering
\includegraphics[height=0.35\textwidth, angle=0]{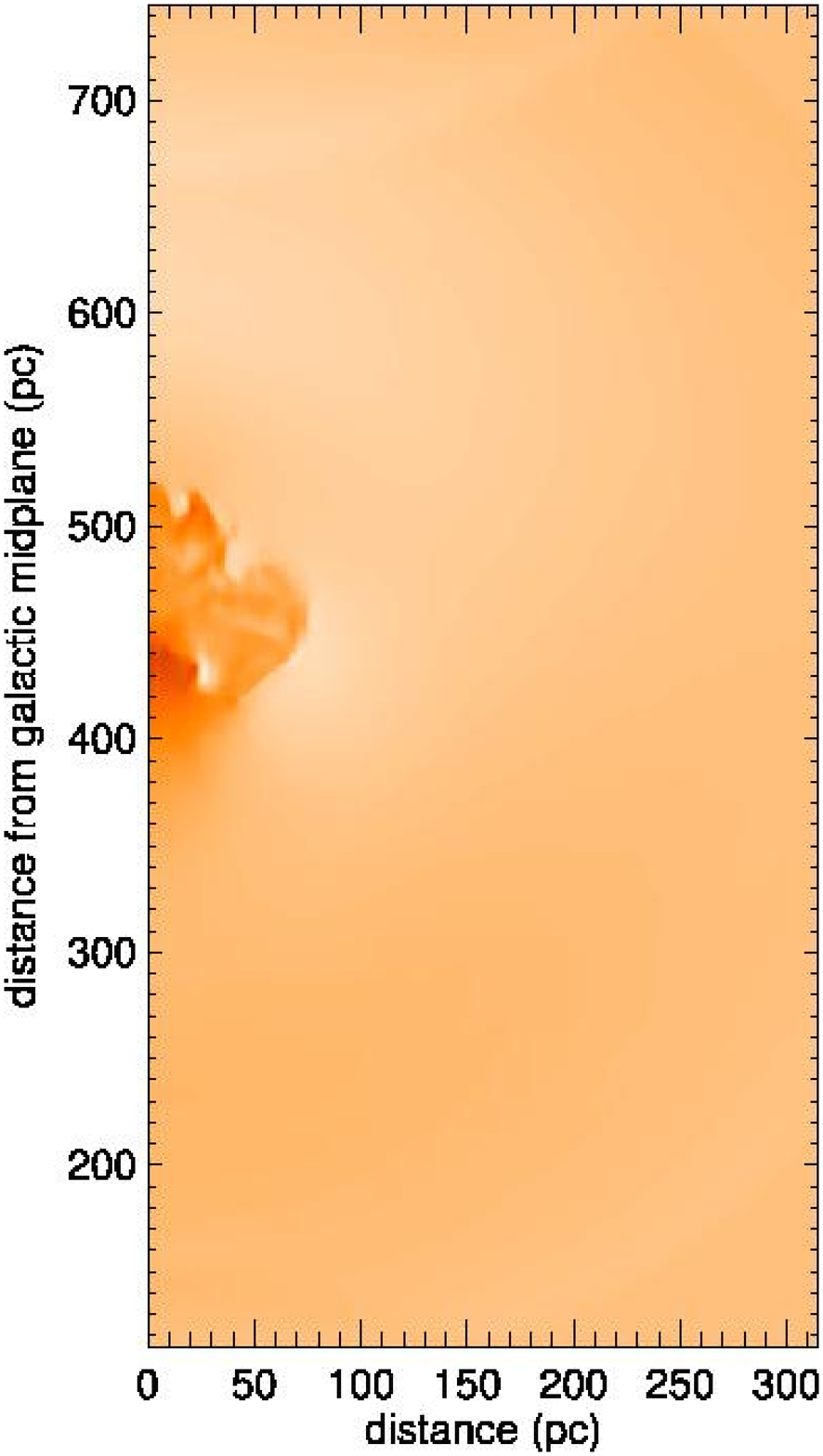}\hspace{0.5mm}\includegraphics[height=0.35\textwidth, angle=0]{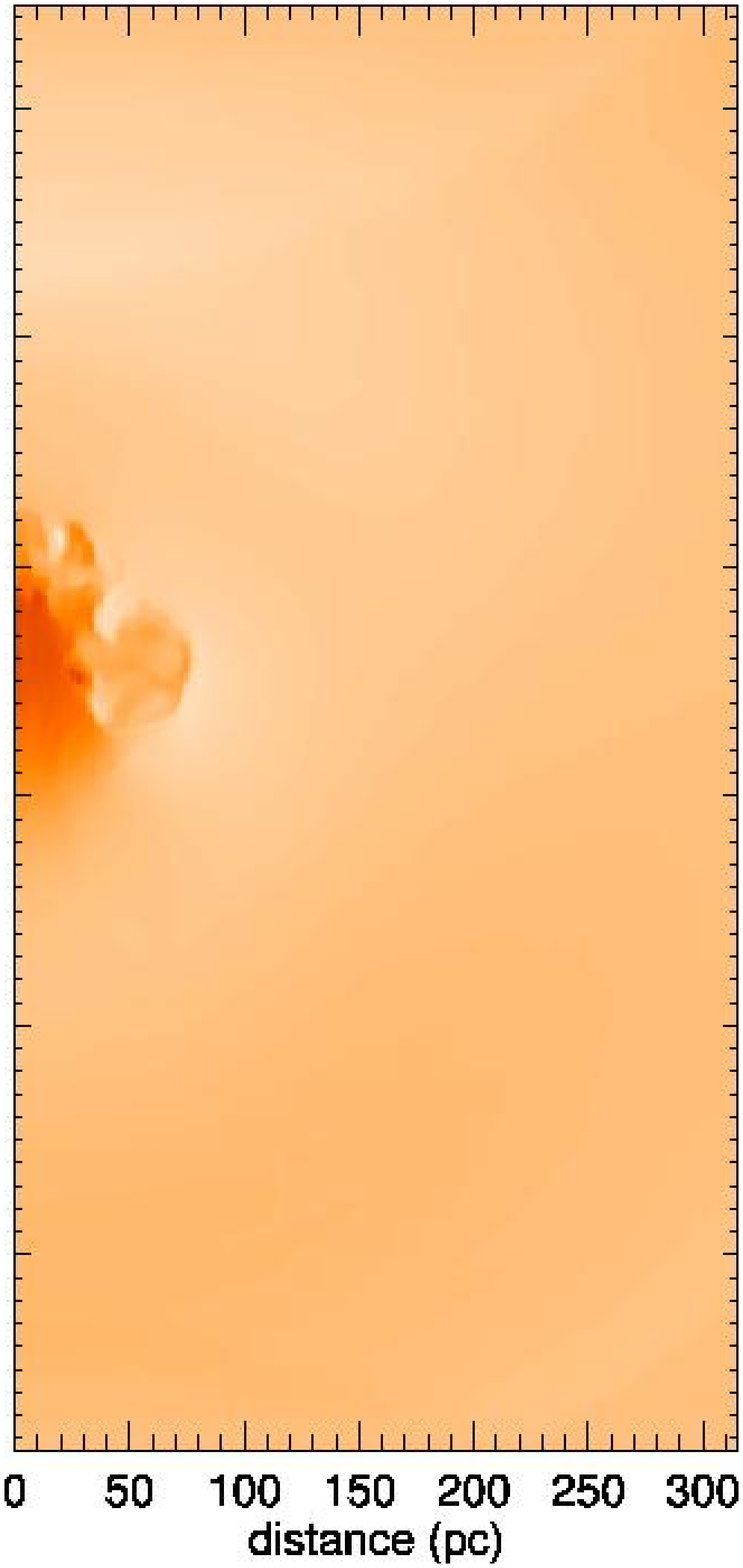}\hspace{0.5mm}\includegraphics[height=0.35\textwidth, angle=0]{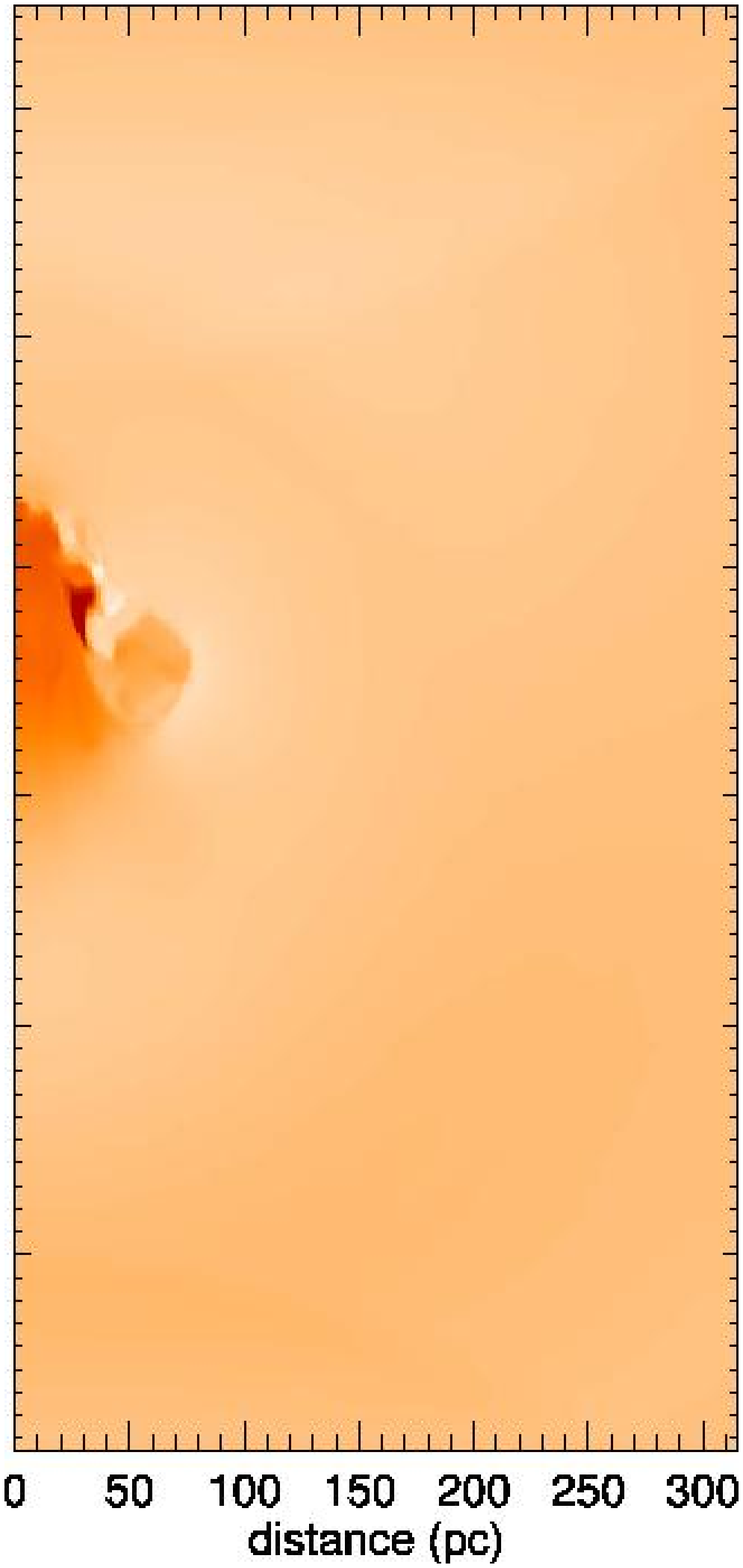}\hspace{0.5mm}\includegraphics[height=0.35\textwidth, angle=0]{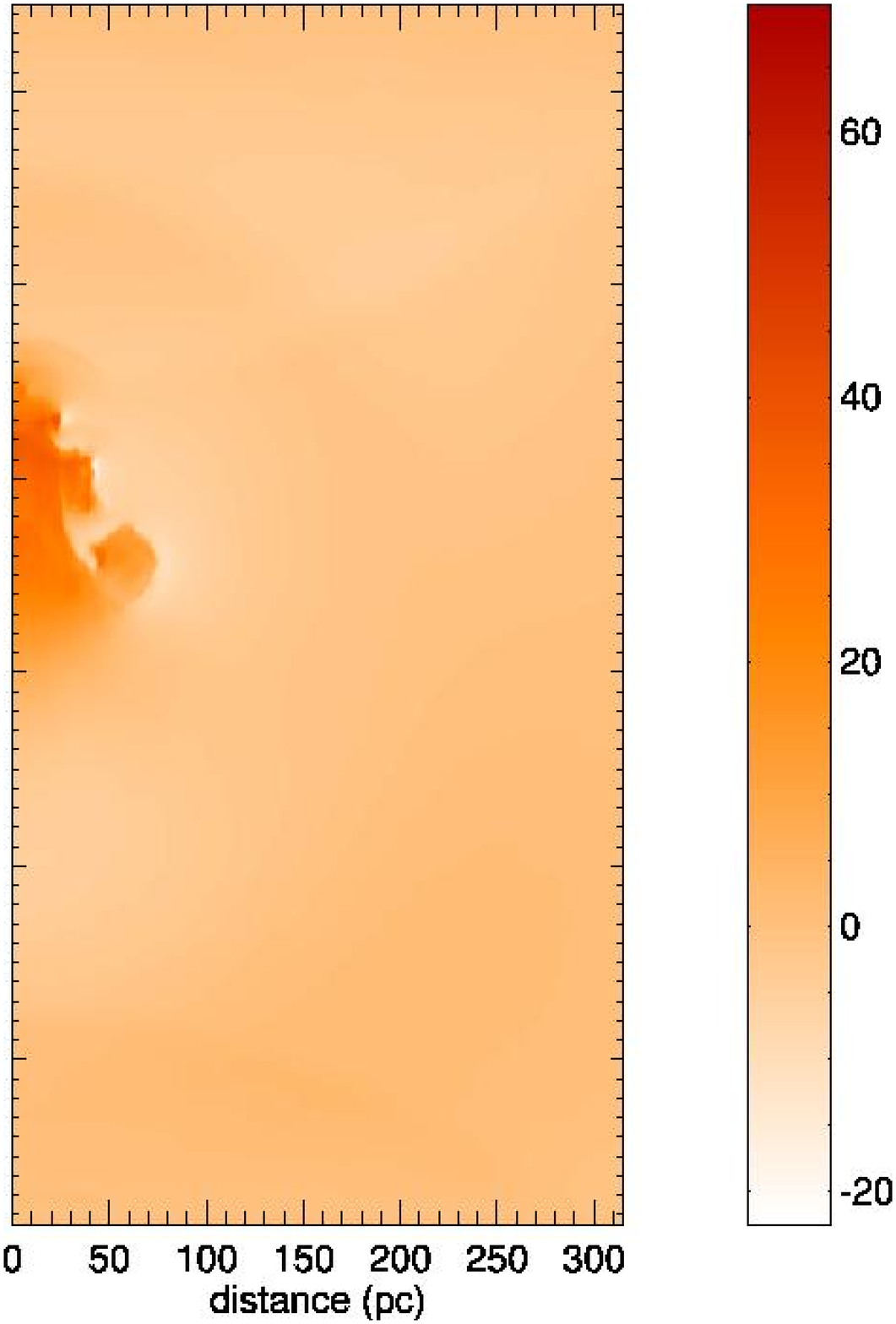}

\caption{Velocity (km/s) component perpendicular to the galactic midplane , every million years (left to right) from 9 Myrs to 12 Myrs for Model A (no magnetic field).  Notice that the areas with highest velocity coincide with the 'stem' of cool, high density material in the last four panels of Figure \ref{mushroom1} and Figure \ref{mushroom2}.  These figures show the x=0 (y-z plane). }
\label{mushroom_vel}
\efig

\bfig 
\centering

\includegraphics[height=0.15\textwidth, angle=0]{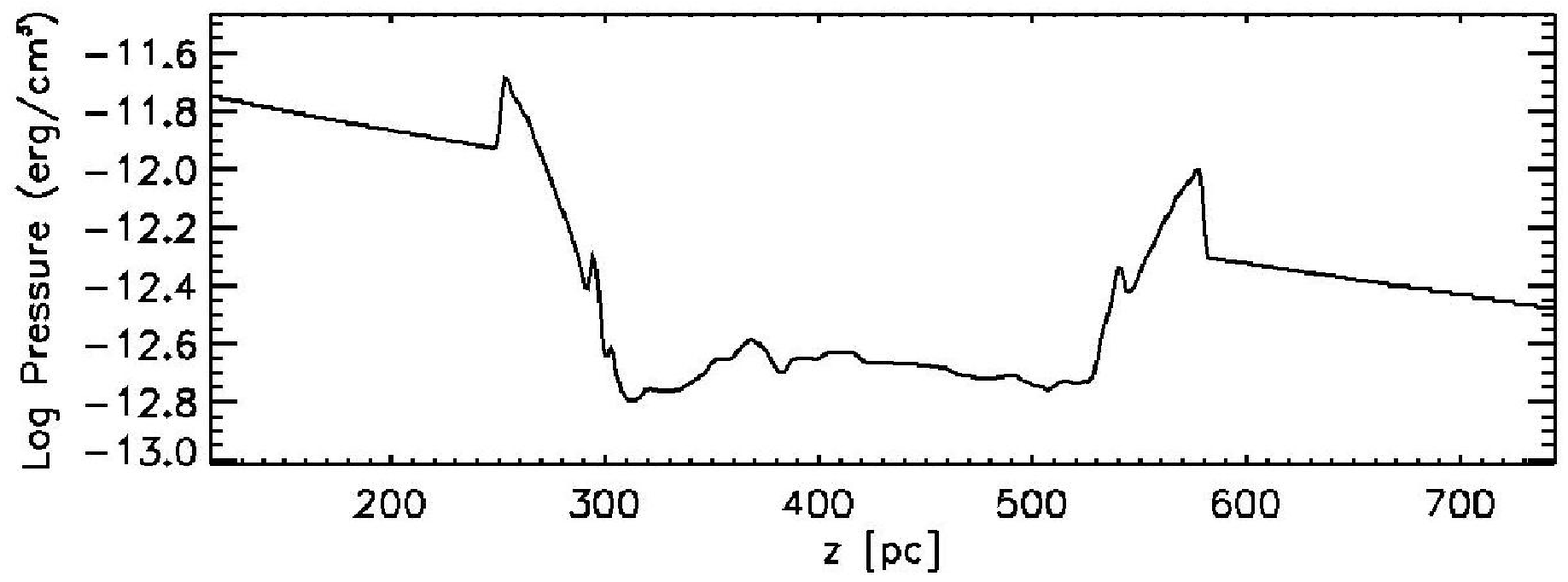}\hspace{0.5mm}\includegraphics[height=0.15\textwidth, angle=0]{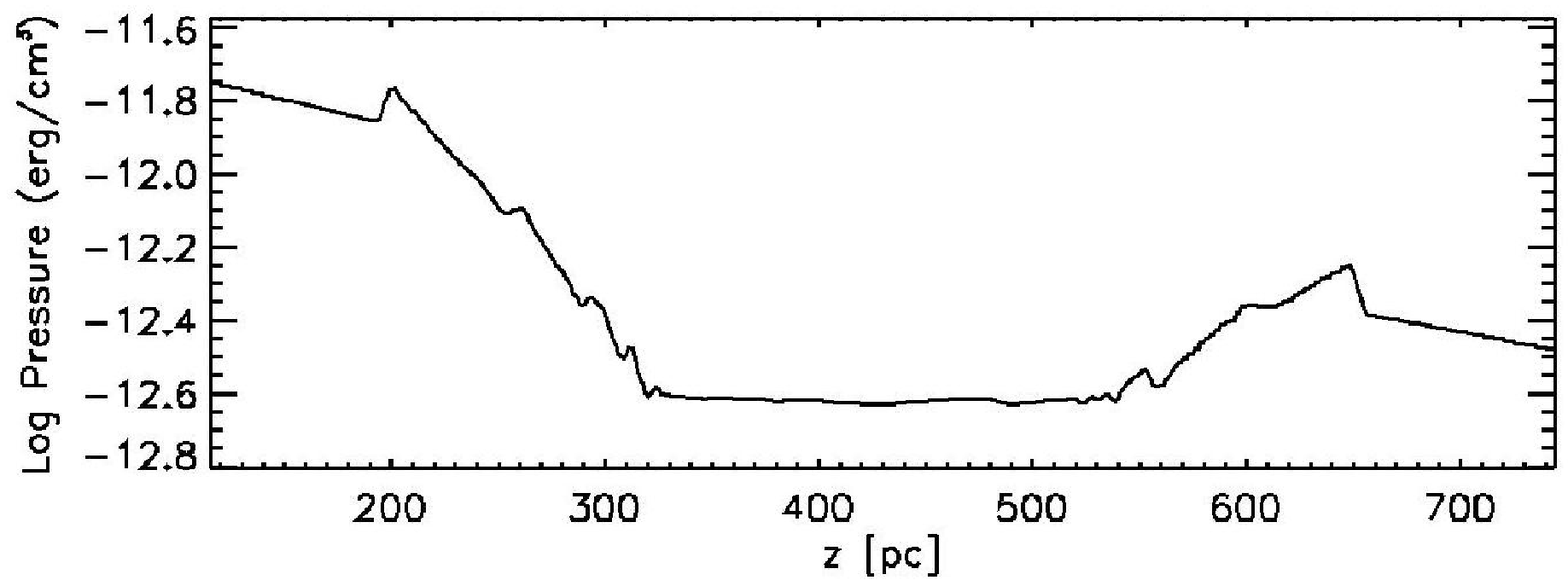}\hspace{0.5mm}\includegraphics[height=0.15\textwidth, angle=0]{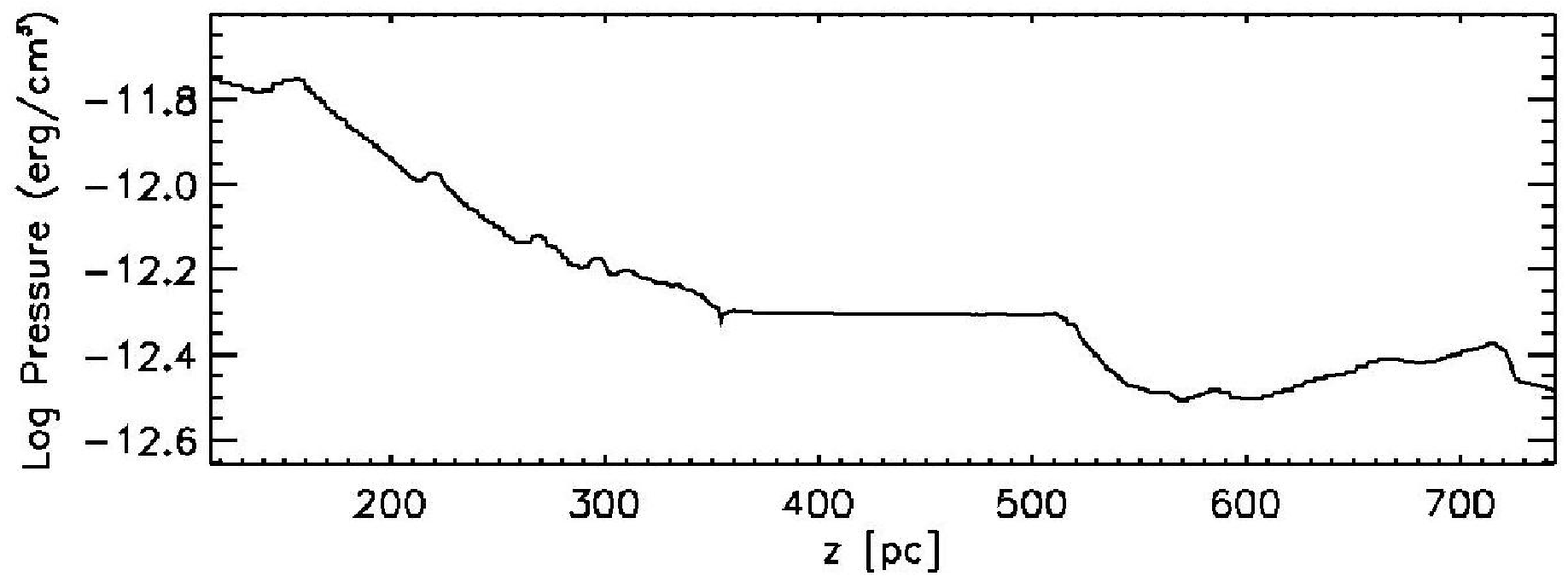}\hspace{0.5mm}\includegraphics[height=0.15\textwidth, angle=0]{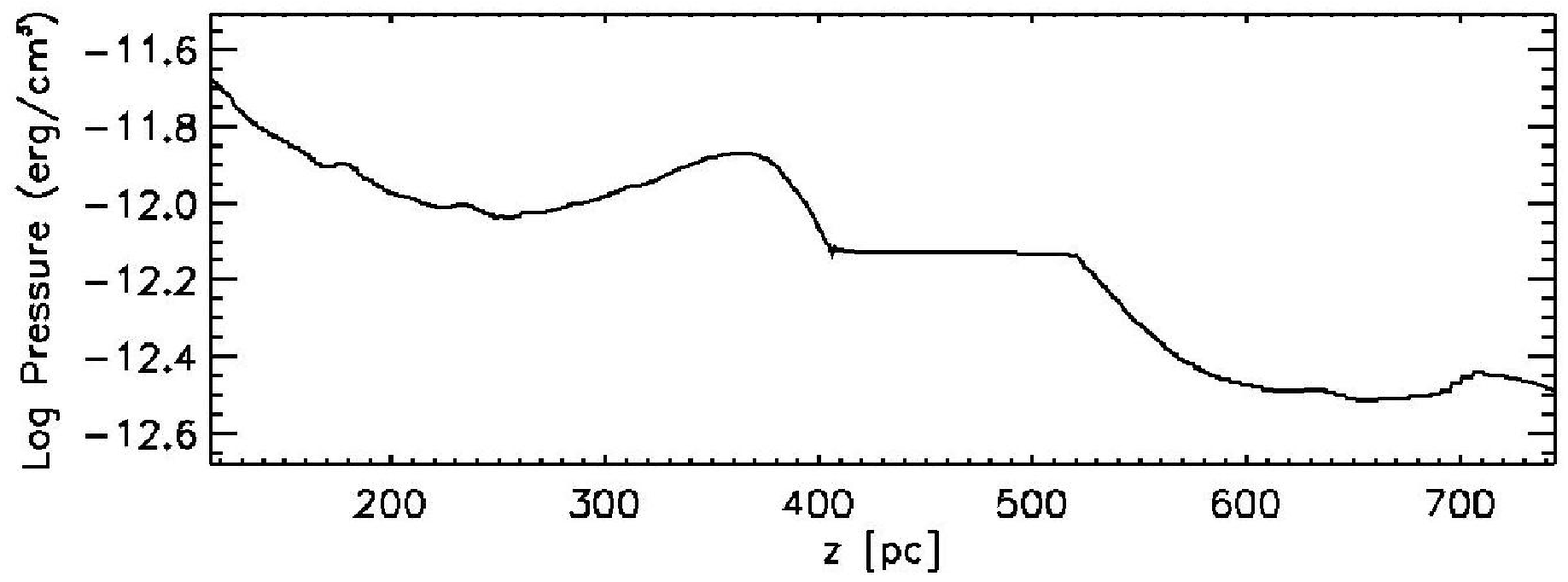}\hspace{0.5mm}\includegraphics[height=0.15\textwidth, angle=0]{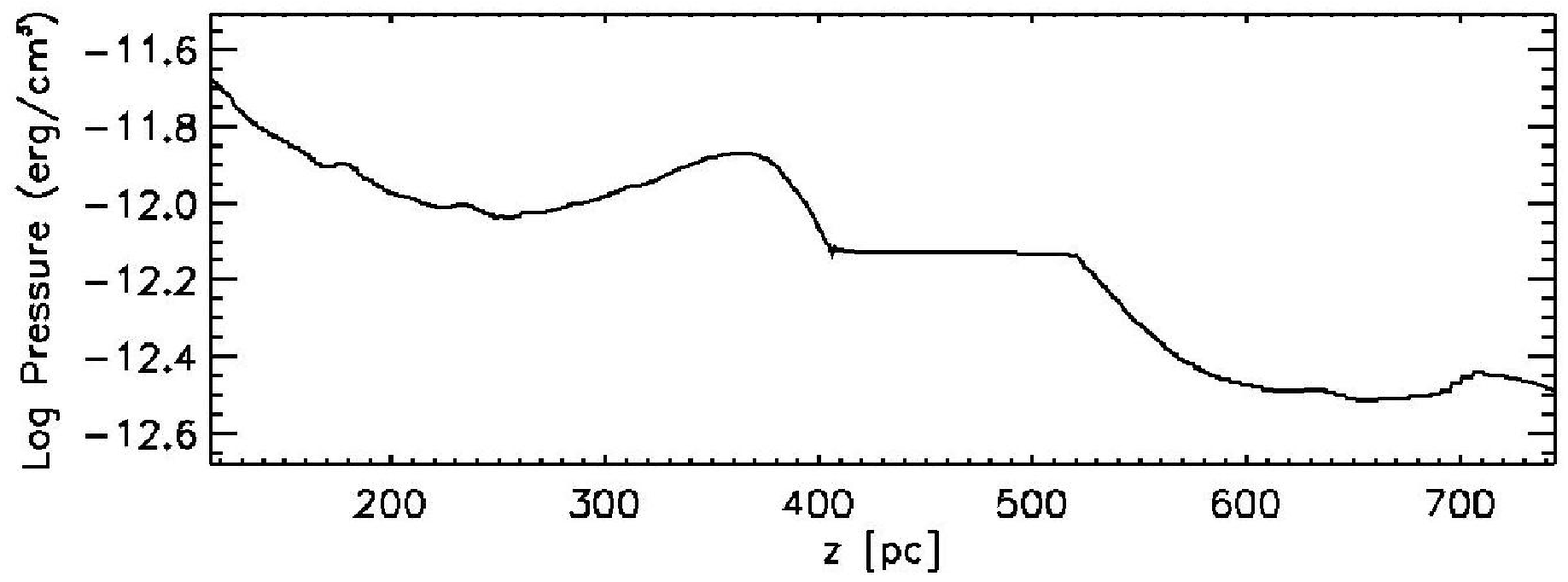}\hspace{0.5mm}\includegraphics[height=0.15\textwidth, angle=0]{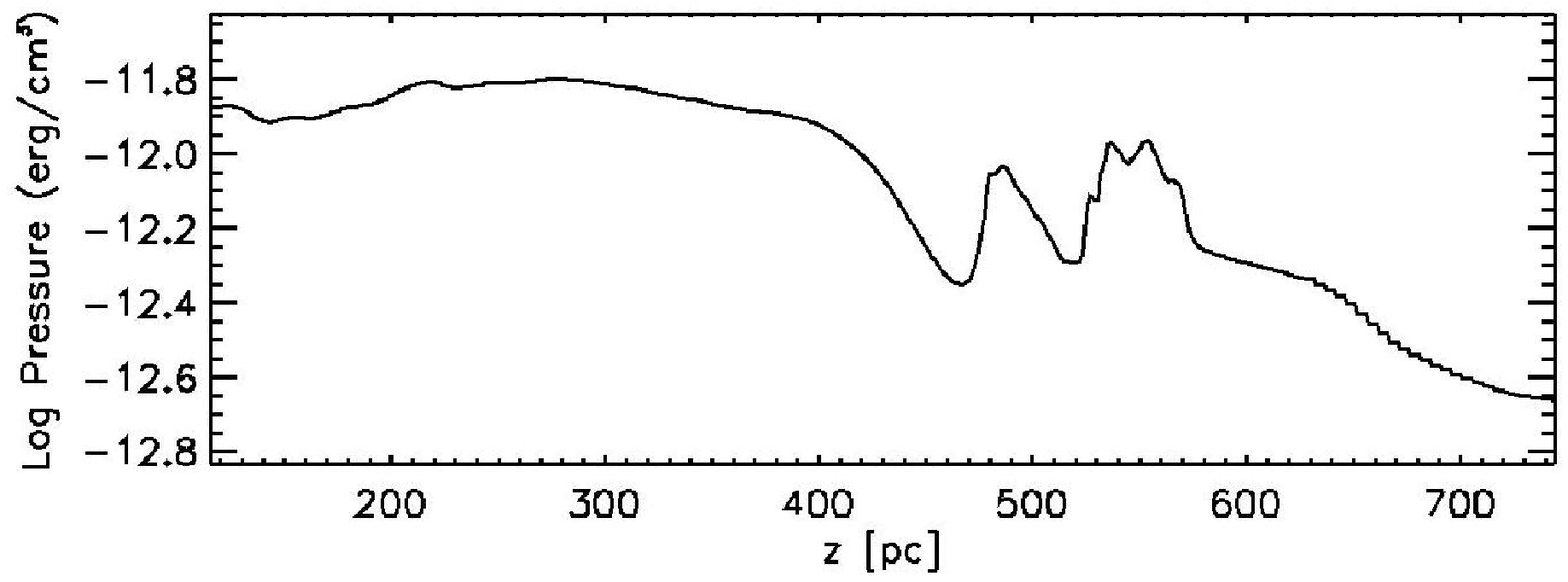}
 \caption{Log of thermal pressure for 2 Myrs, 4 Myrs, 6 Myrs, 8 Myrs, 10 Myrs, and 12 Myrs (left to right, top to bottom).  The thermal pressure in the hot bubble and the surrounding cool shell decreases with time before 2 Myr.  In the cool shell below the bubble, the pressure gradient becomes stronger than the gradient needed for hydrostatic equilibrium and remains like this until $\sim$6Myrs.   As a result, portions of the shell begin to rebound, building up a bulge of high density, high pressure gas just beneath the hot bubble.   This zone moves upwards into the hot bubble, modifying the hot bubble into a "mushroom cap" shape between 8 and 10 Myr. }
\label{mushroom_pres}
\efig 

\subsection{With Magnetic Field}
\subsubsection{Model B: 4$\mu$G magnetic field parallel to the galactic mid-plane}
Here we examine the evolution of an SNR in an ambient medium that is threaded by a magnetic field running parallel to the midplane.  The temperature and density structures for this model can be seen in Figures \ref{y_mhd_temp} and \ref{y_mhd_dens} respectively, for times up to 8 Myrs in increments of 1 Myr.  
Two prominent differences between this model and Model A are that, here, the shock front and hot bubble expand asymmetrically. 
The shock front travels faster in the two directions perpendicular to the magnetic field ($\pm\hat{x}$ and $\pm\hat{z}$), than in the directions parallel to the magnetic field ($\pm\hat{y}$).
In contrast, the hot bubble expands more quickly along the direction parallel to the magnetic field, especially at times $>$ 300,000 years. The bubble stays narrow in the directions perpendicular to ${\bf{B}}$.  At earlier times, the approximately spherically symmetric ram force from the explosion dominated, causing the bubble to expand roughly spherically symmetrically.  As the bubble expanded, the outward ram force decreased allowing the magnetic effects to become more manifest.  The magnetic field is denser in the remnant's shell than in the hot rarefied bubble. Thus the magnetic pressure in the shell helps to compress the hot bubble in the  directions perpendicular to the field.  As a result the hot rarefied bubble becomes elongated along the direction of the magnetic field, an effect that becomes more pronounced with time.  The cool shell also thickens in the direction perpendicular to ${\bf{B}}$.

The morphological anisotropies are more extreme than in the simulation of \cite{ferriere91}.  They performed a two-dimensional simulation of a SNR in a medium of constant mass density and threaded by a 3 $\mu$G magnetic field of constant density and running parallel to the galactic midplane. The hot portion of their SNR is only slightly elongated with respect to the outer part of its shell.  Our results show much more elongation.  There are three reasons for this (1)our magnetic field is stronger at 4 $\mu$ G, (2) their  number density (0.32 cm$^{-3}$) is much higher than ours ($<$0.1 cm$^{-3}$ at 400 pc), and (3) the maximum evolution time for the SNR presented in their figures (1 Myrs) is much shorter than ours (8 Myrs). As demonstrated in Figure \ref{y_mhd_dens} bubbles evolving in a uniform magnetized media become more elongated with time.  For a detailed explanation of the effects of a magnetic field on a SNR, please see \cite{ferriere91}.

Bubbles elongated along the direction of the magnetic field, such as W44 and 3C58, have been observed, though in W44's case, other physics may have contributed to the elongation \citep{cox}.  W44 has an aspect ratio of 7 to 5, and 3C58 has an aspect ratio of 10 to 7 \citep{kundu_velusamy}.  W44 is thought to have evolved to the stage where it is forming a cool shell.  Our remnant forms a cool shell between 100,000 and 300,000 years.  At 300,000 years our aspect ratio is virtually 1 to 1, thus even less extreme than W44.  If we increase the magnetic field from 4 $\mu$G to 7.1 $\mu$G, at 300,000 years our aspect ratio is 7 to 6, thus even less extreme than W44.  In the simulation, the aspect ratio increases as the remnant ages.  We expect that comparatively few of the very old and elongated remnants have been or will be observed because their bubbles' soft X-ray emission and their shells' synchrotron emission are much dimmer than those of younger remnants.
This model is also different than Model A in that Model B exhibits less vertical rise.
With the addition of our $4~\mu \rm{G}$ magnetic field, the bubble is effectively pinned by magnetic effects.  The center of the bubble moves only about 23 pc over the course of 8 Myrs.

A fourth difference between this model and Model A is its structure.  By 300,000 years, Model A exhibited eddies in all three planes, the x-y, y-z, and x-z planes.  In Model B, we see less intricate structure in the x-z plane (see Figure \ref{xy}), and no sign of eddies in the y-z and x-y planes.  In Model B, turbulence is constrained by the magnetic tension, as mentioned by \cite{parker}.  Furthermore, Model A developed a mushroom-like shape, seen in both y-z cuts (Figure \ref{mushroom1}) and x-z cuts near the end of its life.  Although Model B develops an elongated mushroom cross section (it is highest in the center) in the y-z plane, near the end of its life, it does not develop such a shape in the x-y plane.

We do not see a mushroom like structure in the x-y plane throughout our simulation time of 8 Myrs.

Interestingly, unlike Model A where the bubble shrinks in all directions after 2 million years, the bubble in this model continues to expand in the direction parallel to the magnetic field throughout the simulated eight million years.  Even at late times it expands in this direction at a remarkable rate.  Between 7 million and 8 million years, the bubble expands to a width of about 500 pc in the $\hat{y}$-direction (250 pc on each side of the reflective axis at its widest point) from about 460 pc (230 pc on each side).  

\bfig 
\centering
\includegraphics[height=0.3\textwidth, angle=0]{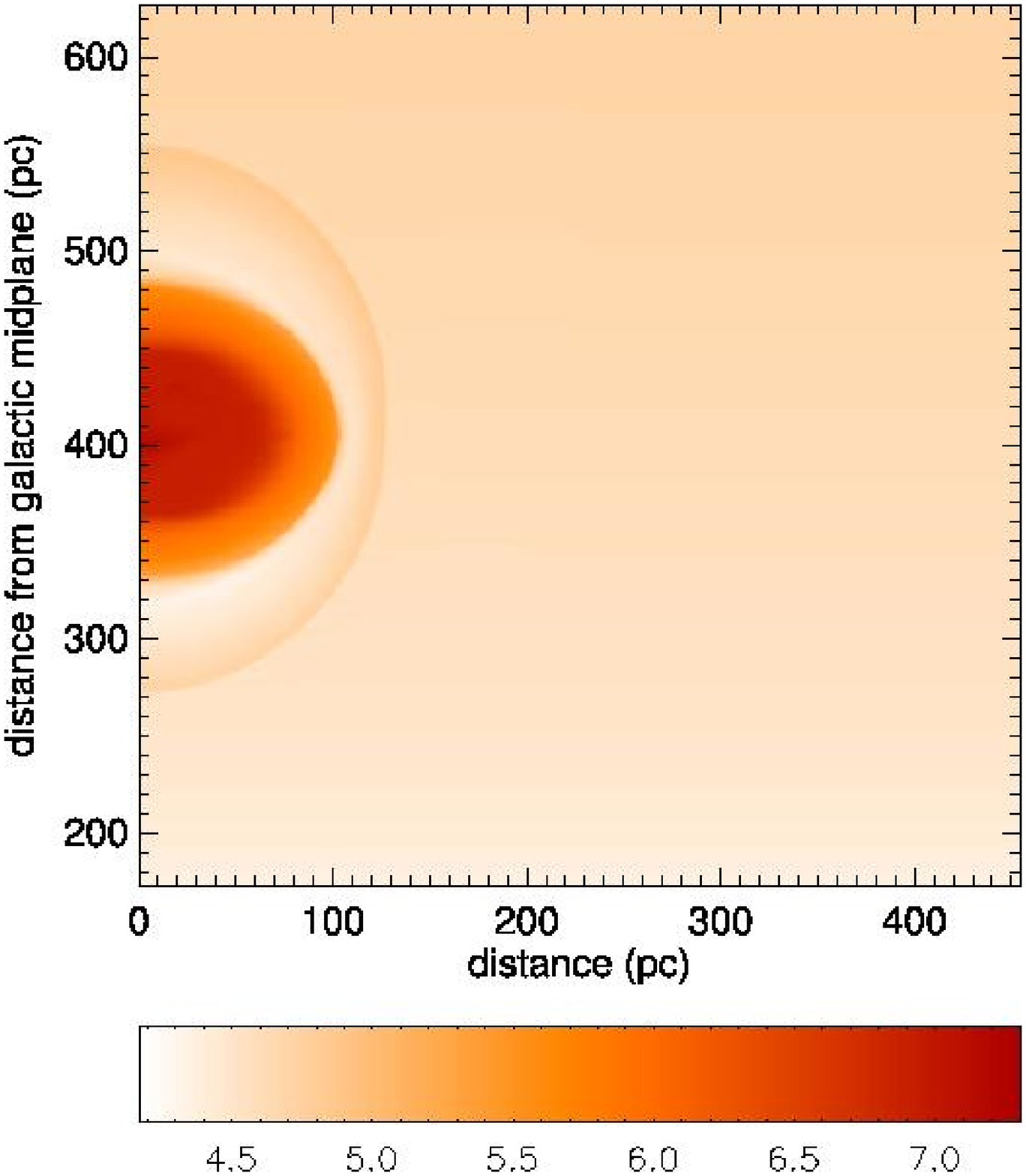}\includegraphics[height=0.3\textwidth, angle=0]{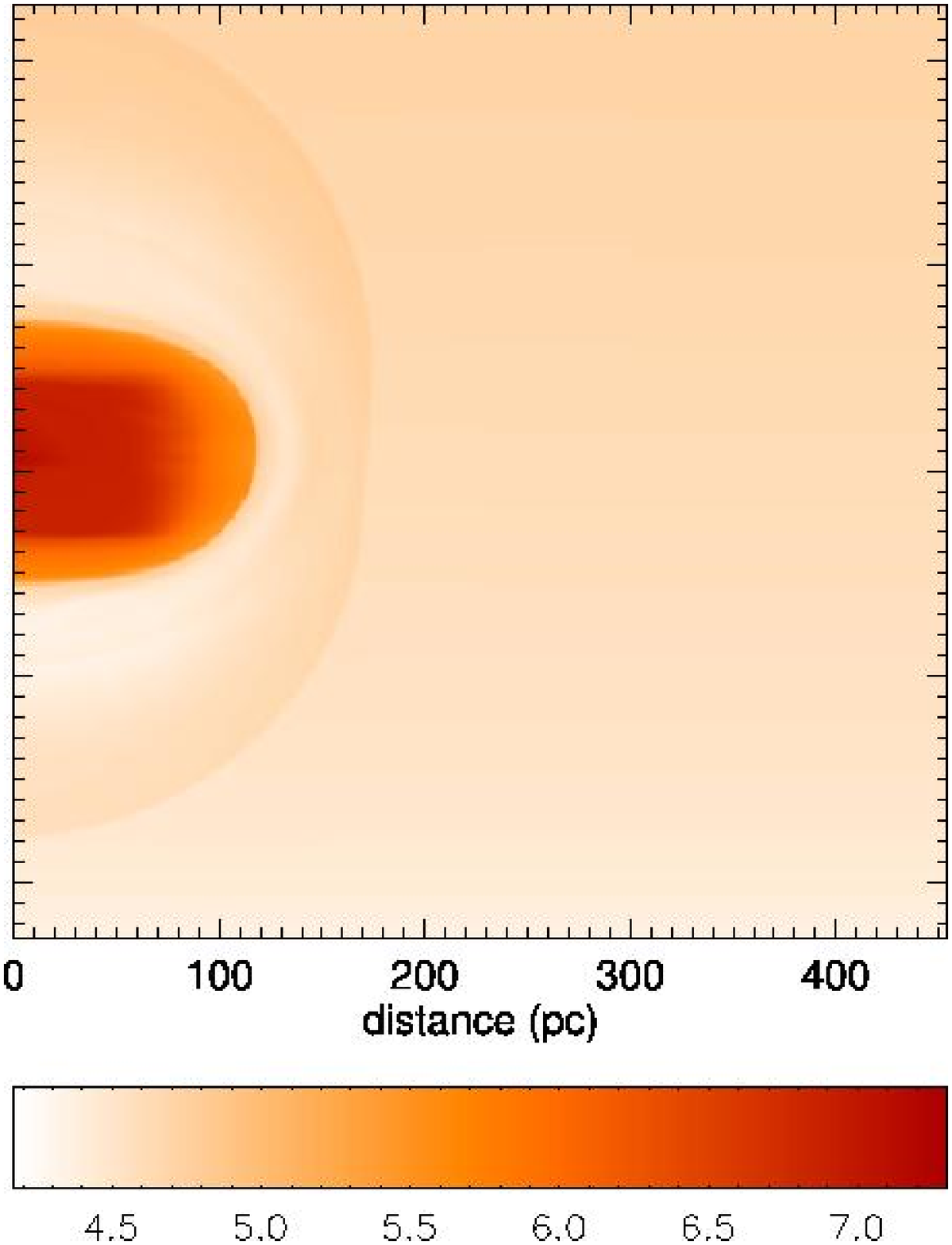}\includegraphics[height=0.3\textwidth, angle=0]{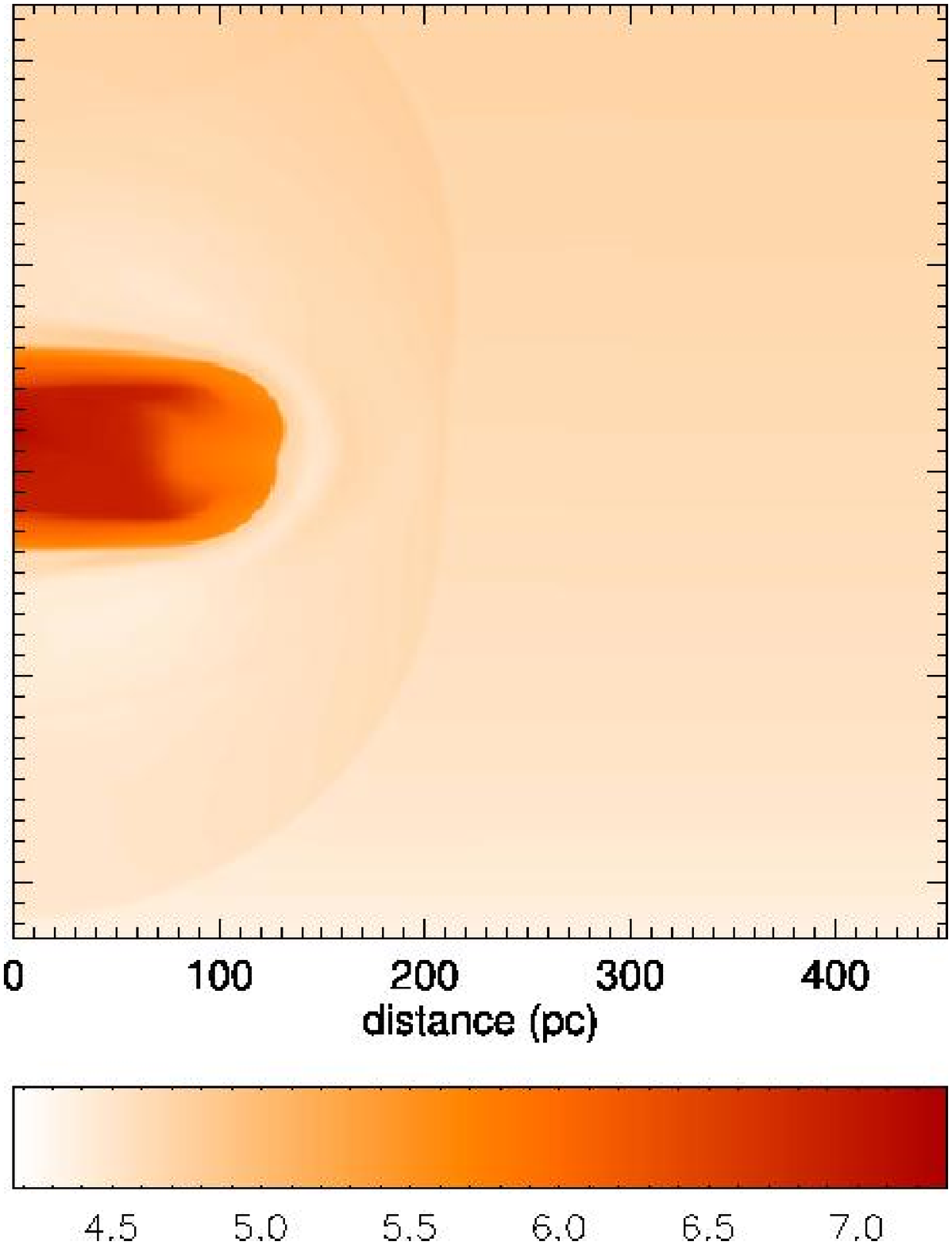}\includegraphics[height=0.3\textwidth, angle=0]{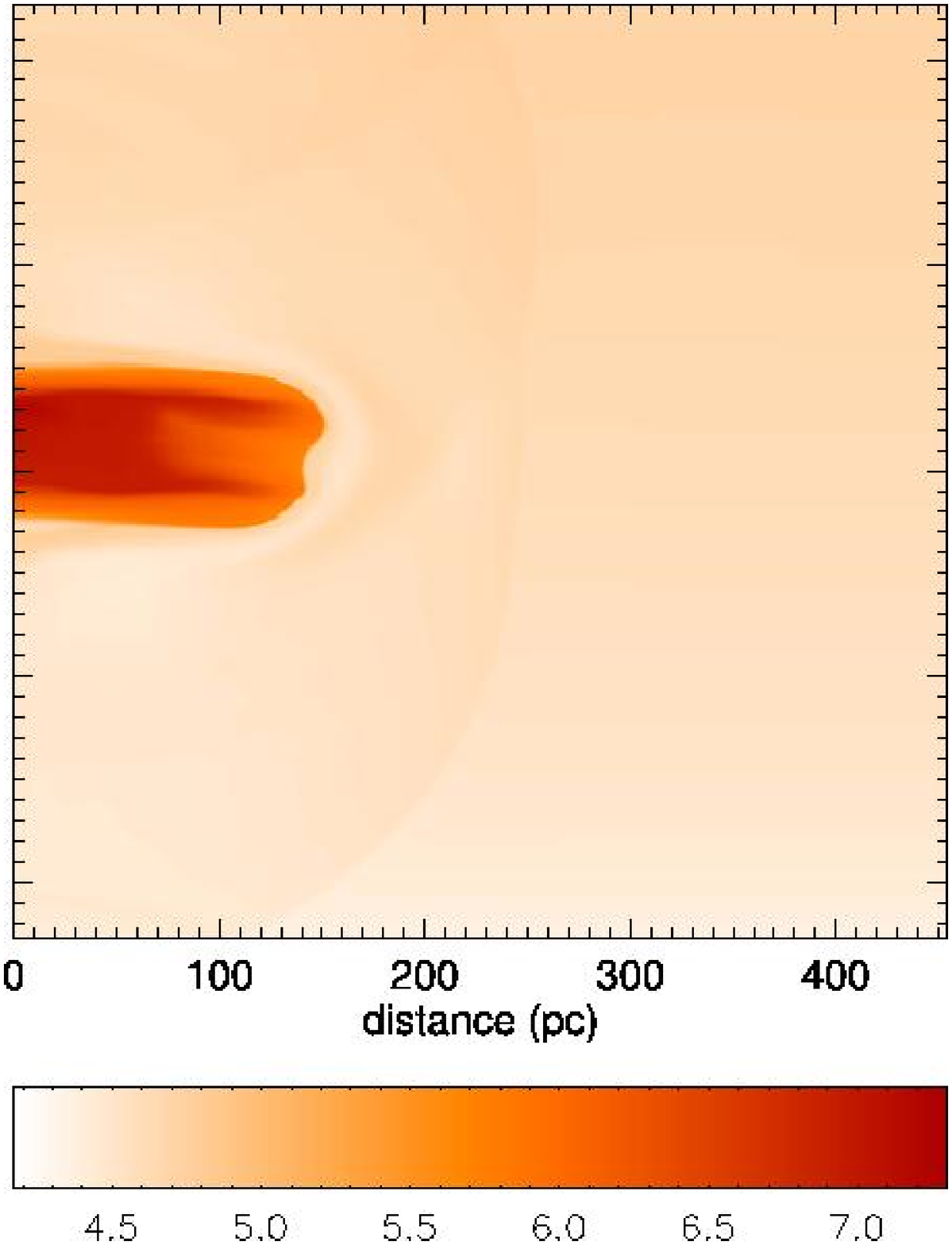}
\vspace{1.5mm}

\includegraphics[height=0.3\textwidth, angle=0]{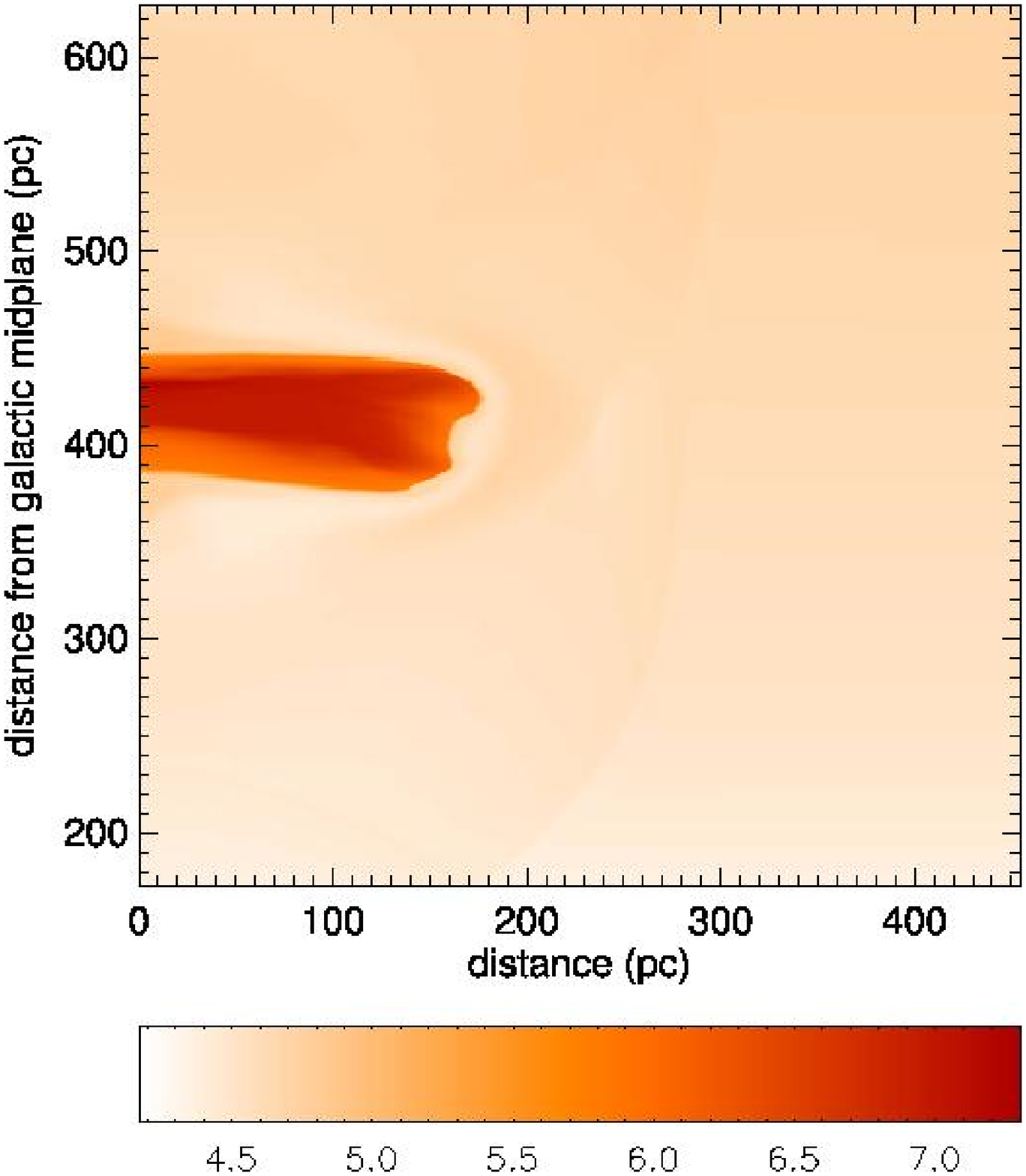}\includegraphics[height=0.3\textwidth, angle=0]{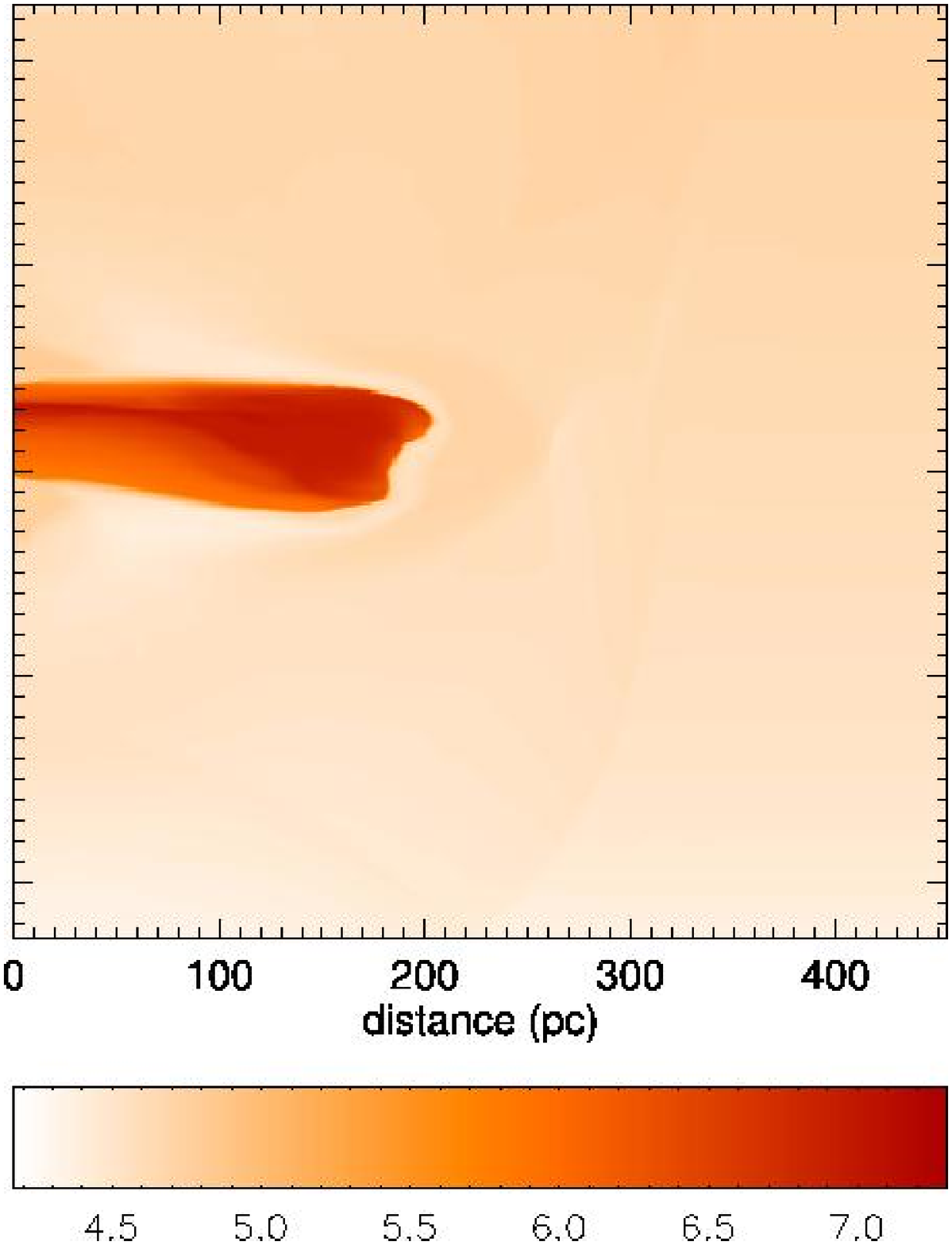}\includegraphics[height=0.3\textwidth, angle=0]{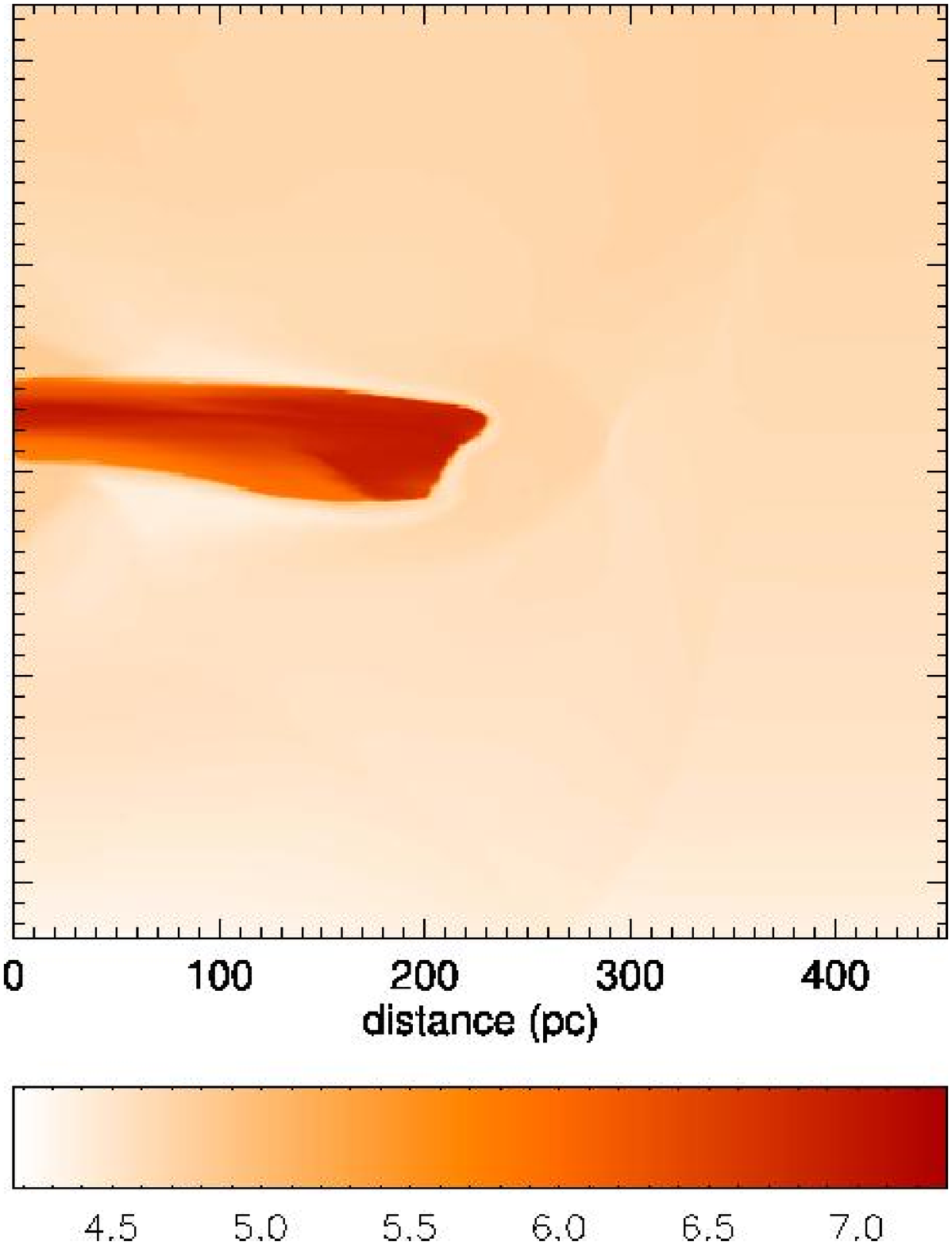}\includegraphics[height=0.3\textwidth, angle=0]{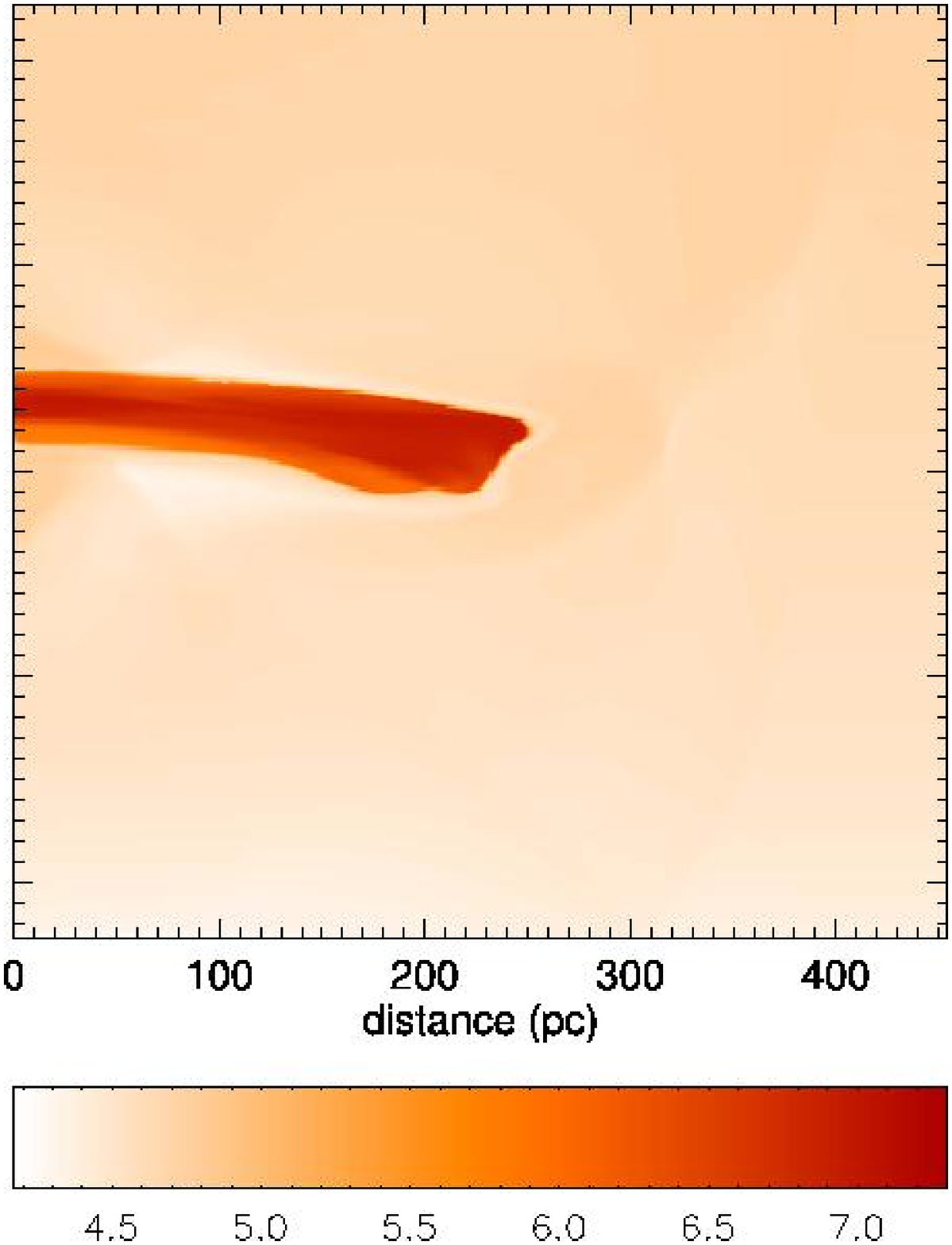}

\caption{Time sequence of temperature (Kelvin) plots for Model B.  The log of temperature is shown (left to right, top to bottom) every million years to 8 Myrs.  The SNR is pinned by the magnetic field which decreases its vertical rise to 23 pc above the initial explosion height of 400 pc.  As in the density plots, Figure \ref{y_mhd_dens}, we can see that the hot bubble becomes increasing long and thin with time.  The bubble is similarly thin in the $\hat{x}$-direction as it is in the $\hat{z}$-direction.  This yields a bubble which is cigar shaped (long in the $\hat{y}$-direction and short in the $\hat{x}$ and $\hat{z}$-directions). These figures show the x=0 (y-z plane).}
\label{y_mhd_temp}
\efig

\bfig 
\centering

\includegraphics[height=0.3\textwidth, angle=0]{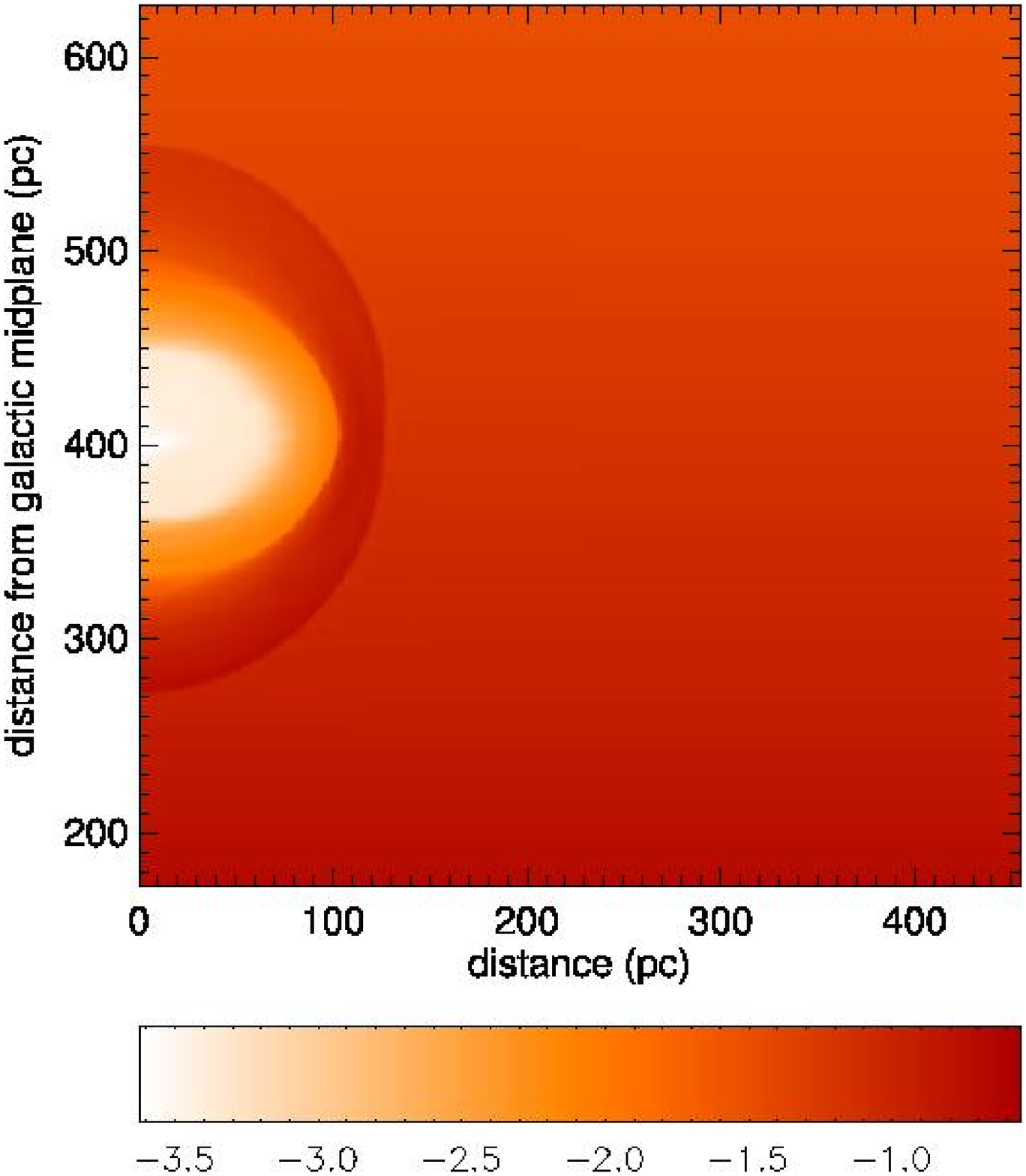}\includegraphics[height=0.3\textwidth, angle=0]{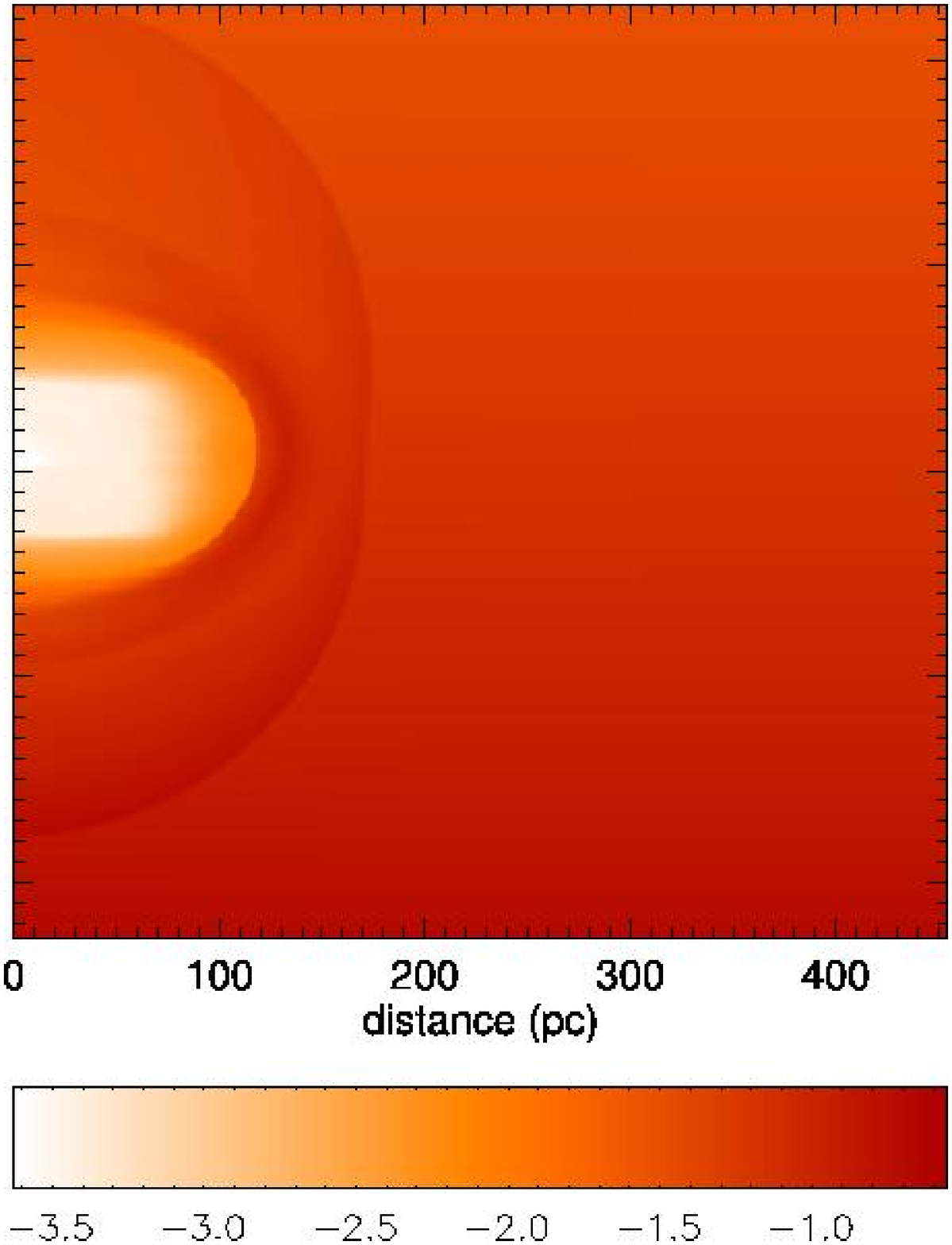}\includegraphics[height=0.3\textwidth, angle=0]{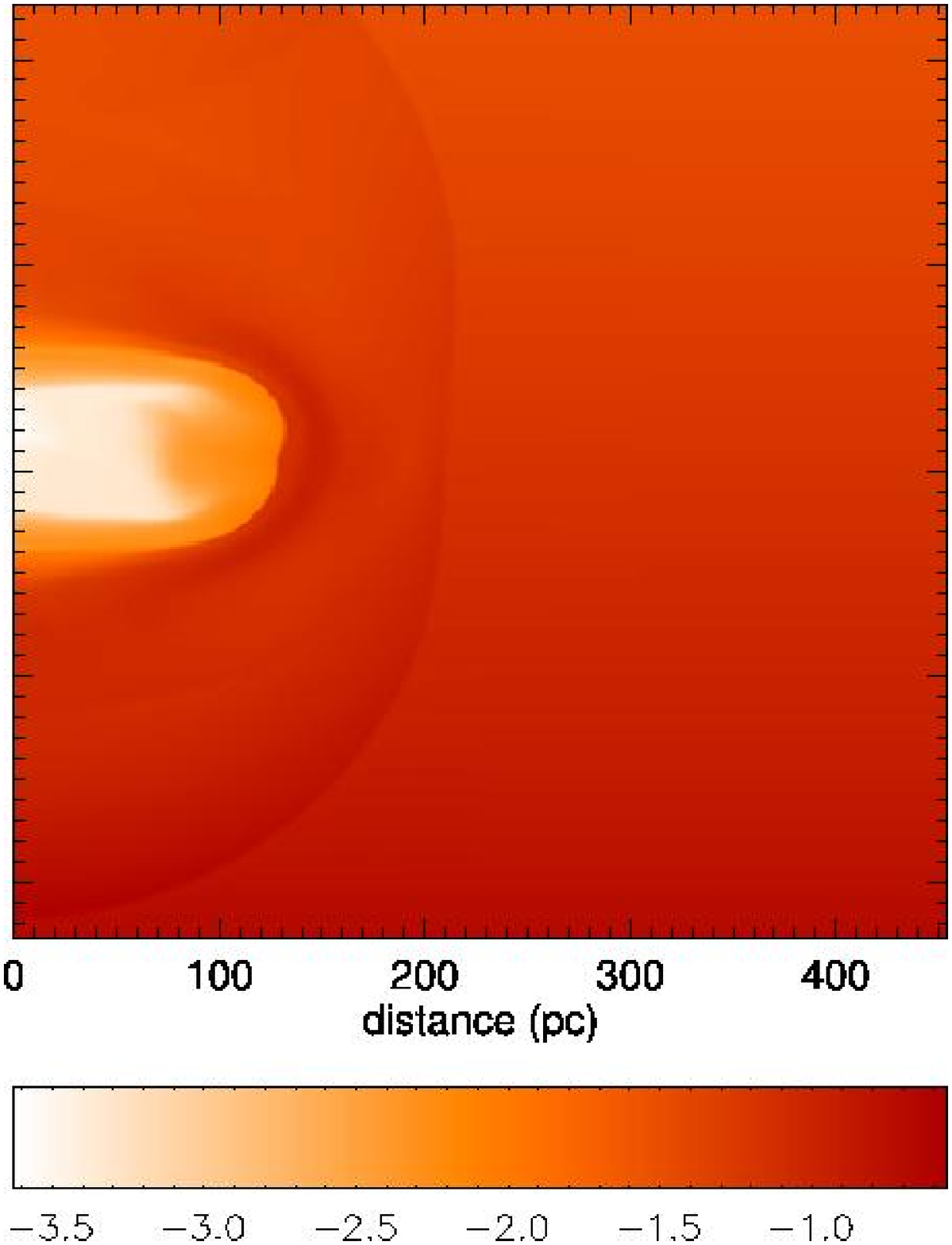}\includegraphics[height=0.3\textwidth, angle=0]{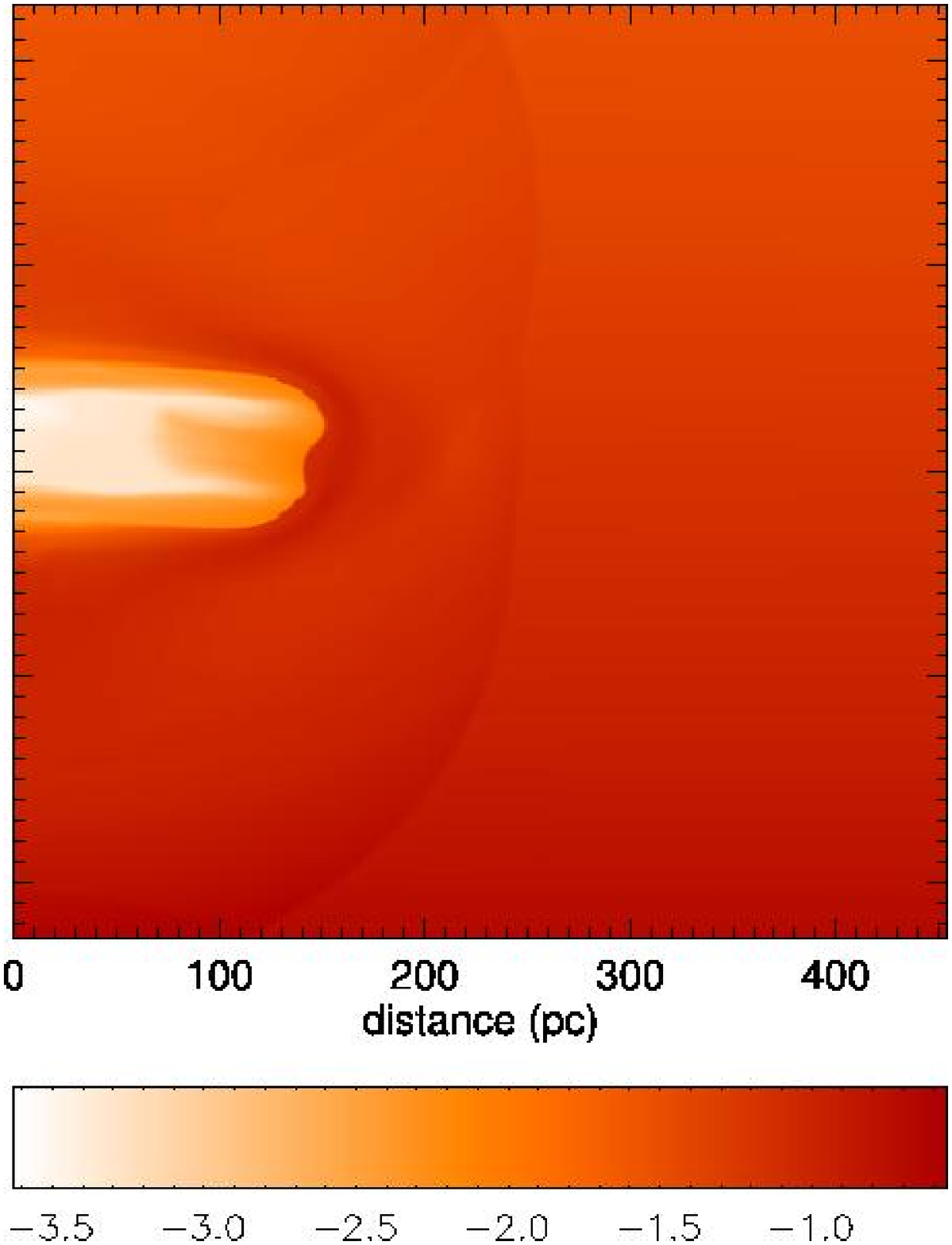}
\vspace{1.5mm}

\includegraphics[height=0.3\textwidth, angle=0]{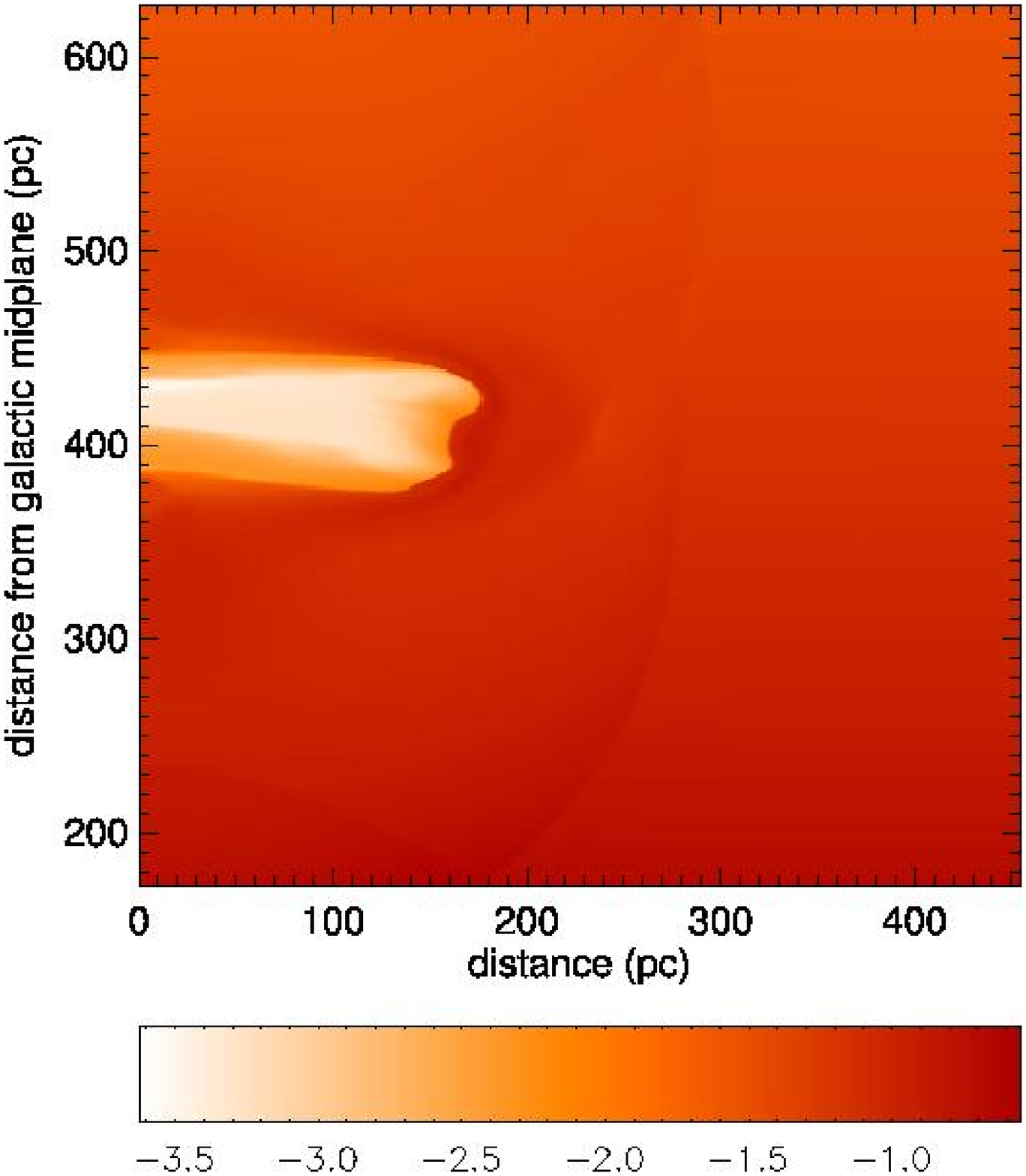}\includegraphics[height=0.3\textwidth, angle=0]{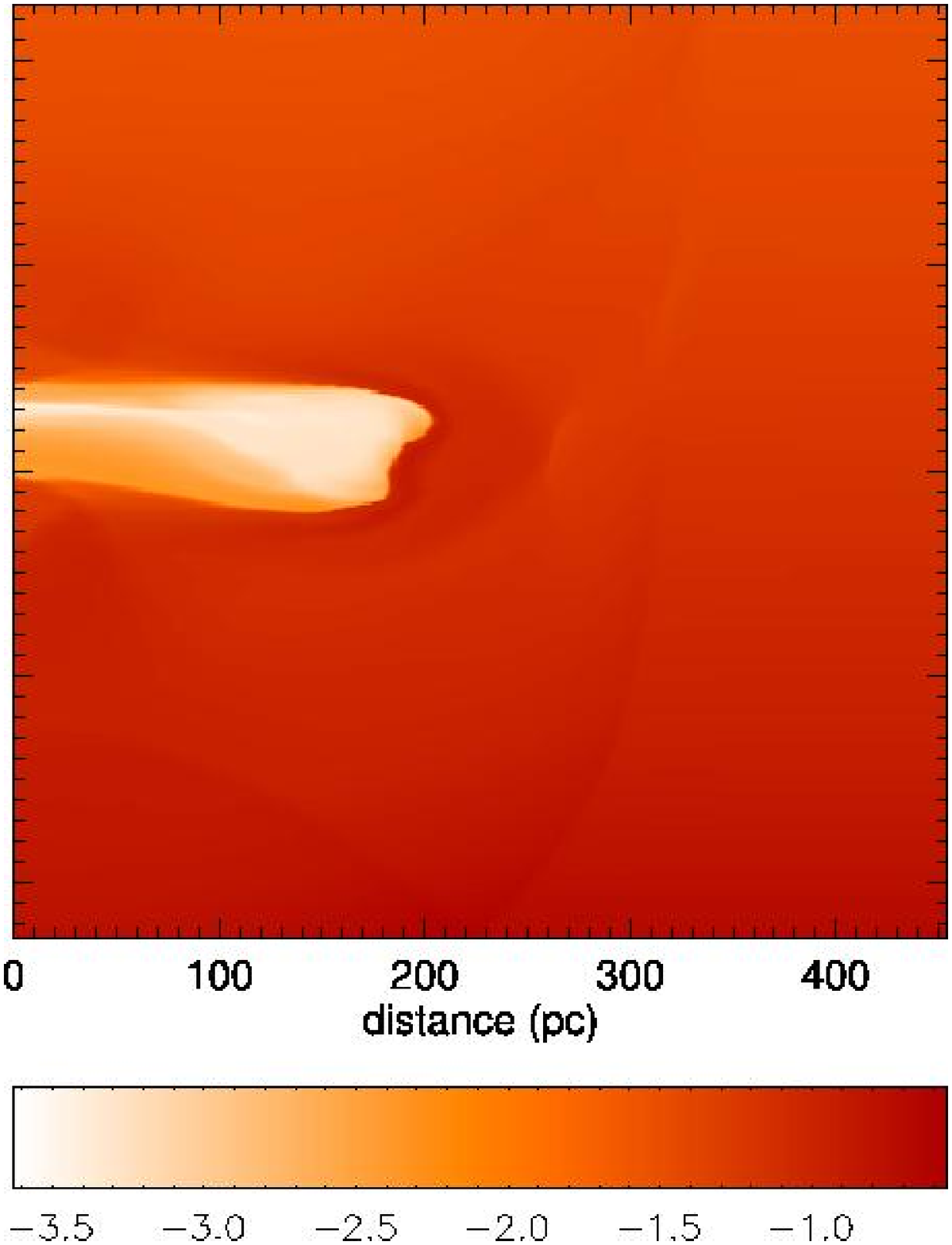}\includegraphics[height=0.3\textwidth, angle=0]{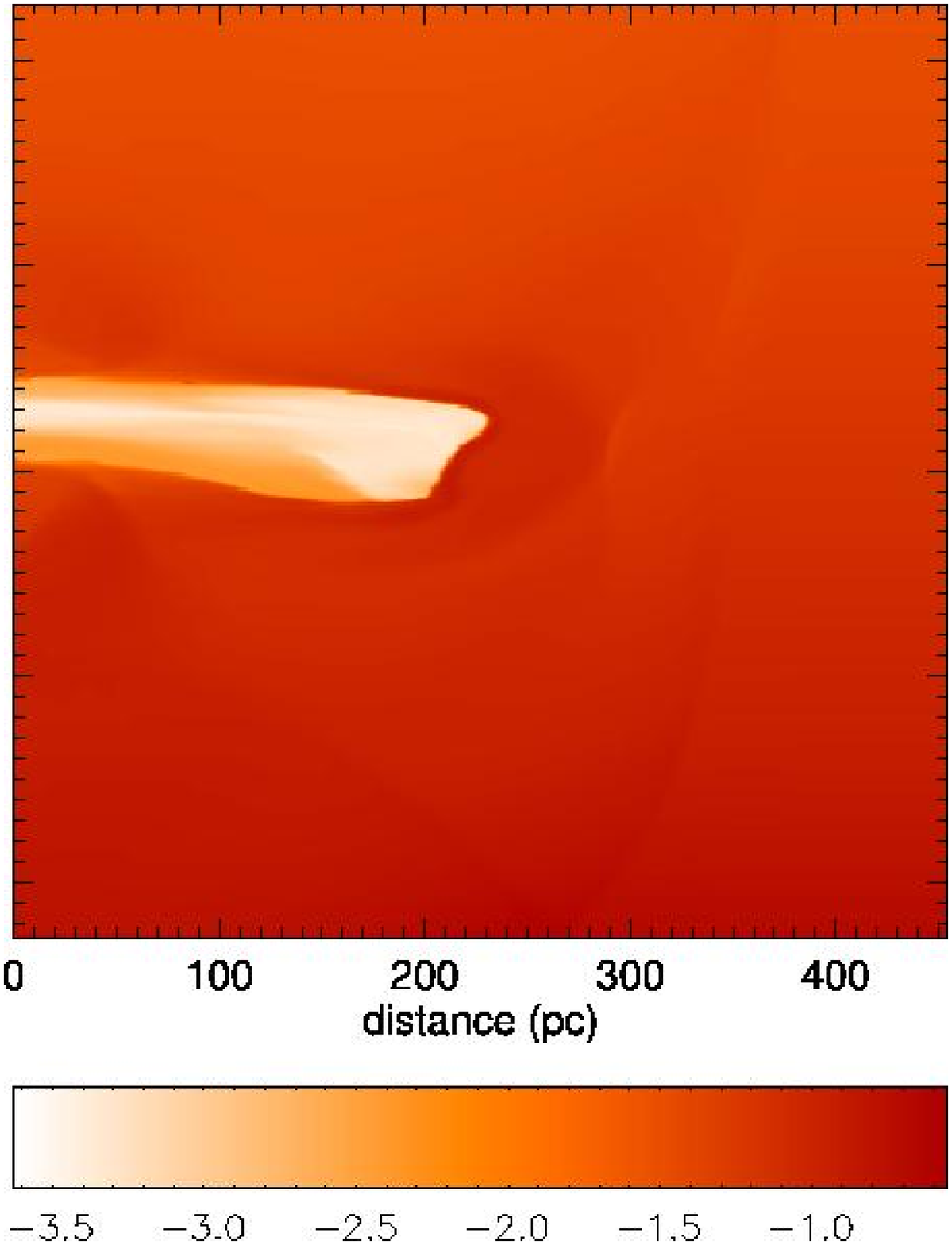}\includegraphics[height=0.3\textwidth, angle=0]{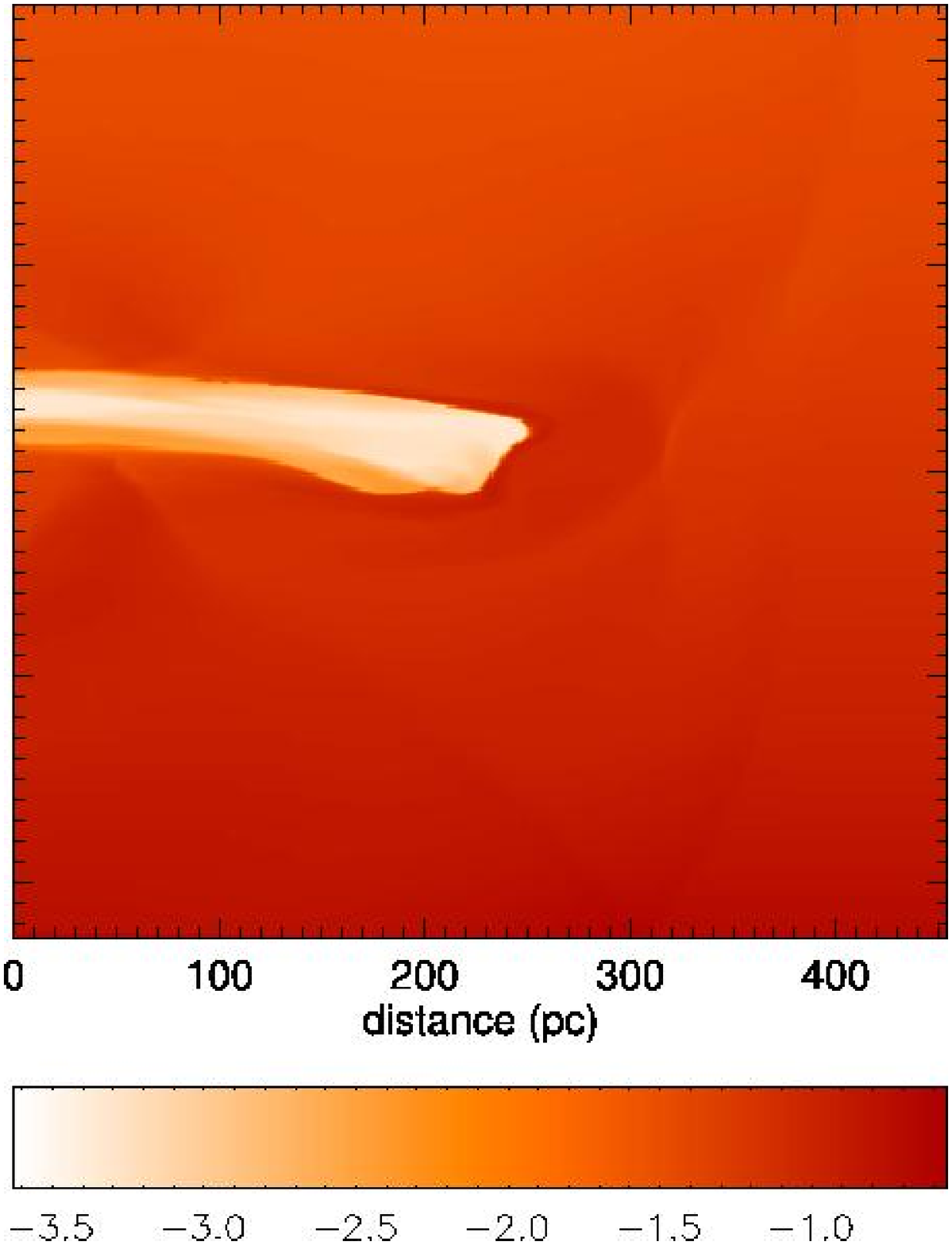}

  \caption{Time sequence of density (atoms/$\rm{cm}^3$) plots for Model B.  The log of density is shown (left to right, top to bottom) every million years up to 8 Myrs. These figures show the x=0 (y-z plane).}
\label{y_mhd_dens}
\efig 

\bfig
\centering
\includegraphics[height=0.3\textwidth, angle=0]{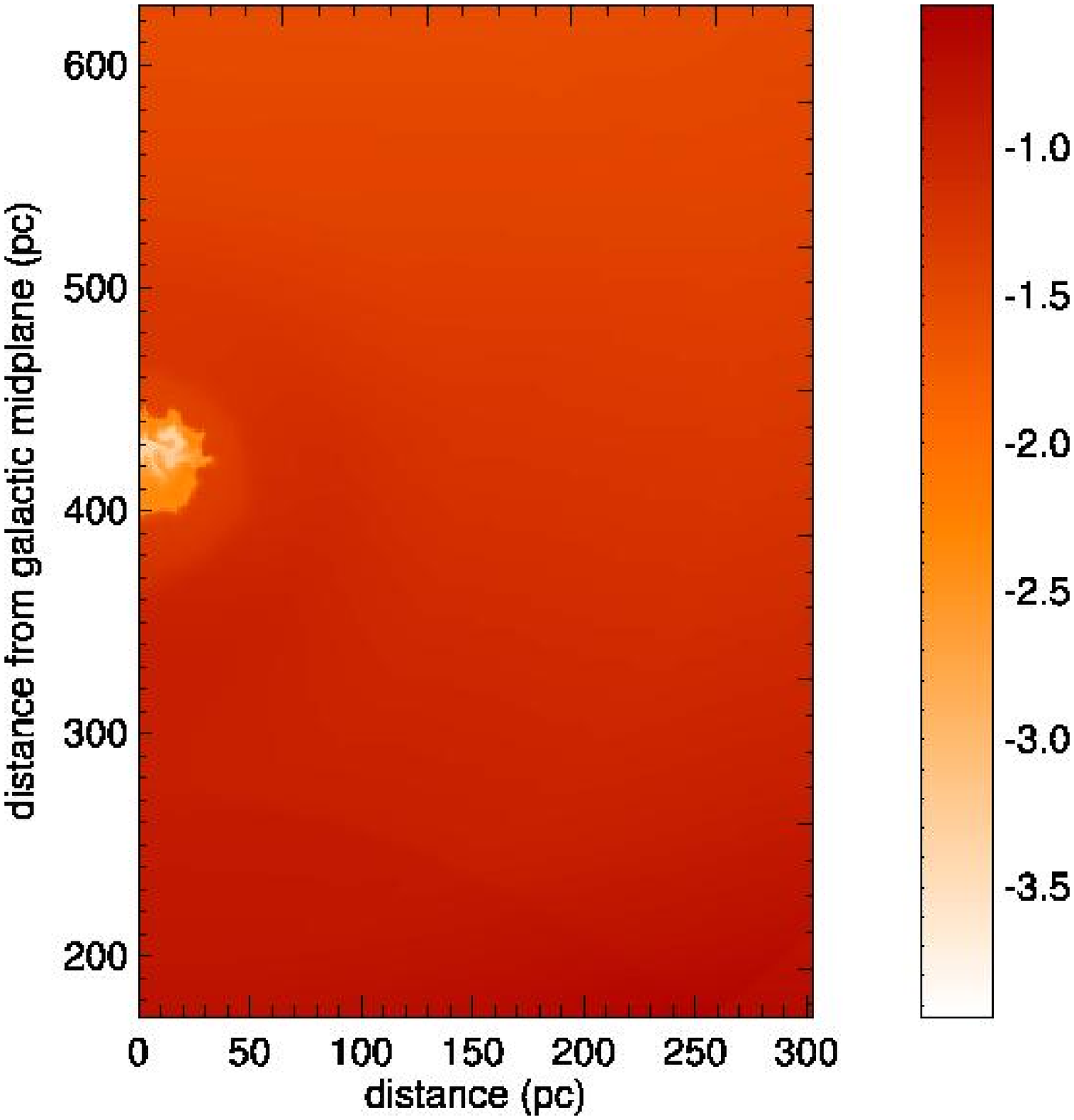}
\caption{X-y slice (magnetic field is into the page) of density (atoms/cm$^3$) for Model B at at 6 Myrs.  We see some structure in this plane that is not visible in the y-z plane.  }
\label{xy}
\efig

\subsubsection{Model C: 4$\mu$G magnetic field perpendicular to the galactic mid-plane}
Here, we examine the evolution of an SNR in the presence of an ambient magnetic field perpendicular to the midplane.  In Figures  \ref{z_mhd_temp} and \ref{z_mhd_dens}, we show the evolution of the remnant's temperature and density structure through 8 Myrs in increments of 1 Myr .
This simulated remnant is similar to Model B in two ways: (1) there are no eddies in the y-z plane (there are eddies in the x-y plane), and (2) the bubble is elongated along the magnetic field.  It is different from model B in that the magnetic field and gravitational field (and hence density and pressure gradients) are now parallel to each other.  In model B, the horizontal magnetic field 'pins' the SNR, effectively preventing it from rising, whereas now the vertical direction is the one direction in which the bubble can expand freely.  As a result, the hot bubble grows slightly larger in the direction of the field than in model B.  By 8 Myrs the Model B bubble has expanded to a total length is 500 pc (250 pc on either side of the y=0 plane at its widest point), but the Model C bubble has expanded to a total length of 470 pc (360 pc above and 110 pc below the explosion height).  

By 8 Myrs the top of the bubble has reached a maximum height of about 760 pc.  We calculate the height of the center of the bubble by finding the volume-weighted average height of all of the rarefied material, that with a density less than $0.0043~ {\frac{\rm{atoms}}{\rm{cm}^3}}~(10^{-26}~ {\frac{\rm{g}}{\rm{cm}^3}})$.  The center's height is 579~pc at 8 Myrs.  The maximum rise of the center of the bubble is 179~pc at 8~Myrs compared with a rise of 23 pc at 8 Myrs for Model A or a rise of 59~pc at 12~Myrs for Model A.
Having the magnetic and gravitational fields in the same direction has increased the distance the bubble has risen, since the direction in which the hot material is most free to move (parallel to the magnetic field, z-direction) is the same as the direction of the preferential expansion due to the thermal pressure gradient and the buoyant force (positive z-direction). 

Because the gravitational potential and thermal pressure gradient in Model C are in the same direction as the magnetic field, the bubble flattens even more quickly than in Model B.  The combination of squeezing in the $\hat{x}$ and $\hat{y}$ directions by the magnetic field and relative freedom in the direction parallel to the magnetic field, the buoyant force, and preferential expansion causes a higher velocity in the positive z-direction than we have seen in any direction for these three simulations.

\bfig 
\centering
\includegraphics[height=0.4\textwidth, angle=0]{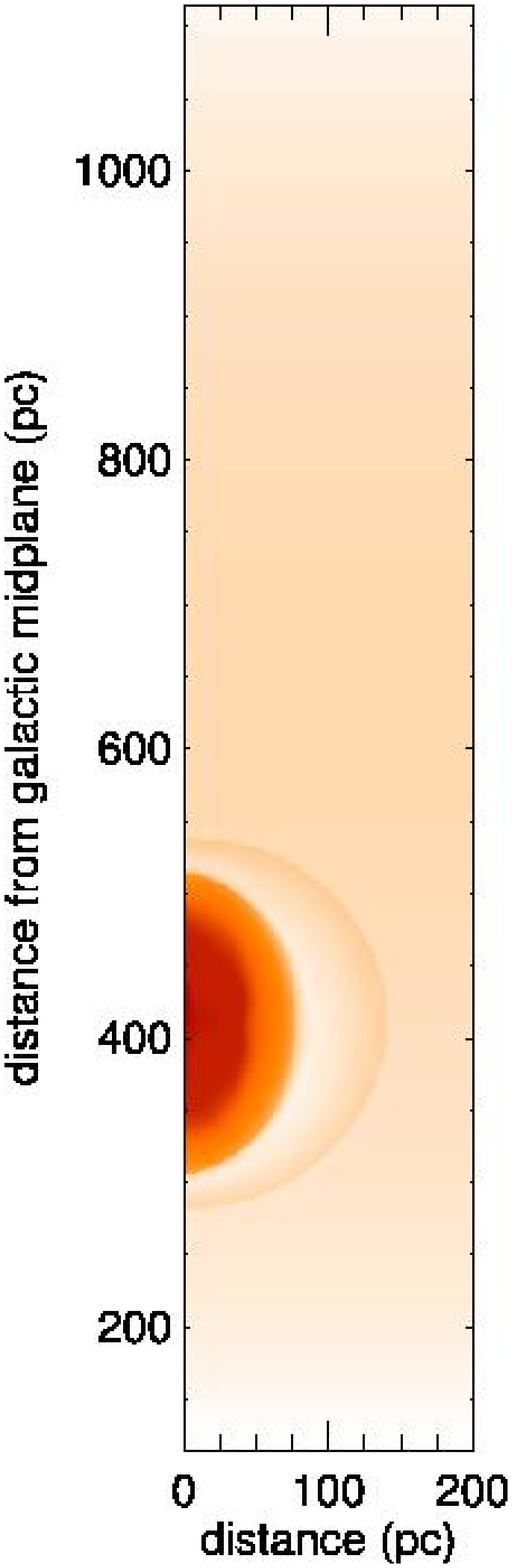}\hspace{0.5mm}\includegraphics[height=0.4\textwidth, angle=0]{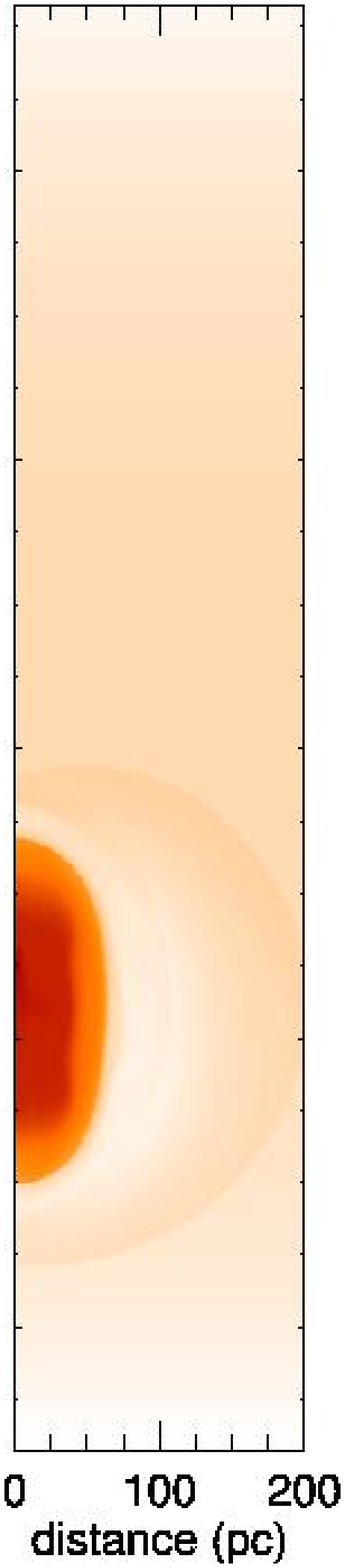}\hspace{0.5mm}\includegraphics[height=0.4\textwidth, angle=0]{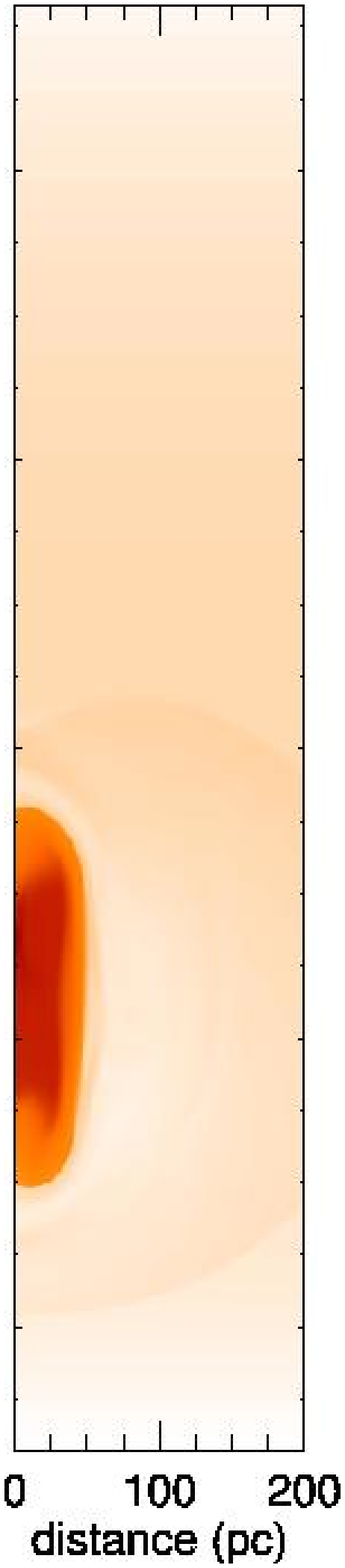}\hspace{0.5mm}\includegraphics[height=0.4\textwidth, angle=0]{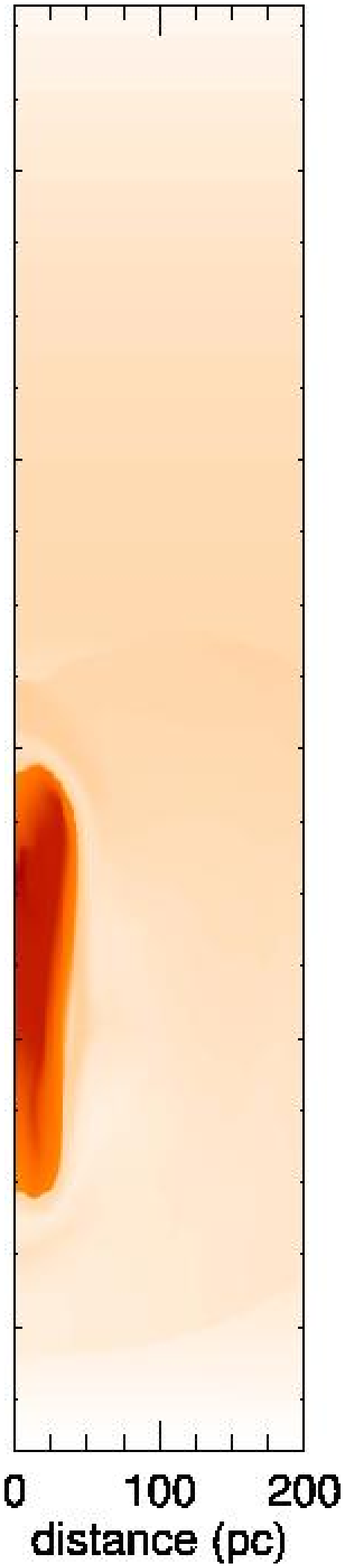}\hspace{0.5mm}\includegraphics[height=0.4\textwidth, angle=0]{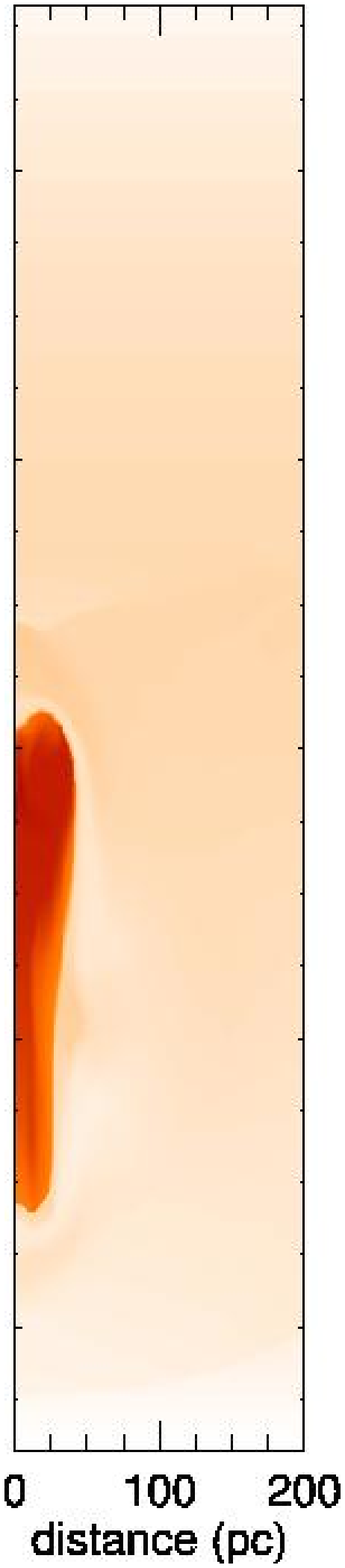}\hspace{0.5mm}\includegraphics[height=0.4\textwidth, angle=0]{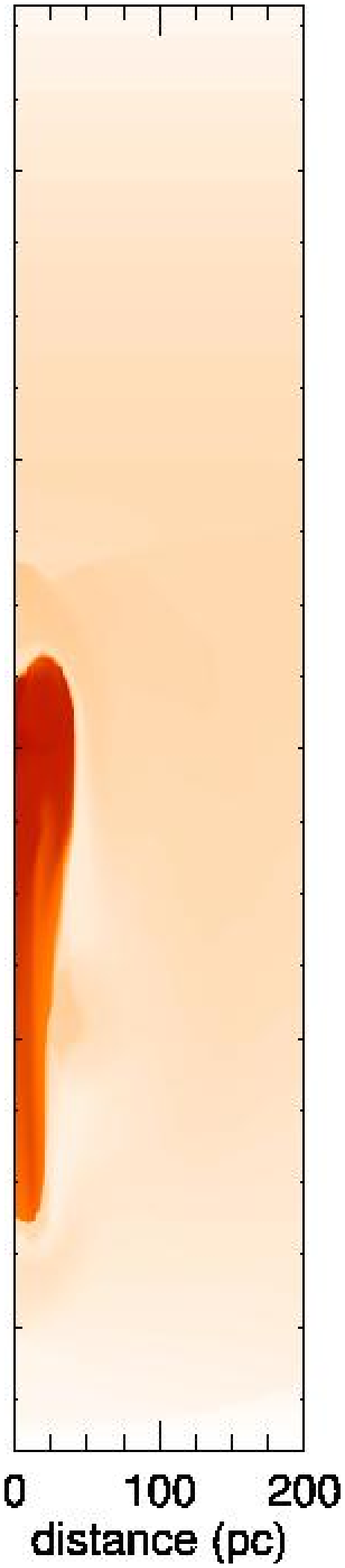}\hspace{0.5mm}\includegraphics[height=0.4\textwidth, angle=0]{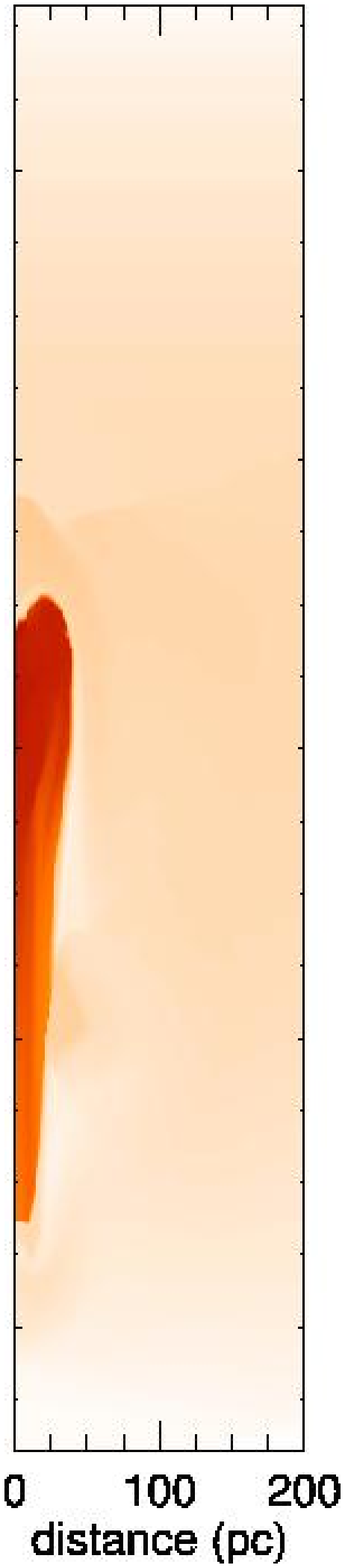}\hspace{0.5mm}\includegraphics[height=0.4\textwidth, angle=0]{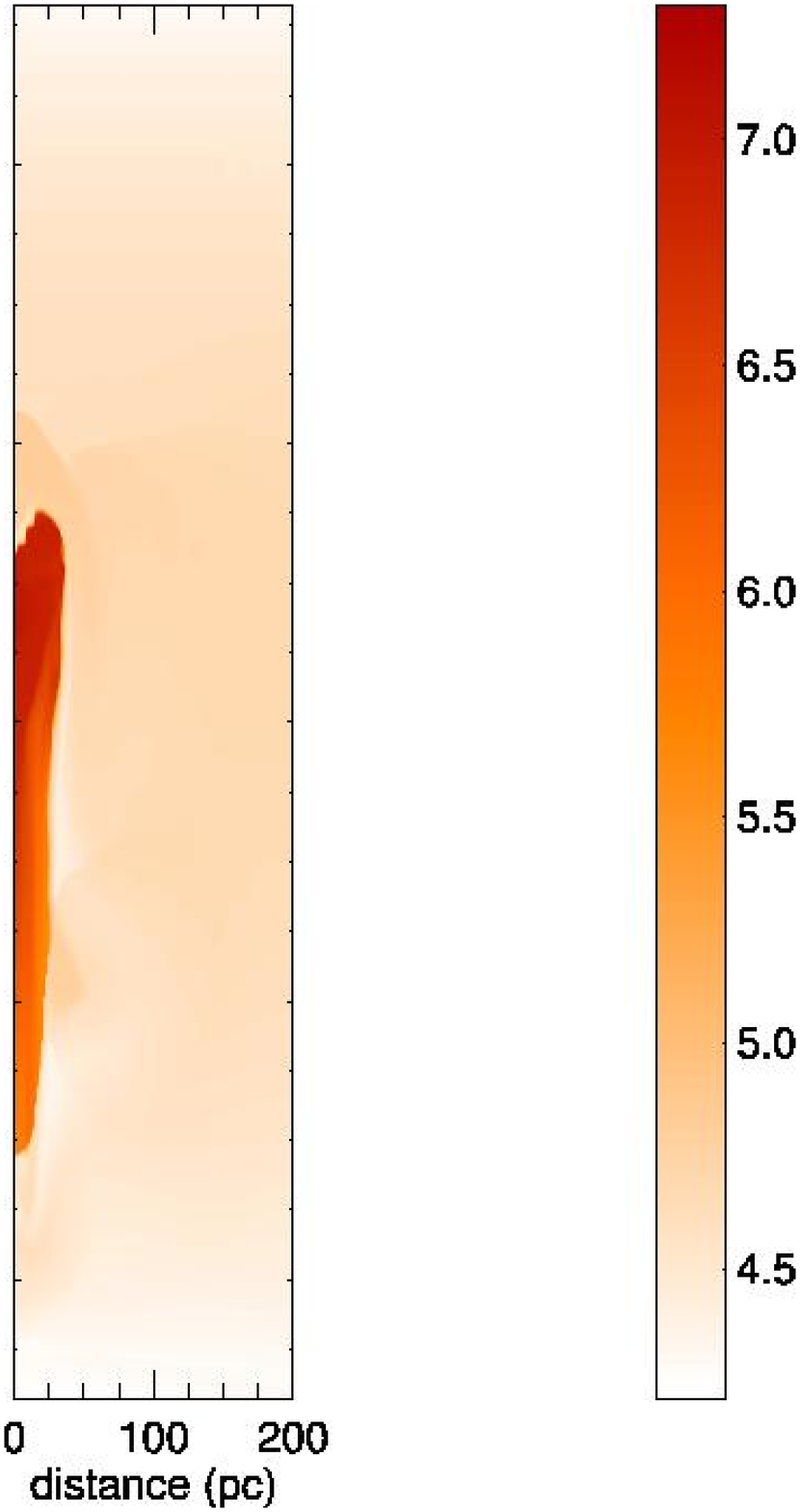}\hspace{0.5mm}

 \caption{Time sequence of temperature (Kelvin) plots for Model C.  The log(temperature) is shown (left to right) every million years up to 8 Myrs. Here we can see that some hot gas has risen (from the explosion height of 400~pc above the midplane) to about 760~pc above the midplane.  In this case, the magnetic field and the gravitational potential are both perpendicular to the midplane.  The two directions parallel to the midplane are identical so cuts through the y-z or x-z plane are virtually indistinguishable. These are slices through x=0 (y-z plane).}
\label{z_mhd_temp}
\efig

\bfig 
\centering
\includegraphics[height=0.4\textwidth, angle=0]{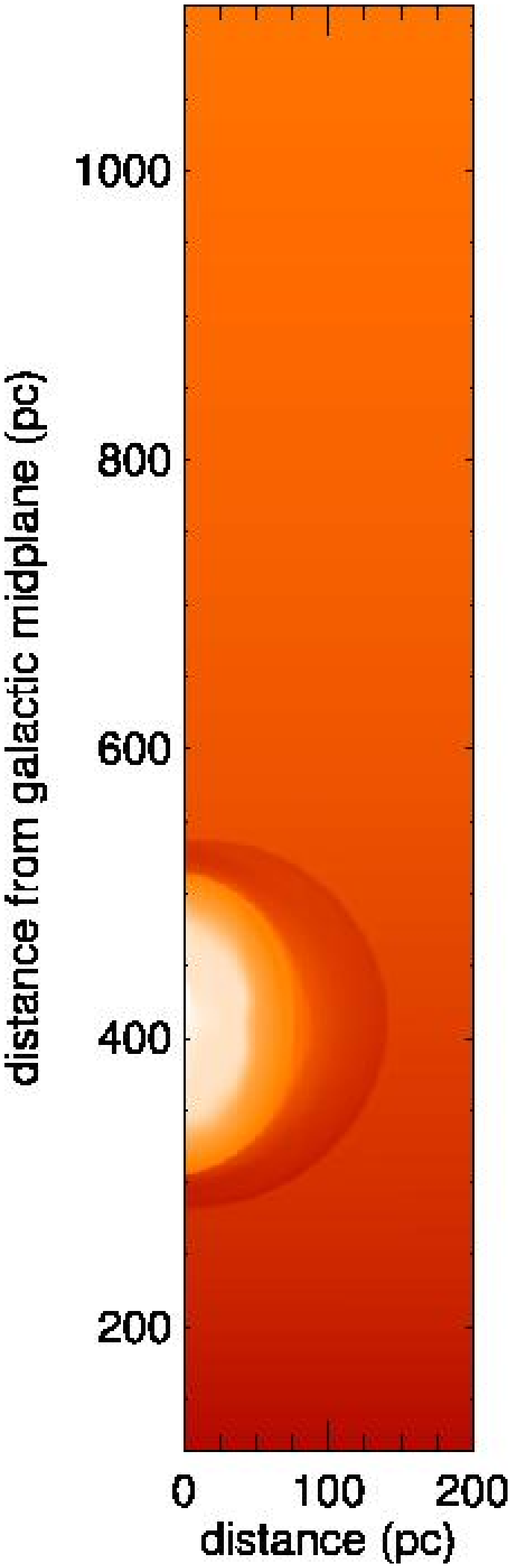}\hspace{0.5mm}\includegraphics[height=0.4\textwidth, angle=0]{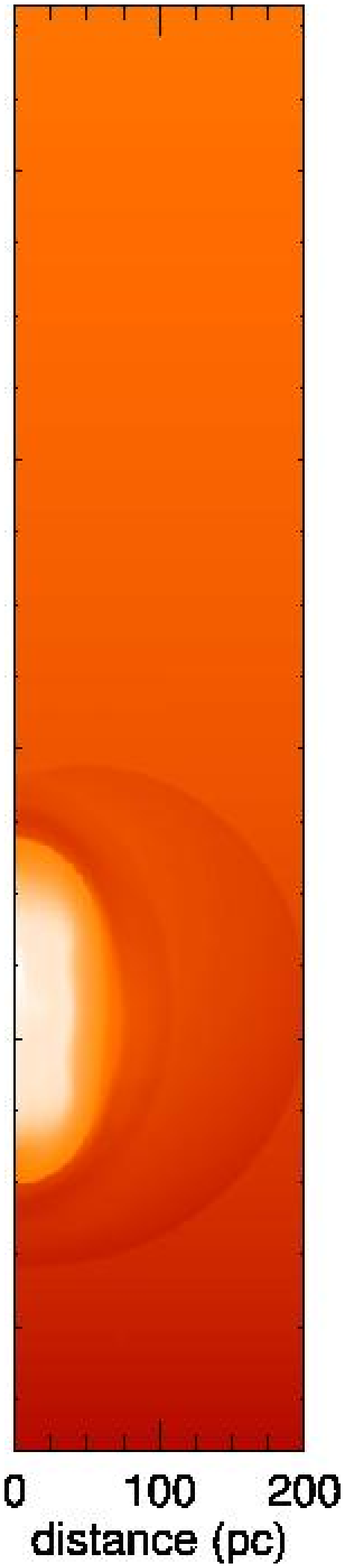}\hspace{0.5mm}\includegraphics[height=0.4\textwidth, angle=0]{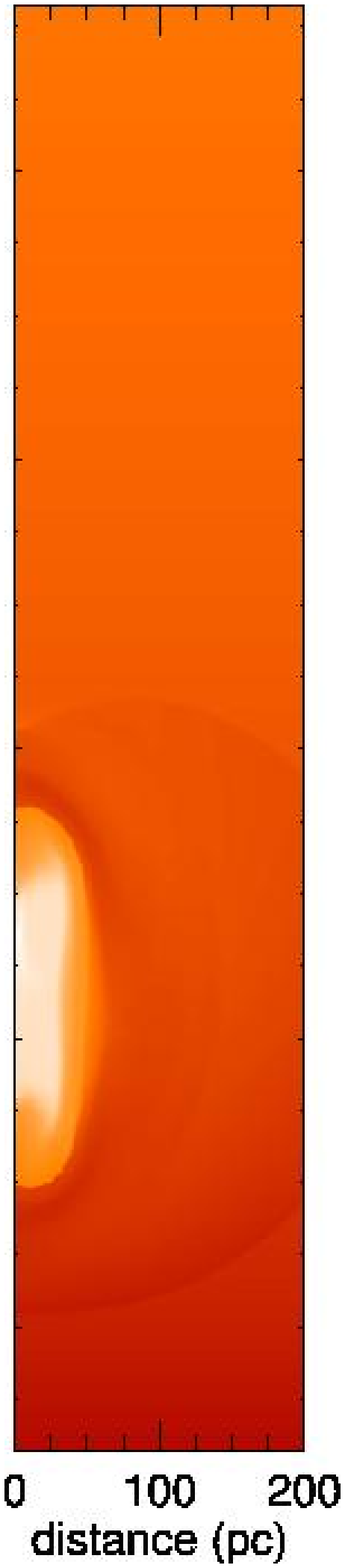}\hspace{0.5mm}\includegraphics[height=0.4\textwidth, angle=0]{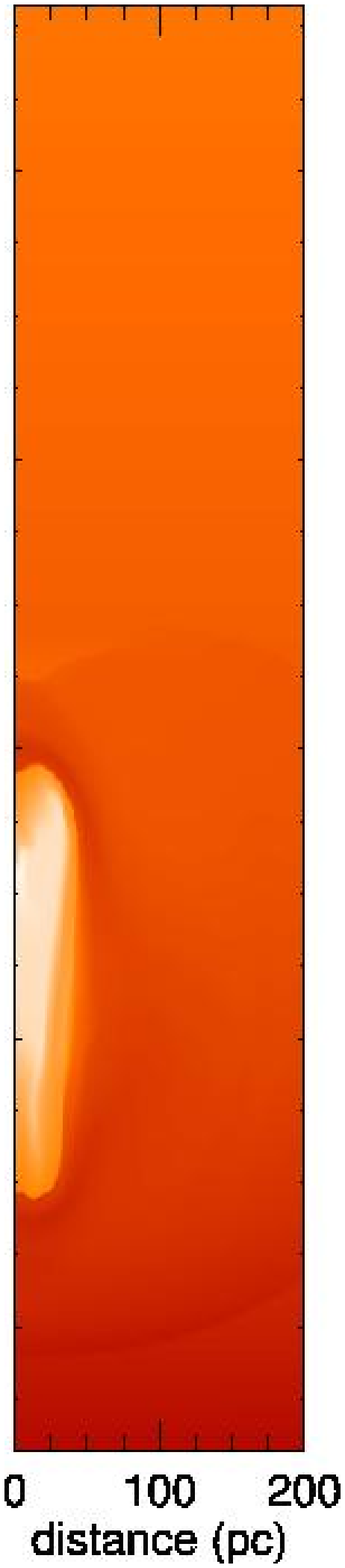}\includegraphics[height=0.4\textwidth, angle=0]{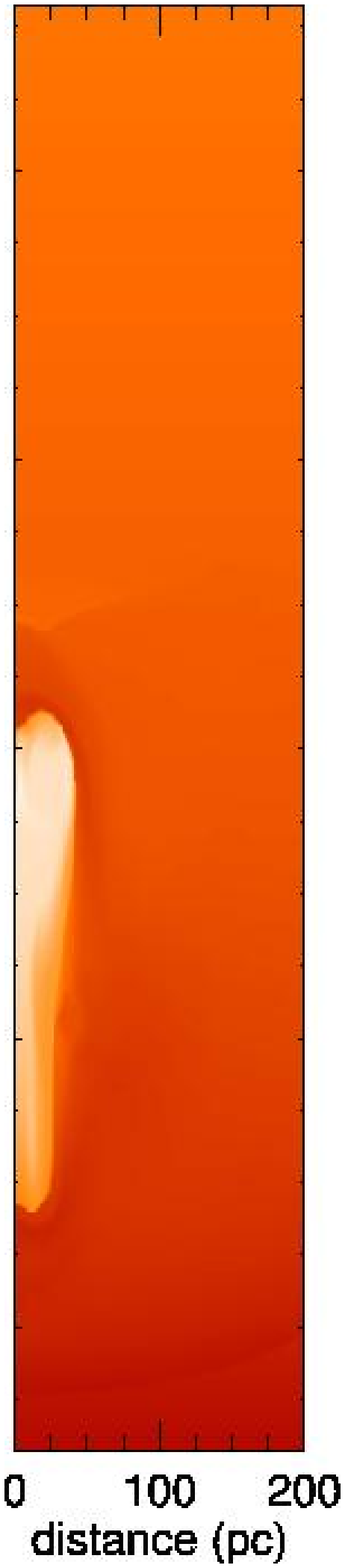}\hspace{0.5mm}\includegraphics[height=0.4\textwidth, angle=0]{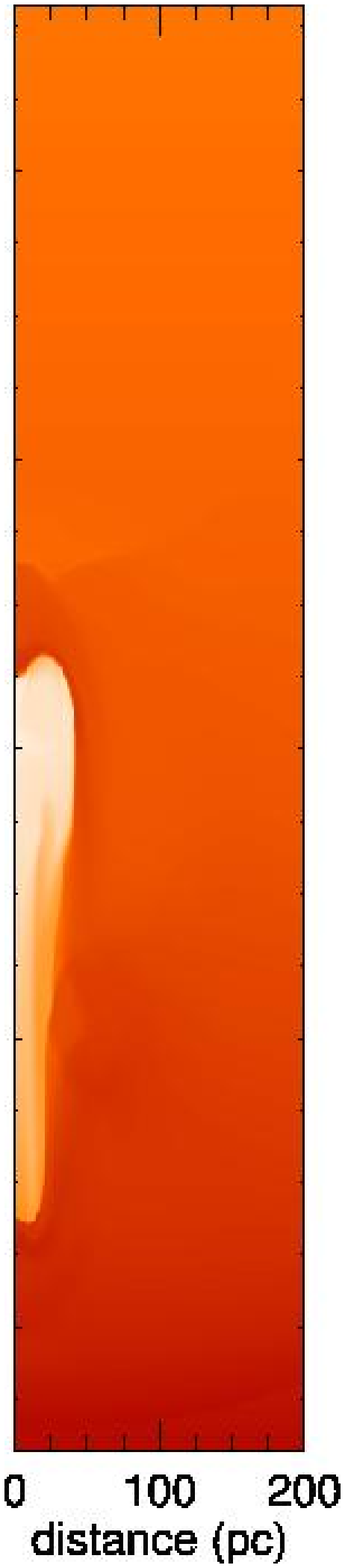}\hspace{0.5mm}\includegraphics[height=0.4\textwidth, angle=0]{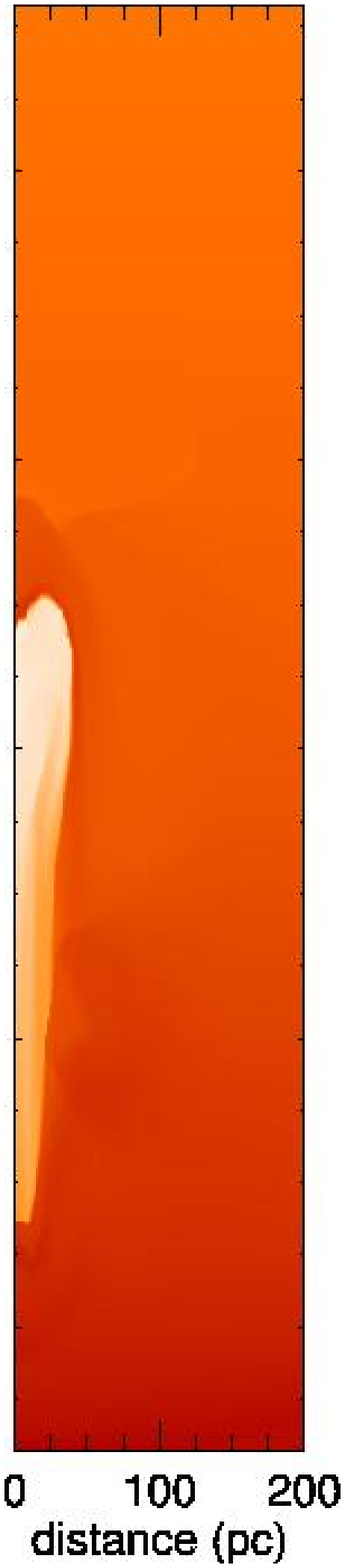}\hspace{0.5mm}\includegraphics[height=0.4\textwidth, angle=0]{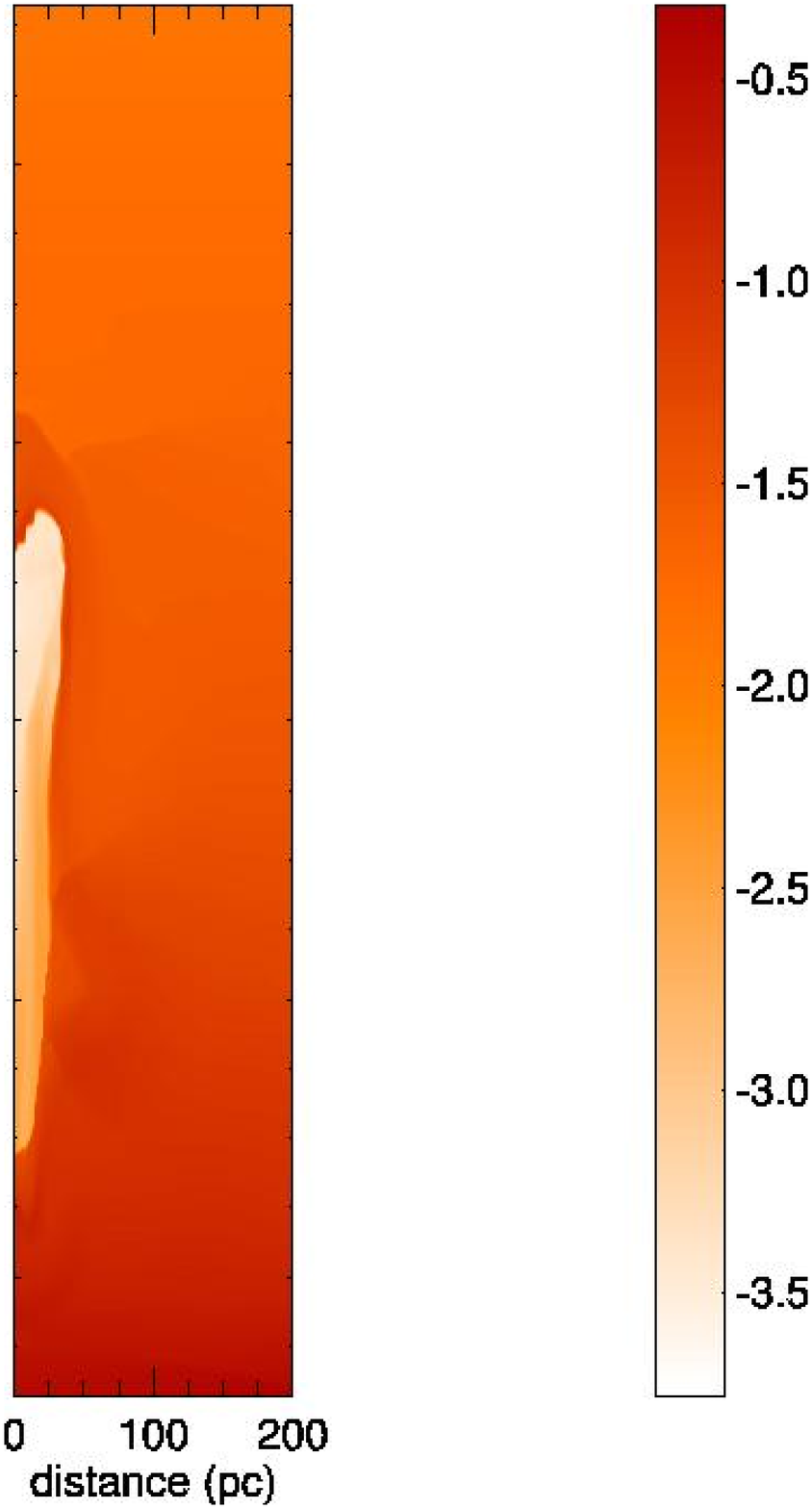}\hspace{0.5mm}

\caption{Time sequence of density (atoms/$\rm{cm}^3$) plots for Model C.  The log(density) is shown (left to right) every million years up to 8 Myrs. The center of the bubble is located at 518~pc at 8 Myrs. The top of the bubble reaches $\sim$ 760~pc at 8 million years.These are slices through x=0 (y-z plane).}
\label{z_mhd_dens}
\efig 

\section{Discussion}
\subsection{Galactic Shear}
Because we simulate SNRs until they are several million years old, it is interesting to consider what effects galactic shear may have on the remnants.  Given the difference in Galactic rotation speed across the remnant, we calculated the maximum distance that galactic shear could stretch the SNR and compare it with the size of the SNR.  Let us first consider our simulation without a magnetic field.  For this simulation, the maximum bubble diameter is about 230~pc, which occurs at 2 Myrs.  From the galactic rotation curves of \citet{Maciel_Lago} we see that the greatest possible difference in galactic rotation velocity (considering the scatter between observed data points) from one side of our remnant to the other would be 20 km/s ($\approx$~20~pc/Myr).  Because our bubble consistently shrinks after 2 Myrs and is very small at late times, we will only consider the shear for the first 8 Myrs and approximate that the bubble stays the same size.  From this we find that, at most, galactic shear would stretch our bubble $(20 {\frac{\rm{pc}}{\rm{Myr}}})\times 8 \rm{Myrs}= 160 \rm{pc}$, which is much smaller than the diameter of our bubble.  Thus, the differential galactic rotation would not tear apart the bubble.

Of our simulated remnants, the one having the greatest width in the direction parallel to the plane is model B.  Its maximum width is about 500~pc. Because of the flatness in the galactic rotation curve, over this distance a liberal estimate of the difference in velocity between the sides of the bubble would still be $\approx$ 20~pc/Myr \citep{Maciel_Lago}.  So, we find that at most the galactic shear would stretch this bubble $(20~ {\frac{\rm{pc}}{\rm{Myr}}})\times 8~ \rm{Myrs}= 160~ \rm{pc}$, which is much smaller than the 500~pc width of the bubble.  Therefore, we expect that a bubble would be disturbed by the shear but not torn apart since the upper limit on the shear is still smaller than the bubble.

\subsection{Buoyancy}
Even for the simulation without a magnetic field, the center of the bubble only rose 59~pc from the initial explosion height.  Adding a magnetic field of 4$\mu$G parallel to the mid-plane pins the cool shell around bubble down and lessens its vertical rise to 23 pc above the explosion height.  However, if the ambient magnetic field is directed perpendicular to the mid-plane then the vertical distribution of the bubble increases.  For this case we see hot gas as high as 760~pc above the galactic mid-plane at 8~Myrs, though the explosion went off at 400~pc.  The center has risen a maximum of about 179~pc at 8~Myrs.  

\begin{figure}
\begin{center}
\includegraphics[width=0.5\textwidth, angle=0]{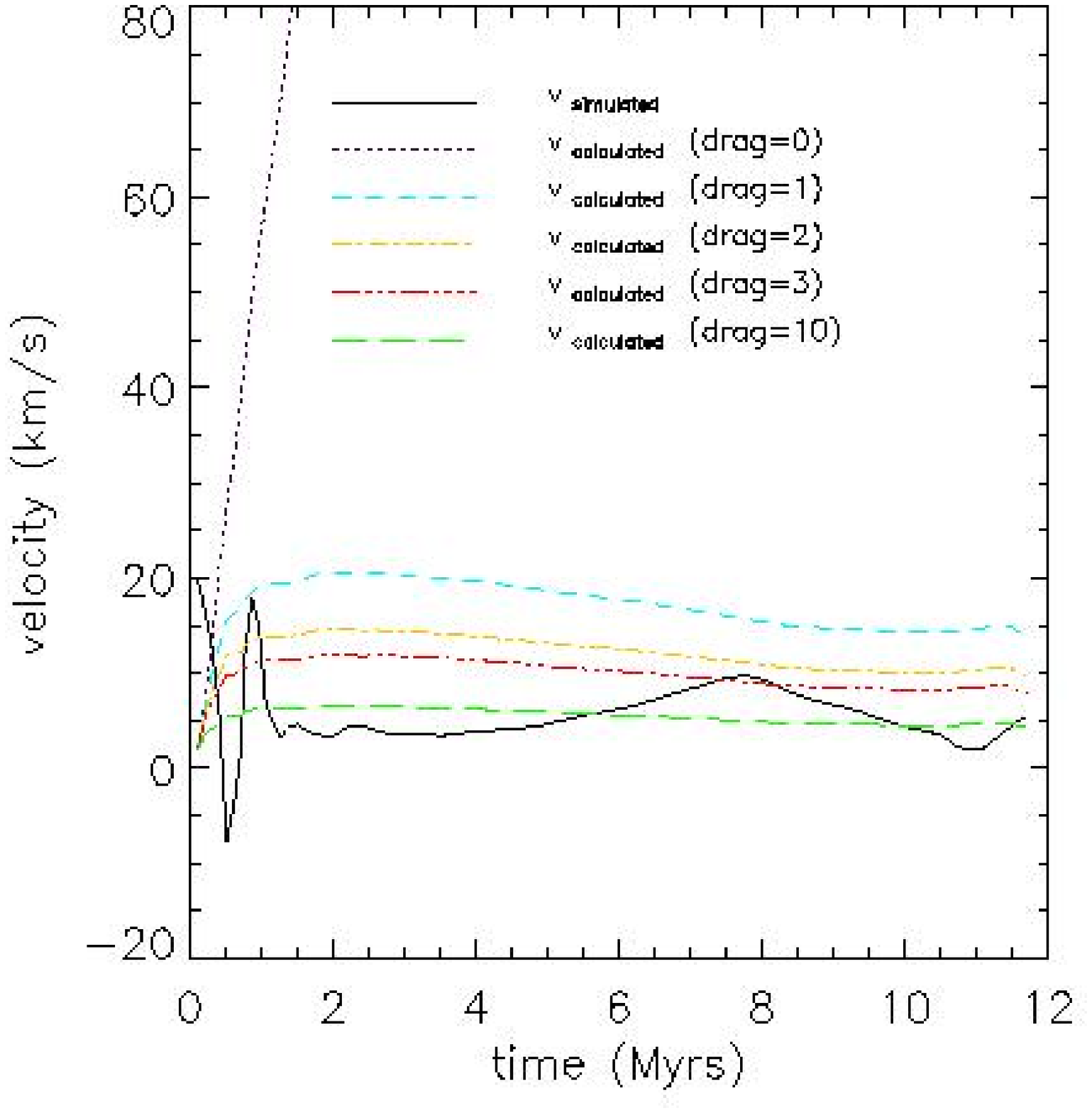}
\caption{The bubble velocity from the Model A simulation (black, solid line) is shown as are the velocities calculated from equations \ref{buoy1} and \ref{buoy2} for various drag coefficients.  }
\label{buoyancy}
\end{center}
\end{figure}

Without a magnetic field to restrain it, Model A could have been expected to rise more than was observed.  Buoyancy, preferential expansion in the direction of lower density and pressure, and faster cooling on the side of the remnant having higher density would have conspired to increase the height of the center of the remnant's hot bubble.  Drag however would act to slow the bubble's motion.  
We estimate the effective drag on the hot bubble by comparing the  simulated (volume-weighted) average velocity of the bubble to the expected velocity of the bubble given various drag coefficients. We calculate the change in the velocity of the bubble per small timestep ($dt$) by adding the acceleration due to buoyancy (${\bf{a}}_{buoy}$) and the drag acceleration (${\bf{a}}_{drag}$) and then multiplying by $dt$.  We use the standard equations for buoyant and drag accelerations:
\begin{equation}
{\bf{a}}_{buoy}={\bf{g}}\left({1-{\frac{\rho_a}{\rho_b}}}\right)
\label{buoy1}
\end{equation}

\begin{equation}
{\bf{a}}_{drag}={\frac{1}{2}}CA_b {v}_b^2 \rho_a /m_b 
\label{buoy2}
\end{equation}
where $g$ is the gravitational acceleration, given by equation \ref{gravity}, at the hot bubble's average height above the midplane, $\rho_a$ is the ambient density at the height of the hot bubble, $\rho_b$ is the average density of the hot bubble, $v_b$ is the average velocity of the hot bubble, $A_b$ is the maximum cross-sectional area of the hot bubble, $m_b$ is the mass of the hot bubble and $C$ is the drag coefficient.  ${\bf{a}}_{drag}$ is oriented in the opposite direction as the bubble's velocity.  We approximate the maximum cross-sectional area of the bubble by calculating the average cross-sectional area and multiplying by 1.5 (eg. we make the approximation that the bubble is spherical), $A_{b_{max}}=1.5\times A_{b_{ave}}$. 
We calculate the expected velocity of the bubble without drag and for various drag coefficients.
Figure \ref{buoyancy} shows the results of our buoyancy calculations for the SNR born at 400~pc above the midplane without a magnetic field, Model A.  We plot the average velocity of our bubble along with the expected velocity due to buoyancy and drag for several drag coefficients.  \citet{jones1996} suggests from simulations that the drag coefficient for hot bubbles in the ISM may be around one.  However, our results, shown in Figure \ref{buoyancy}, suggest that the drag coefficient may be as high as ten for our model without a magnetic field, indicating that there is much less buoyancy than expected.  

\subsection{Mushroom-Shaped Clouds}
The existence of mushroom-shaped clouds outlined by high density gas was suggested by \citet{English}, and discussed again in \citet{avillez_maclow}.  The simulations of \citet{English} showed that the cool dense gas surrounding a remnant evolving in a non magnetized medium, developed
 a mushroom shape at 8 Myrs.  In our simulations of a remnant evolving in a non magnetized medium, we see
 a mushroom shaped cloud begin to develop at 9 Myrs.  
A shell of cool dense gas outlines the mushroom cap and is filled with hot rarefied gas.  The mushroom stem is made up of cool, dense gas.   A further developed 
structure is seen at 10 Myrs.  As the bubble ages past 10 Myrs the high-density, 
low-temperature 'stem' is pushed up through the cap of the mushroom leaving behind a torus of hot gas. 
For Model B the bubble resembles and elongated mushroom cap (in the y-z plane), and the stem is less apparent.  Model C resemble a long tubes that is thicker in the direction away from the midplane; there is no resemblance to a mushroom cloud.  

\subsection{Bubble Shape for Different $\bf{B}$ Strengths}
As the magnetic field increases, the bubble gets more elongated in the direction parallel to the magnetic field.  For Model B, when $\bf{B}_y$=4$\mu$G the bubble has a slight mushroom shape cross section, (the bubble is lower on the edges than at the center in the $\hat{y}$-direction.  When $\bf{B}_y$ is increased to 7.1 $\mu$G, the bubble is further elongated and the edges are tapered.  Figure \ref{By strength} shows the temperature for y-z slices for Models having no magnetic field (left), a magnetic field directed parallel to the midplane with a strength of 4 $\mu$G (center), and a magnetic field directed parallel to the midplane with a strength of 7.1 $\mu$ G (right).  For Model C, when $\bf{B}_z$=4$\mu$G the bubble is fuller in the direction away from the midplane (also the direction of decreasing thermal pressure and density).  When $\bf{B}_z$ is 7.1 $\mu$G  the bubble is even more elongated, and it is tapered at the ends. Figure \ref{Bz strength} shows y-z temperature slices for the models having no magnetic field (left), a magnetic field perpendicular to the midplane of 4 $\mu$G (center), or a magnetic field perpendicular to the midplane of 7.1 $\mu$ G (right).
\bfig
\centering
\includegraphics[height=0.3\textwidth, angle=0]{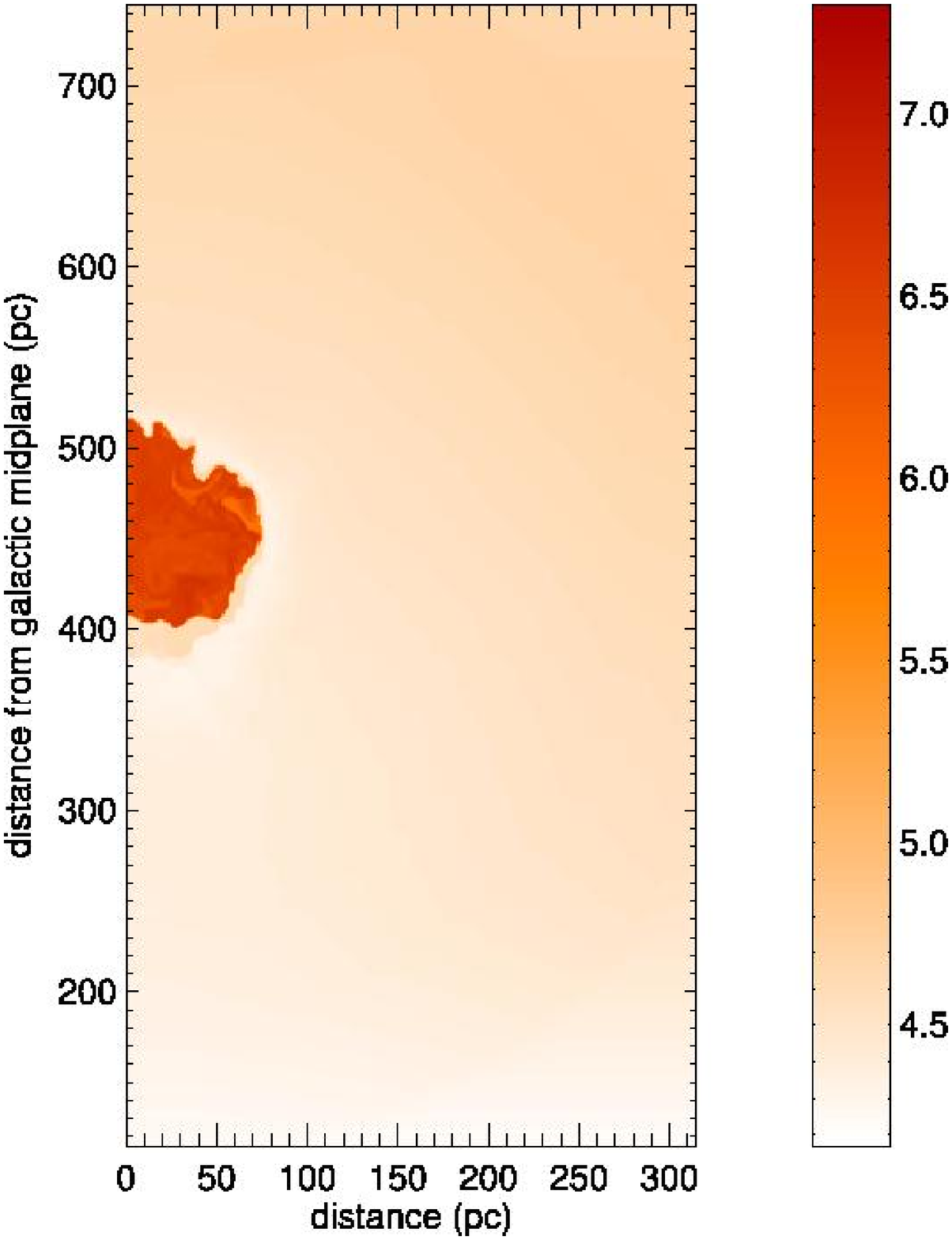}\hspace{0.5in}\includegraphics[height=0.3\textwidth, angle=0]{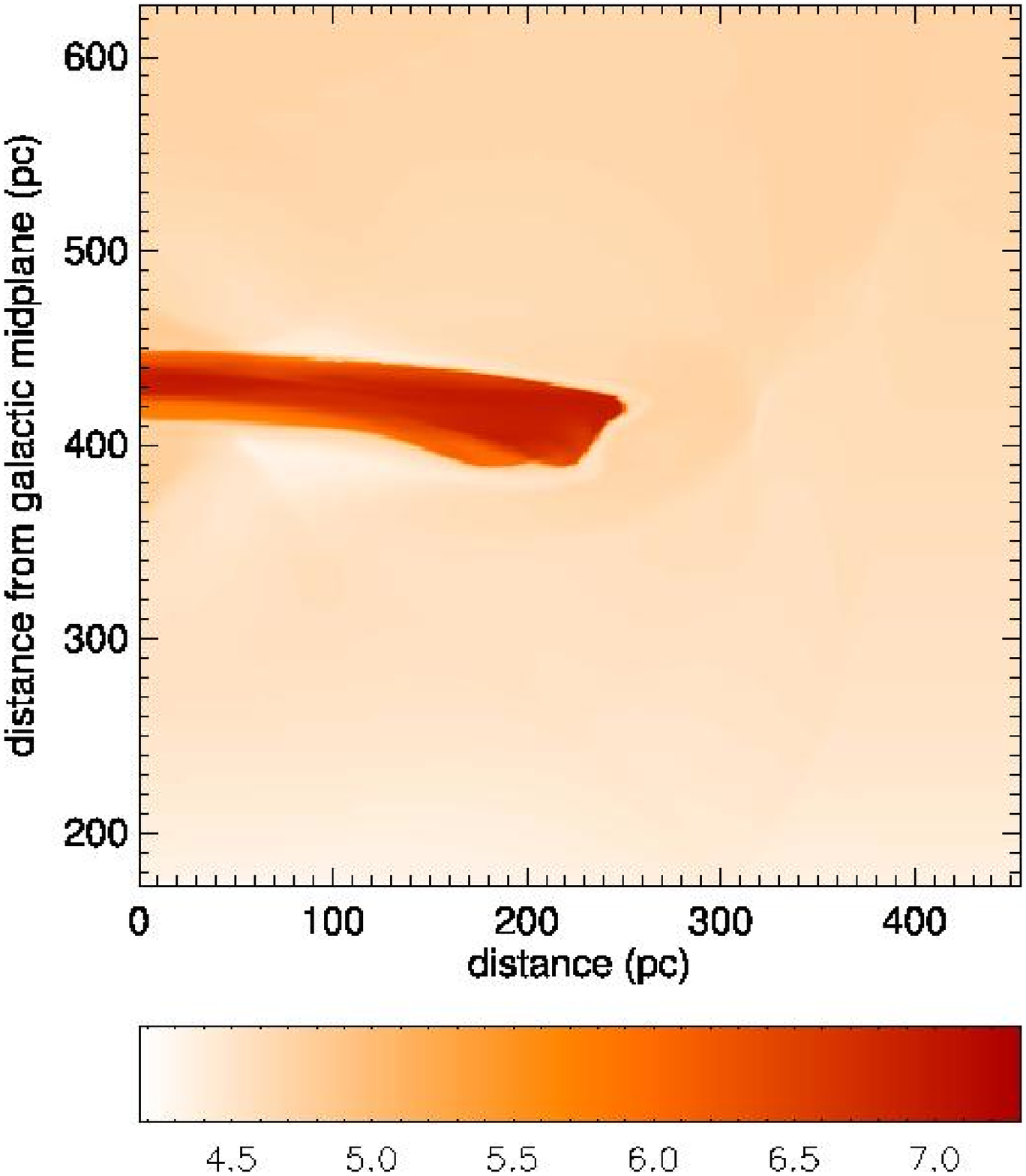}\includegraphics[height=0.3\textwidth, angle=0]{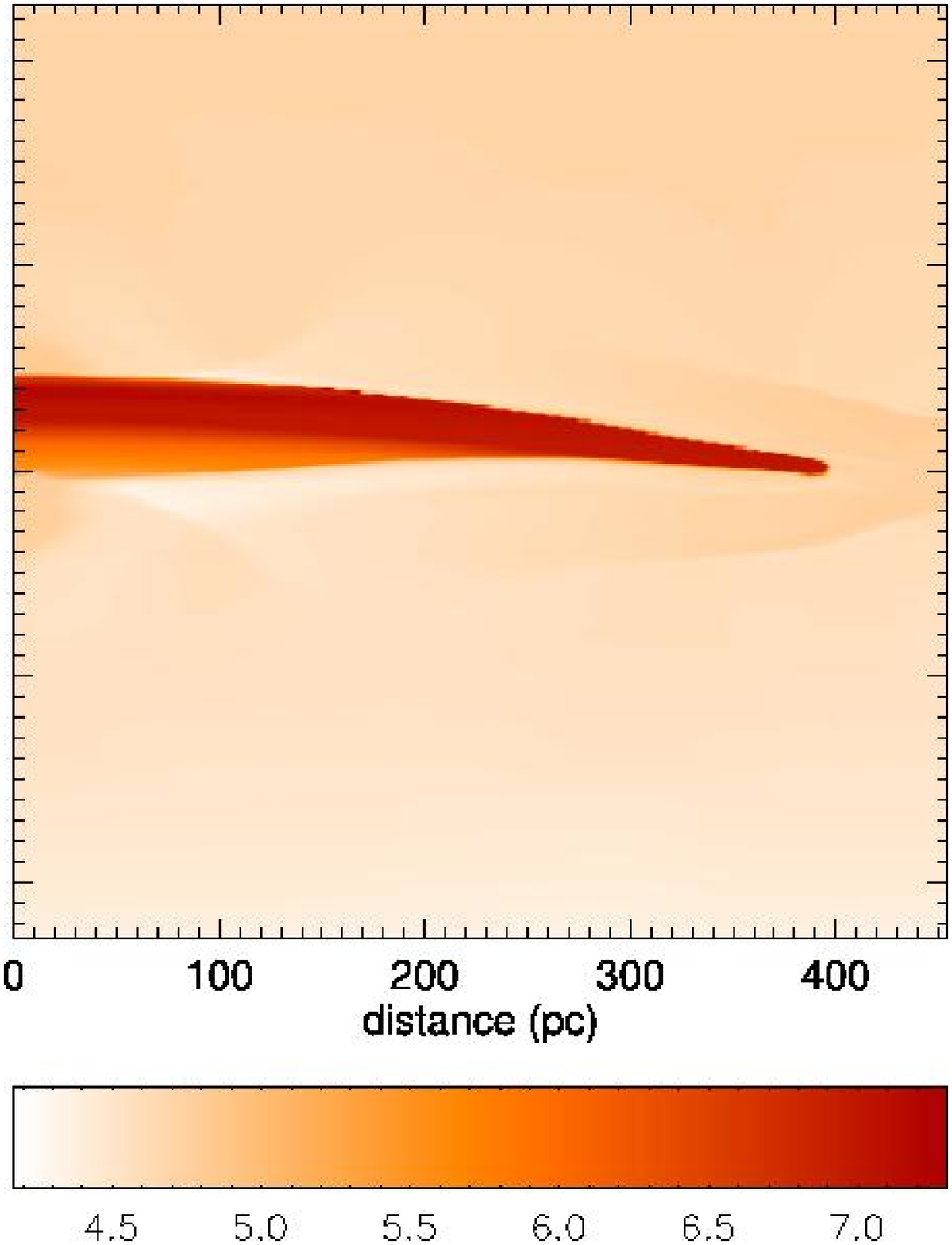}

\caption{Y-z slices showing temperature for Models A, B, and like model B, but with $\bf{B}_y$=7.1$\mu$G, at 8 Myrs.  Left: With no magnetic field, the bubble starts to form a mushroom cap shape by 8 Myrs.  Center: With a magnetic field of 4 $\mu$G, the bubble is longer and thinner, but has the rough shape of an elongated mushroom cap (bubble is lower at the edges then in the center).  Right: With a magnetic field of 7.1 $\mu$G, the bubble has a long shape that is tapered at the end; there is no resemblance to a mushroom cap.  }
\label{By strength}

\efig
\bfig
\centering
\includegraphics[height=0.3\textwidth, angle=0]{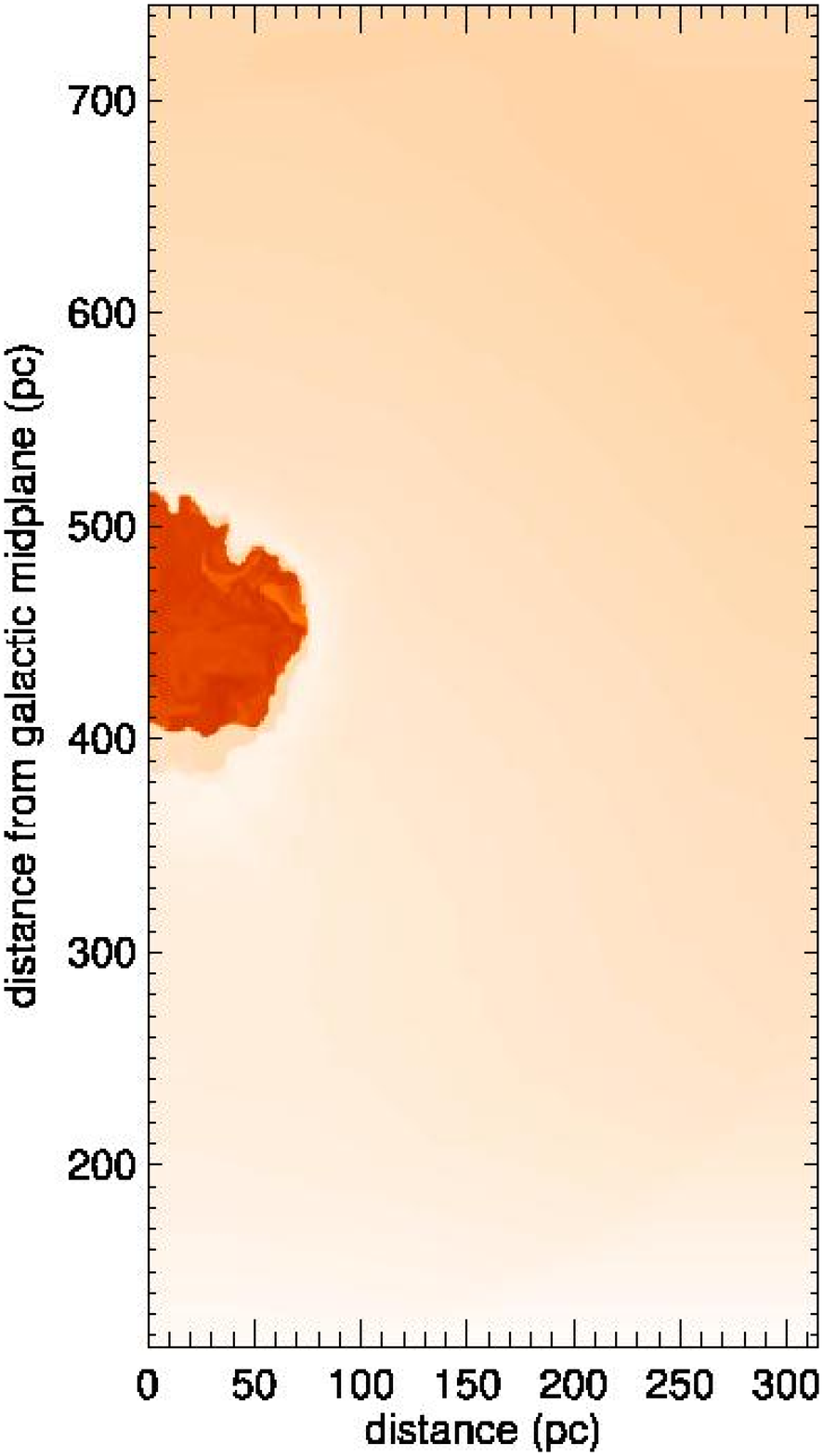}\hspace{3.mm}\includegraphics[height=0.3\textwidth, angle=0]{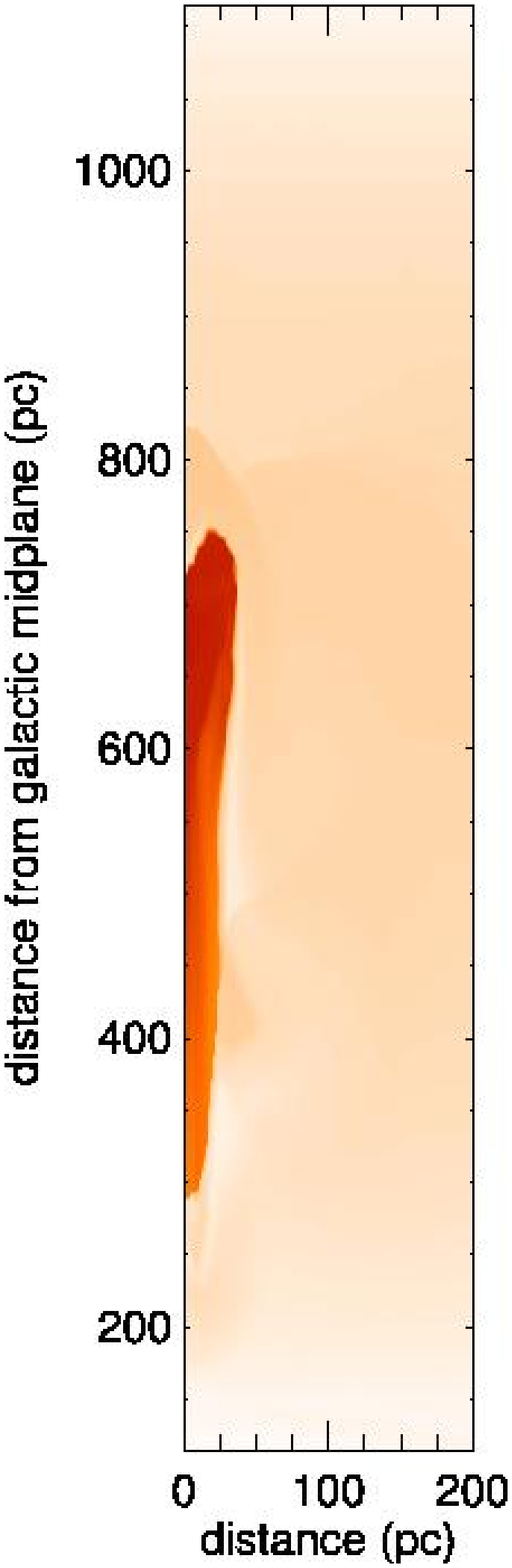}\hspace{3.mm}\includegraphics[height=0.3\textwidth, angle=0]{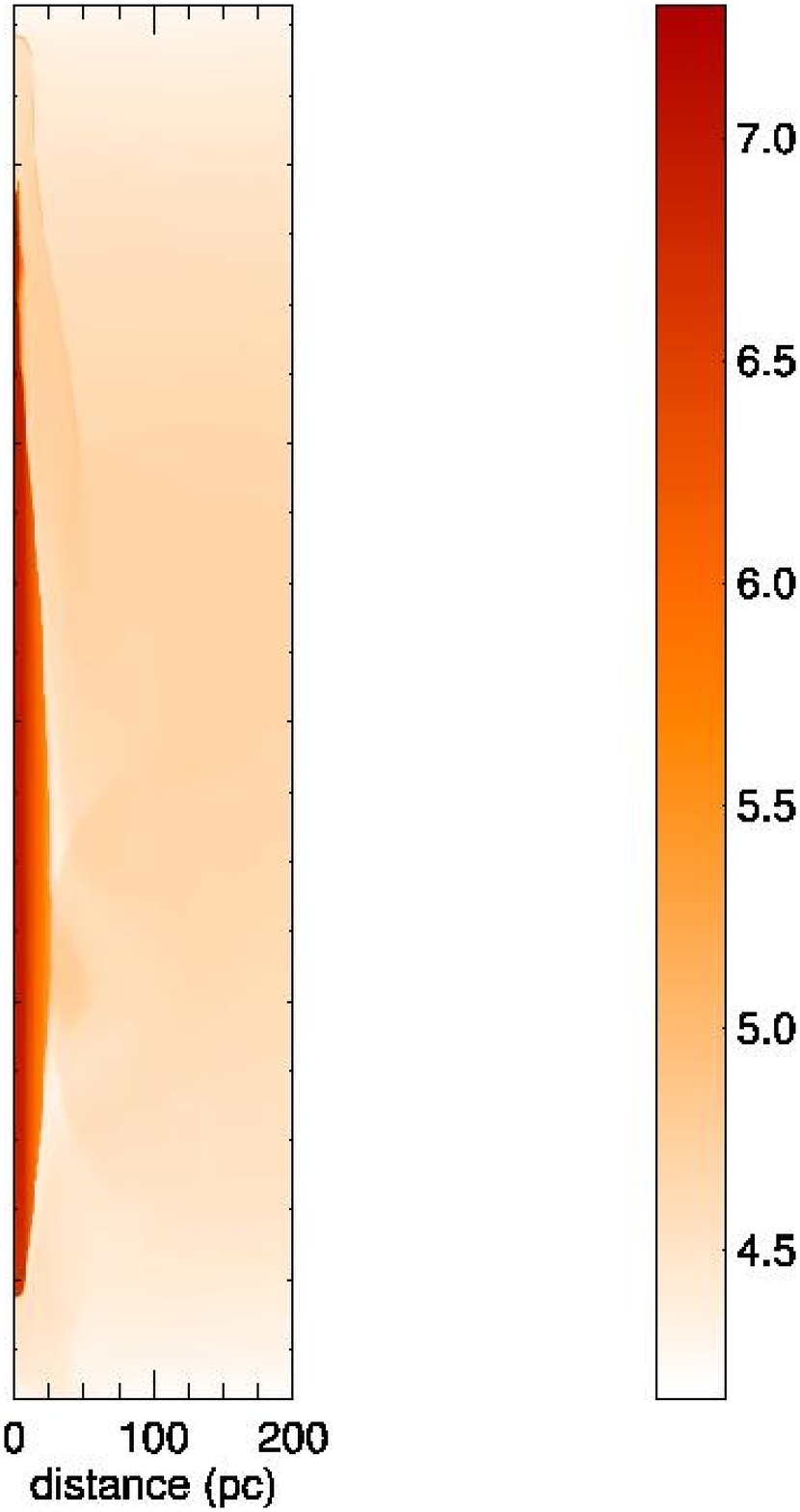}
\caption{Y-z slices showing temperature for Models A, C, and like model C, but with $\bf{B}_z$=7.1$\mu$G, at 8 Myrs. Left: With no magnetic field, the bubble forms a mushroom-cap shape.  Center: With a magnetic field of 4$\mu$G perpendicular to the midplane, the bubble is elongated vertically, and fuller at the top of the bubble (in the direction of decreasing thermal pressure and decreasing ambient density).  Right:  With a magnetic field of 7.1 $\mu$G the bubble is more elongated and is tapered at the ends. }
\label{Bz strength}
\efig

\subsection{Caveats}
In the Milky Way, the strength of the ordered component of the magnetic field decreases slightly with distance from the galactic mid-plane.  The resulting gradient in magnetic pressure would allow the bubble to expand preferentially away from the galactic mid-plane.  Computationally, the gradient in magnetic field strength gives rise to a gradient in the magnetic pressure that would be taken into consideration when calculating the ambient thermal pressure for HSE, effectively allowing us to use a temperature distribution having less variation. Furthermore, \cite{tomisaka98} found that in his superbubble simulations the variation with height of the magnetic field strength was important to whether the superbubble could ``blow out'' of the disk. 
			
The magnetic field of the galaxy contains random components which are significant with respect to the average field strength.  The random component of the magnetic field in external galaxies is slightly larger than the mean galactic field and contains about one and a half times the energy \citep{Zweibel and Heiles}.  Adding a random magnetic field would allow for the possibility that gas might be able to escape through regions of lower magnetic field. 
Adding a random magnetic field would also reduce the magnetic tension since the initial magnetic field lines would be longer, and would increase the magnetic pressure in all directions which could change the structure of the bubble and the direction in which the gas preferentially moves.  

\section{Summary}
We have examined three magnetic field backgrounds for our SNR explosion born 400 pc above the galactic midplane.  For the case of no magnetic field we see cauliflower-like eddies develop within the bubble in the first several million years.  We see a modest rise of 59 pc over the 12 Myr simulation time.  A mushroom structure forms by 9 Myrs, with the stem made of cool low, density material and a cap outlined in cool, low density material and filled with hot, low density material.  When we add a magnetic field of strength 4$\mu$G directed parallel to the midplane, we find that the bubble becomes elongated in the direction of the field.  The bubble still somewhat resembles a mushroom cap at late times.  There is no cauliflower-like structure in y-z slices of the bubble, but there is cauliflower-like structure in the x-z plane (perpendicular to $\bf{B}$.  The vertical rise of the bubble is reduced to 23 pc above the explosion height.  When our magnetic field strength is 4 $\mu$G and the orientation is perpendicular to the midplane, the bubble is elongated vertically.  It resembles a long tube that increases in thickness with distance from the midplane.  It does not resemble a mushroom shaped cloud.  Using an ambient magnetic field that is perpendicular to the midplane increases the bubble's vertical rise to 179 pc.

\subsection{Acknowledgments}

The authors would like to thank Asif ud-Doula for his help and advice with ZEUS-MP, Timur Linde for his help with the MHD portion of FLASH, and David Henley for his physical insight.  Thanks also to the FLASH users and staff who respond to the mailing list.

The software used in this work was in part developed by the DOE-supported ASC / Alliance Center for Astrophysical Thermonuclear Flashes at the University of Chicago.

This research has been supported by 
the University of Georgia and 
NASA, through grant NAG5-NNGO4GD78G.

\begin{center}
  {\bf Appendix : MHD Reflective Boundary Conditions}
\end{center}
\begin{appendix}

As an exercise in setting the magnetic reflective boundary conditions, we describe the most general case, that having symmetry across the x-z, y-z, and x-y planes.  Such conditions would be appropriate if there were no gravitational or density gradients in the z-direction.

\begin{figure}
\begin{center}
\includegraphics[width=0.4\textwidth, angle=0]{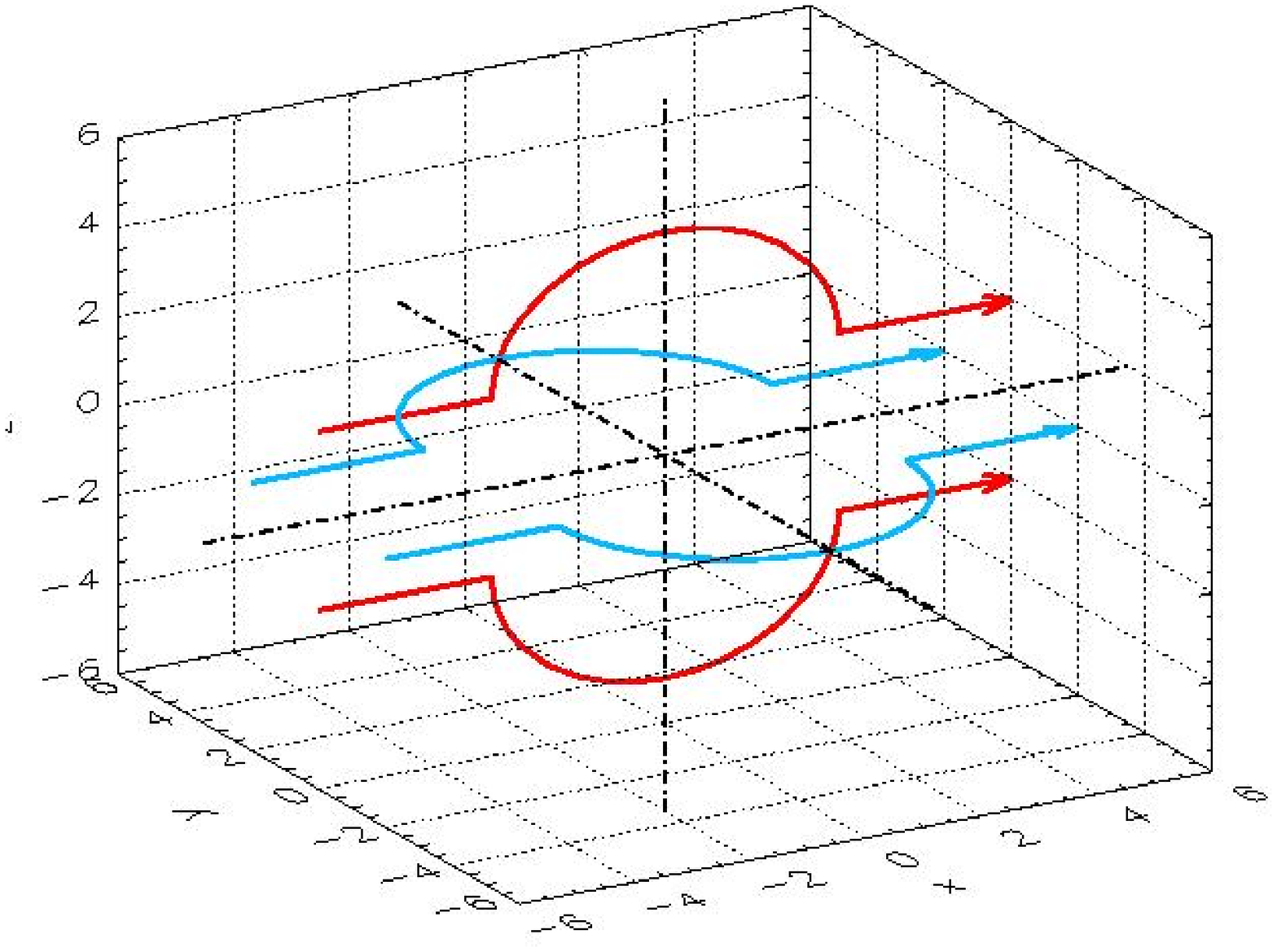}
 \caption{This schematic showing magnetic field lines, initially in the positive x-direction, bending around a whole SN remnant demonstrates the magnetic field's behavior at reflecting boundaries. Consider that one might want to simulate one eighth of this remnant, specifically the upper back right octant. The y-component of the magnetic field changes sign across the $y=0$ (x-z plane) boundary, and the z-component of the magnetic field changes sign across the $z=0$ (x-y plane) boundary.  The x-component does not change sign when crossing these boundaries.}
\label{bfield1}
\end{center}
\end{figure}

Consider Figure \ref{bfield1}, a schematic of space threaded by magnetic field lines.  A SN explosion at (x,y,z)=(0,0,0) has pushed the field lines (originally in the $+\hat{x}$ direction) away from the center.  Due to symmetries only one eighth of this space need be modeled with a magnetohydrodynamic code.  We will consider the octant with positive x,y, and z to be the simulation region.  It is subdivided into rectilinear volumes called ``real cells''.  Because the MHD calculations require knowledge of the physical conditions in the space neighboring the real cells, the codes also tracks a layer or a couple of layers of cells just outside each of the simulation region's planar boundaries.  These are called ``ghost cells''.  If we wish to simulate only the portion of the grid where x,y, and z are positive, then we must determine the conditions in the ghost cells.

The temperature and density distributions are symmetric across the x=0, y=0, and z=0 planes, so the ghost cells outside of these boundaries are set to the temperature and densities of the first real cells inside of these boundaries.  The boundary conditions for the magnetic field are not as simple.
Let us first consider the magnetic field along the x-z plane (red field lines in Figure \ref{bfield1}).   We see that when the z-component of the top field line is positive, the z-component of the bottom field line is negative.  The x-component of both field lines is always positive. 

Let $B_x(i,j,k)$ and $B_z(i,j,k)$ denote the magnitude of the x and z components of the magnetic field in the i, j, kth cell, i.e. in the cell whose indices in the $\hat{x}$, $\hat{y}$ and $\hat{z}$ directions are i, j, and k respectively.  Let * denote all values of a particular index.  Let $\mathcal{K}$ denote the real cell abutting the inner z boundary, (the first real cell along the $\hat{z}$-direction) and let $\mathcal{K}'$ denote the ghost cell abutting the inner z boundary (the first ghost cell along the $\hat{z}$-direction.)  Then 
\begin{eqnarray}
B_z(*,*,\mathcal{K}')=-B_z(*,*,\mathcal{K})\\
B_x(*,*,\mathcal{K}')=B_x(*,*,\mathcal{K}) \label{two}.
\end{eqnarray} 
When simulating magnetized plasmas, the FLASH code typically uses 2 ghost cells.  Thus, for the second ghost cell,
\begin{eqnarray}
B_z(*,*,\mathcal{K}'-1)=-B_z(*,*,\mathcal{K}+1)\\
B_x(*,*,\mathcal{K}'-1)=B_x(*,*,\mathcal{K}+1).
\end{eqnarray}
It is common for FLASH to number the cells such that k=1, 2, 3, and 4 denote the second ghost cells, first ghost cells, first real cells, and second real cells, respectively.  The boundary between the ghost and real cells lies between the k=2 and k=3 cells.  The same logic applies to the inner x and inner y boundaries as well.  More generally, $\mathcal{K}=2G-\mathcal{K}'$ where $G$ is the boundary between the real and ghost cell.  For example, for an inner reflective boundary with 2 ghost cells G would be 2.5 (halfway between the last ghost zone, 2, and the first real zone, 3).  This notation also works for an outer reflective boundary which we have not used.

From the x-z plane we can also look at the inner $\hat{x}$-boundary (i.e. the x=0 plane).  We can see that the z-component of the field lines changes sign on either side of $x=0$, and that the x-component keeps the same sign.  This translates into boundary conditions as:

\begin{eqnarray}
B_z(\mathcal{I}',*,*)=-B_z(\mathcal{I},*,*)\\
B_x(\mathcal{I}',*,*)=B_x(\mathcal{I},*,*), \label{six}
\end{eqnarray}
where $\mathcal{I}$ indicates a real cell and $\mathcal{I}'$ indicates the corresponding ghost cell.

Now let us look at the x-y plane (blue field lines in Figure \ref{bfield1}).  Notice that the y-component of the field line has the opposite sign when it is on the opposite side of the $y=0$ boundary.  The x-component is always positive.  This would give the following boundary conditions for the y-boundary.
\begin{eqnarray}
B_y(*,\mathcal{J}',*)=-B_y(*,\mathcal{J},*)\\
B_x(*,\mathcal{J}',*)=B_x(*,\mathcal{J},*),
\end{eqnarray}
where $\mathcal{J}$ indicates a real cell and $\mathcal{J}'$ indicates the corresponding ghost cell.
We can also examine the $x=0$ boundary by examining the lines plotted in the x-y plane.  The y-component changes signs across the boundary, and the x-component stays the same across the boundary, yielding the following boundary conditions:
\begin{eqnarray}
B_y(\mathcal{I}',*,*)=-B_y(\mathcal{I},*,*)\\
B_x(\mathcal{I}',*,*)=B_x(\mathcal{I},*,*).  \nonumber
\end{eqnarray}
The $B_x(\mathcal{I}',*,*)$ condition is the same as in the x-z plane (equation \ref{six}) as expected.

\begin{figure}
\begin{center}
\includegraphics[width=0.4\textwidth, angle=0]{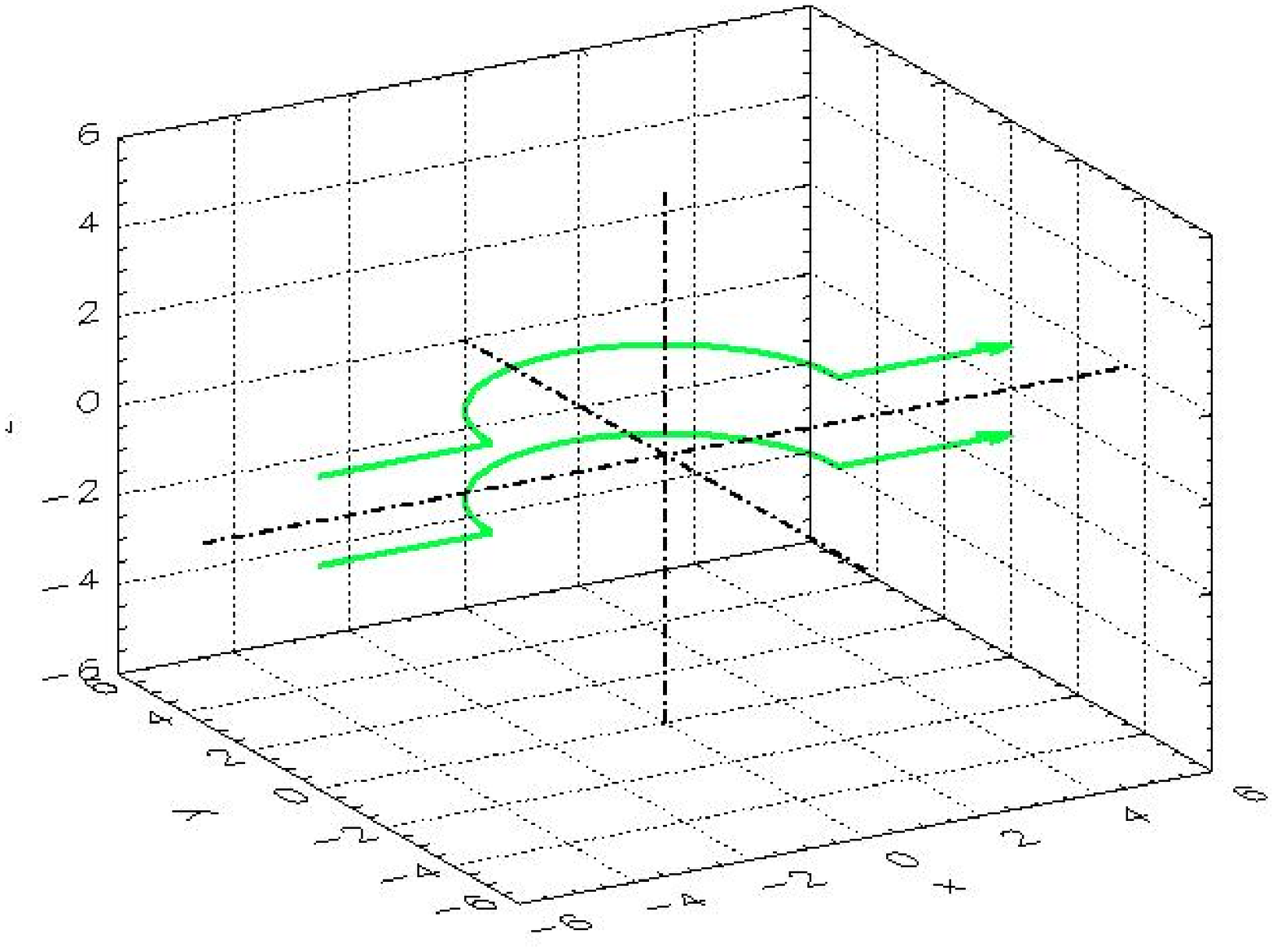}
\caption{This schematic showing magnetic field lines initially in the positive x-direction bending around a SNR, demonstrates that the y-component of the magnetic field lines keeps the same sign as it crosses the $z=0$ boundary.}
\label{bfield2}
\end{center}
\end{figure}

Next, consider how the y-component changes across the $z=0$ boundary.  As can be seen in Figure \ref{bfield2}, the x-component and the y component stay the same on either side of the z-boundary.  This gives the following boundary conditions:
\begin{eqnarray}
B_y(*,*,\mathcal{K}')=B_y(*,*,\mathcal{K})\\
B_x(*,*,\mathcal{K}')=B_x(*,*,\mathcal{K}).  \nonumber
\end{eqnarray}
We again get the same condition for $B_x(*,*,\mathcal{K}')$ as in equation \ref{two} as expected.
By analogy, the z-component does not change across the $y=0$ boundary, which gives us our last boundary condition:
\begin{equation}
B_z(*,\mathcal{J}',*)=B_z(*,\mathcal{J},*).
\end{equation}
In summary, in the ghost cells along each inner boundary, the (magnitude of the) component of the field that is parallel to the initial ambient field should equal that in the corresponding real cells.  A magnetic field component that is perpendicular to the initial ambient field will change signs between the real and ghost cells only when the boundary is in the direction of the initial ambient field (eg. the x=0 (y-z plane) boundary when the field is in the $\hat{x}$-direction) or if the boundary is the same direction as the component itself (eg. the z=0 (x-y plane) boundary for the z-component).

\end{appendix}

\end{document}